\pdfoutput=1
\documentclass[journal]{IEEEtran}
\usepackage[utf8]{inputenc}

\usepackage{array}
\usepackage{amsthm,amsmath,amssymb}
\usepackage{multirow}
\usepackage{threeparttable}
\usepackage{booktabs}
\usepackage{enumerate}
\usepackage{cite}
\usepackage{graphicx}
\usepackage[table]{xcolor} 
\usepackage{colortbl}
\usepackage{notation}
\usepackage{graphicx}
\usepackage[caption=false,font=footnotesize]{subfig}

\newcommand{\rmv}{\hspace*{-.3mm}}

\newcommand{\nn}{\nonumber}

\DeclareMathAlphabet{\mathpzc}{OT1}{pzc}{m}{it}

\makeatletter
\g@addto@macro\normalsize{\setlength\abovedisplayskip{6pt plus 2pt minus 2pt}\setlength\belowdisplayskip{6pt plus 2pt minus 2pt}\setlength\abovedisplayshortskip{2pt plus 2pt}\setlength\belowdisplayshortskip{4pt plus 2pt minus 2pt}}
\makeatother

\begin{document}

\title{Learning Latent Fields in Multiobject Tracking}

\author{Shaoxiu~Wei,~\IEEEmembership{Student Member,~IEEE,} Bhaskar Rao,~\IEEEmembership{Fellow,~IEEE,}
        and~Florian Meyer,~\IEEEmembership{Member,~IEEE}
        \vspace*{-3mm}
\thanks{This work was supported by the National Science Foundation (NSF) under Award Nos. 2146261 and 2506477 and by the Office of Naval Research (ONR) under Award No. N00014-24-1-2021.  A preliminary version of this paper was presented at FUSION 2026, Trondheim, Norway, June 2026.}
\thanks{Shaoxiu~Wei and Bhaskar~Rao are with the Department of Electrical and Computer Engineering, University of California San Diego, La Jolla, CA 92093, USA (e-mail: shwei@ucsd.edu; brao@ucsd.edu).}
\thanks{Florian~Meyer is with the Scripps Institution of Oceanography and the Department of Electrical and Computer Engineering, University of California San Diego, La Jolla, CA 92093, USA (e-mail: f{}lmeyer@ucsd.edu).}
        }

\providecommand{\bu}{\textcolor{blue}}
\providecommand{\rd}{\textcolor{red}}
\providecommand{\gn}{\textcolor{green!60!black}}
\maketitle

\IEEEpeerreviewmaketitle

\begin{abstract}
Multiobject tracking using measurements from arrays of wireless sensors is a key task in applications ranging from autonomous navigation to public safety. Many existing methods compress raw complex-valued signals into detections or intensity maps during preprocessing, thereby discarding potentially relevant information about the objects and the propagation environment. We propose a Bayesian framework for multiobject tracking that relates object states to raw complex-valued signals via a physics-based superpositional model. The model combines a physics-based array manifold, such as radar steering vectors or the replica vectors of matched-field processing, with a latent field. This latent field is a family of context-dependent signal-power distributions that captures the effects of the objects and the propagation environment and is shared across objects and time. It is parameterized by a conditional normalizing flow, which is trained during runtime without labeled data by maximizing an evidence lower bound (ELBO) derived from the Bayesian model. Learning and inference are thus coupled in a sequential framework. We evaluate the method in simulation and in real-world radio and acoustic experiments, with spatial scales ranging from meters to kilometers, and show that it outperforms detection-based methods.
\end{abstract}
\begin{IEEEkeywords}
	Multiobject tracking, Bayesian framework, superpositional model, normalizing flow, wireless sensor arrays.
\end{IEEEkeywords}

\section{Introduction}
\label{sec:introduction}

\par
\begin{figure}[t]
	\centering
	\includegraphics[scale=1.05]{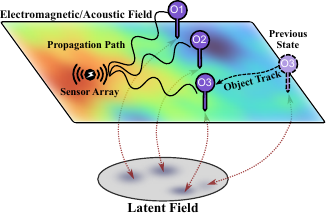}
	\caption{Objects O1, O2, and O3 moving within a shared propagation environment generate raw complex-valued measurements at an array of wireless sensors. Our approach aims to learn a latent field of signal-power statistics that characterize object and propagation effects to improve multiobject tracking performance\vspace{-3.5mm}.}
	\label{fig:scene}
\end{figure}

 Multiobject tracking from received signals is a core problem in wireless sensing systems \cite{PatAshKypHerMosCor:05, WymLieWin:09}. Typically, objects and the shared propagation environment shape the received waveforms through diffraction, scattering, and reflection. As a result, the raw complex-valued signals received by an array of wireless sensors contain signal-strength variations, interference patterns, and background returns that reflect not only object locations but also the propagation environment. Fig.~\ref{fig:scene} illustrates this setting.

\par Conventional methods for multiobject tracking typically process data in two stages. A front-end detector based on beamforming, statistical parameter estimation, or learned regression compresses the high-dimensional raw complex-valued signals into geometric parameters (e.g., delays and angles) \cite{Kay:B93,KriVib:96,Cap:J69} or feature representations (e.g., class and shape) \cite{GaoXinRoyLiu:J21,WanJiaGaoHwaXinLiu:C21}. A downstream tracking method then estimates object states from these detection results using a model of the object dynamics \cite{Mah:B07,BarWilTia:B11}. A full complex-valued fast Fourier transform (FFT) changes the measurement coordinates while preserving the information in the input samples. Information loss can arise from magnitude conversion, selection or projection of the data, and reduction to point detections or other compressed features.

\par A natural way to reduce information loss is to avoid a two-stage processing approach and perform multiobject tracking using richer measurement representations, such as signal intensity maps. This strategy, known as track-before-detect (TkBD), has been explored using both model-based and learning-based approaches \cite{RutGorMas:J05,RisVoVoFar:J13,FisTsaTapNurShl:C24}. BP-TkBD \cite{LiaMey:J25}, a TkBD method based on belief propagation (BP), operates directly on raw complex-valued sensor measurements. Model-based TkBD methods, including BP-TkBD, describe the effects of the propagation environment on the measurements using prescribed models. Modeling the propagation environment can be challenging, making the resulting tracking methods sensitive to model mismatch.

\par To learn propagation statistics during multiobject tracking, we build on the following observations: 
\begin{enumerate}
    \item Objects at nearby positions in a shared environment experience similar signal propagation conditions.
    \item Multiobject tracking implicitly explores the propagation environment by observing objects at many locations.
    \item A statistical surrogate of object and propagation effects on signal power can improve tracking performance.
\end{enumerate}
Motivated by these observations, we combine a physics-based array manifold with learnable signal-power PDFs in a superpositional measurement model. In particular, we learn the statistical surrogate in observation~3 as a family of context-dependent signal-power PDFs that we refer to as the latent field. The context consists of the object's position and possibly sensing parameters and signal features. The parameters of the latent field are shared across objects and time. For a fixed context, the signal-power PDFs summarize the combined effects of propagation and object properties, such as source strength or reflectivity.

\par At each time step, particle-based BP \cite{KscFreLoe:01,DouFreGor:01} approximates the posterior PDFs of object states, existence variables, signal powers, and noise variance. We use the inferred existence probabilities and signal-power PDFs to update a conditional normalizing flow \cite{RezMoh:C15} through an objective derived from the evidence lower bound (ELBO). The normalizing flow represents the latent field, which is thus continuously updated while objects are tracked. More details on latent field modeling are discussed in Section \ref{sec:latent_field_prior}.
\par

\section{Related Work and Contributions}

\par Object detection and multiobject tracking  from wireless signals have been studied extensively using both model-based and learning-based methods\vspace{-3mm}.

\subsection{Object Detection}
Various methods exist for extracting object location information from wireless signals. Model-based approaches include matched filtering \cite{Tur:J60}, beamforming \cite{VanBuc:88}, subspace methods such as multiple signal classification (MUSIC) \cite{Sch:J86}, and sparse Bayesian learning (SBL) \cite{Tip:J01,GerMecChrXenNan:J16}. Constant false-alarm rate (CFAR) detectors \cite{Roh:J83} adapt detection thresholds to local clutter statistics. Array-processing methods such as beamforming, MUSIC, and SBL use physics-based manifolds to relate the sensor response to object location parameters. Some of these model-based approaches also employ a statistical noise model. Learning-based approaches instead establish mappings from signal-derived features to locations. Examples include fingerprinting methods in wireless communications \cite{WanGaoMaoPan:J17,AyyAruWu:C20} and neural networks that detect objects in FFT-derived radar maps. RODNet \cite{WanJiaGaoHwaXinLiu:C21} uses complex range--angle (RA) maps, and RAMP-CNN \cite{GaoXinRoyLiu:J21} combines complex RA maps with magnitude maps in the range--velocity (RV) and velocity--angle (VA) domains\vspace{-3mm}.

\subsection{Multiobject Tracking}
In detect-then-track multiobject tracking \cite{BarLi:95}, object states are estimated from detections over time. Model-based approaches describe data association and state evolution probabilistically. Examples include joint probabilistic data association (JPDA) filters \cite{ForBarSch:83}, methods based on random finite sets (RFSs) \cite{Mah:B07, GarWilGraSve:J18, Mah:J03}, and BP-based tracking methods \cite{MeyKroWilLauHlaBraWin:J18,LiaMey:J23,WeiLiaMey:J26,MeyWil:J21}. Conventional JPDA assumes a known number of objects and requires separate track-management procedures or existence-aware extensions to handle object appearance and disappearance. RFS and BP formulations can explicitly represent uncertain object existence. \par Learning-based architectures have also been developed for visual tracking. The approach in \cite{FeiPinZis:C17} jointly learns detection and across-frame track regression using a convolutional neural network (CNN). Graph neural networks learn associations between image-based detections \cite{BraLau:C20}. The transformer-based MOTR \cite{ZenDonZha:C21} performs end-to-end visual tracking by representing each object with a learned feature vector that is updated using the image features of each new frame. These architectures are trained on labeled images or trajectories and provide precedents for learning-based tracking.

\par For tracking methods that operate on point detections, information omitted by the detector is unavailable for the Bayesian multiobject tracking method. Additional information loss may be the result of additional data reduction or inference approximations. 
\par The alternative strategy, TkBD \cite{RutGorMas:J05,RisVoVoFar:J13,LiaMey:J25,WesModLeiPed:26}, performs inference on lower-level measurements, but it can involve high computational complexity and can be sensitive to model mismatch. Several TkBD methods obtain tractable inference through specific modeling choices. Histogram probabilistic multihypothesis tracking (H-PMHT) \cite{StrGraWal:J02, DavGae:B18} models intensity-image cell energies as a histogram of synthetic point counts. Conventional implementations add procedures for track initialization and termination. The method in \cite{VoVoPhaSut:10} derives tractable image-based updates by assuming that the regions influenced by different objects do not overlap. The Poisson/multi-Bernoulli approach in \cite{KroWilMey:21} supports scalable Bayesian TkBD inference using a measurement model in which each object contributes to one cell. Many TkBD methods have been evaluated mainly in simulation, so their performance in complex real-world propagation environments is less explored. BP-TkBD \cite{LiaMey:J25} uses a superpositional model of raw complex-valued sensor measurements and jointly infers object kinematics, existence, signal powers, and noise variances through particle-based BP with Gaussian moment matching. Its object-specific signal-power priors are propagated using prescribed state-transition models. We retain the superpositional model with conditionally Gaussian amplitudes but replace these prescribed state-transition models by the latent field. The latent field is learned during runtime and shares information about object and propagation effects across the tracked objects.

\subsection{Contributions}
\par Learning statistical representations of signal propagation has precedents in wireless localization. Gaussian-process models learn location-dependent signal-strength likelihoods from calibration data \cite{FerHahFox:C06}. WiFi simultaneous localization and mapping (WiFi-SLAM) jointly estimates a device trajectory and signal-strength maps from unlabeled measurement sequences, while using kernel parameters learned beforehand from labeled data \cite{FerFoxLaw:C07}. Multisnapshot SBL also estimates source-power and noise-variance hyperparameters from complex array measurements through evidence maximization \cite{GerMecChrXenNan:J16}.
The proposed framework combines a physics-based superpositional model of raw complex-valued sensor data with Bayesian multiobject tracking and a shared, context-dependent latent field represented by a conditional normalizing flow.  We learn this field during runtime through an ELBO-derived objective without labeled data or offline pretraining. To the best of our knowledge, this combination has not been proposed in the multiobject tracking literature.

\par Whereas existing formulations generally treat the signal amplitude or power as part of the object states or features, in our formulation each object contributes information to the latent field, which can also be evaluated at contexts not encountered during tracking.

\par Our main contributions are summarized as follows:
\begin{enumerate}

	\item We formulate a superpositional model of raw complex-valued array signals with a physics-based array manifold and the shared latent field.

	\item We combine BP tracking with learning the latent field, represented by a conditional normalizing flow, during runtime without labeled data or offline pretraining.
	
	\item We develop an efficient likelihood evaluation that exploits low-rank and Kronecker structure, with a complexity that scales linearly in the measurement dimension.
		
	\item We demonstrate improved tracking performance on real radar and acoustic data and quantify the effect of individual model components through ablation studies.

\end{enumerate}
\par This paper advances over the preliminary account of our method provided in the conference publication \cite{WeiRaoMey:C26} by (i) presenting a detailed derivation of the learning objective that accounts for the uncertain existence of objects, (ii) developing an efficient likelihood evaluation that exploits low-rank and Kronecker structure, together with a complexity and runtime analysis, (iii) demonstrating performance advantages using real radar and acoustic data, (iv) comparing the learned latent field with physics-based predictions of the KRAKEN normal-mode program, and (v) assessing the design of the signal-power prior in a controlled ablation study.

\par The remainder of this paper is organized as follows. Section~\ref{sec:method} formulates the multiobject tracking problem and introduces the physics-based superpositional model, the statistical model of the latent field, and its parameterization by a conditional normalizing flow. Section~\ref{sec:inference_learning} presents Bayesian inference and the learning objective with its parameter updates. Section~\ref{sec:experiments} evaluates the proposed method in a simulated radar scenario and using real radar and acoustic data, including a controlled acoustic simulation. Section~\ref{sec:conclusion} concludes the paper. Due to space limitations, detailed derivations of the learning objective and of the BP-based inference, the efficient likelihood evaluation and its complexity analysis, as well as additional numerical results have been omitted and can be found in the supplementary material \cite{WeiRaoMey:SM27}.

\par \textit{Notation}: Scalars are denoted by italic letters, vectors by bold lowercase letters, e.g., $\V{x}$ and $\bm{\gamma}$, and matrices and tensors by bold uppercase letters, e.g., $\bm{\Sigma}_k$ and $\V{Z}_k$. The sets of real, positive real, and complex numbers are denoted by $\mathbb{R}$, $\mathbb{R}_{>0}$, and $\mathbb{C}$, respectively. The superscript $(\cdot)^{\mathrm{H}}$ denotes the conjugate transpose, $\M{I}_C$ the $C \times C$ identity matrix, and $\mathrm{vec}(\cdot)$ the vectorization of a tensor. The sequence $\{\V{z}_k\}_{k=1}^{K}$ is written as $\V{z}_{1:K}$. Probability density functions (PDFs) and probability mass functions are denoted by $f(\cdot)$, approximate posterior PDFs by $q(\cdot)$, and a subscript $\bm{\theta}$ indicates dependence on the learnable parameters. The zero-mean proper complex Gaussian PDF with covariance matrix $\bm{\Sigma}$ is denoted by $\mathcal{CN}(\V{0},\bm{\Sigma})$, and its PDF evaluated at $\V{z}$ by $\mathcal{CN}(\V{z};\V{0},\bm{\Sigma})$. The standard Gaussian PDF is denoted by $\mathcal{N}(0,1)$. Finally, $\mathbb{E}_{q}[\cdot]$ denotes the expectation with respect to $q(\cdot)$, and $\mathrm{KL}(\cdot\,\|\,\cdot)$ denotes the Kullback--Leibler divergence.

\section{Problem Formulation and Statistical Model}
\label{sec:method}

We formulate the multiobject tracking problem, specify the superpositional measurement model, and introduce the statistical model of the latent field.

\subsection{Problem Formulation}
\label{sec:problem}
\paragraph{Scenario} We consider a sequence of sensor measurements received over discrete time steps $k=1,\ldots,K$. Objects may appear and disappear over time. We represent the unknown and time-varying number of objects using $N_k$ potential objects. Potential object $n \in \{1,\ldots,N_k\}$ has a kinematic state $\V{x}_{k,n} \in \mathbb{R}^{d}$ (e.g., position and velocity) and a binary existence variable $r_{k,n}\in\{0,1\}$. The number of existing objects is $L_k=\sum_{n=1}^{N_k}r_{k,n}$. In contrast to the known number $N_k$ of potential objects, $L_k$ is an unknown random variable. At each time step $k$, the sensor array produces a complex-valued measurement tensor $\V{Z}_k \in \mathbb{C}^{C_1 \times \cdots \times C_m}$. We denote the vectorized representation of the measurement tensor by $\V{z}_k = \mathrm{vec}\!\big(\V{Z}_k\big) \in \mathbb{C}^{C}$, where $C = \prod_{j=1}^{m} C_j$.

\paragraph{Goal} At each time step $k$, our goal is to infer the number of objects and their states from the raw complex-valued measurements $\V{z}_{1:k}$ received up to that time. Because sequential estimation of object states implicitly explores the propagation environment, it provides the information needed to learn the latent field, which in turn can improve subsequent inference.

\subsection{Physics-Based Superpositional Model}
\label{sec:superposition} 

We review the single-dictionary, single-snapshot case of the measurement model in \cite[Sec.~II-A]{LiaMey:J25}, omitting the dictionary and snapshot indices. Here, a dictionary specifies the state-dependent array response, a snapshot is one of several measurement vectors acquired at the same time step, and vectorization collects the samples within one measurement block. The extension to multiple dictionaries and snapshots follows the conditional block-likelihood factorization described therein.

At each time step $k$, each existing object contributes a complex array response determined by the physics-based array manifold, while propagation effects are captured through a stochastic complex amplitude. The raw signal $\V{z}_k$ is modeled as a coherent superposition of the contributions from existing objects, represented by the $N_k$ potential-object terms, plus noise, i.e.\vspace{-1mm},
\begin{equation}
	\V{z}_k \;=\; \sum_{n=1}^{N_k} r_{k,n}\;\rho_{k,n}\; \V{a}\!\left(\V{x}_{k,n}\right) \;+\; \V{w}_k.
	\label{eq:meas_model}
\end{equation}
Here, the physics-based array manifold $\V{a}(\cdot)$ maps an object state $\V{x}_{k,n}$ to the corresponding complex array response $\V{a}(\V{x}_{k,n}) \in \mathbb{C}^{C}$. The noise $\V{w}_k$ is zero-mean proper complex Gaussian with covariance $\eta_k\M{I}_C$, with $\eta_k>0$ and unknown. The diagonal covariance models a white effective background across measurement entries.

\par The complex amplitude $\rho_{k,n}\in \mathbb{C}$ captures the unknown gain and
phase of object $n$, representing aggregate gain effects compatible with the array response prescribed by the array manifold. It is modeled
as a zero-mean proper complex Gaussian random variable, i.e.,
\begin{align}
	\rho_{k,n} \mid \gamma_{k,n} \sim \mathcal{CN}(0, \gamma_{k,n})
	\label{eq:rho_gamma}
\end{align}
where the positive random variable $\gamma_{k,n}$ is the conditional mean squared magnitude of the complex amplitude, which we refer to as the signal power. For an existing object, $\gamma_{k,n}$ is given by
the latent field evaluated at the object's context, as detailed in
Section~\ref{sec:latent_field_prior}. For an indexed quantity $u_{k,n}$, we write $u_{k,1{:}N_k}=(u_{k,1},\ldots,u_{k,N_k})$ for its ordered collection. In particular, $\V{x}_k=\V{x}_{k,1{:}N_k}$, $\V{r}_k=r_{k,1{:}N_k}$, and $\bm{\gamma}_k=\gamma_{k,1{:}N_k}$ collect the kinematic states, existence variables, and signal powers. Conditional on these variables and $\eta_k$, the amplitudes $\rho_{k,n}$ are mutually independent with the PDFs in \eqref{eq:rho_gamma}, and the noise is independent of all amplitudes with the PDF stated above. Marginalizing the amplitudes therefore yields the joint likelihood of the measurement entries,
\begin{align}
	&f(\V{z}_{k}\mid \V{r}_{k},\V{x}_{k},\bm{\gamma}_{k},\eta_k)
	=
	\mathcal{CN}\!\big(\V{z}_{k}; \V{0}, \bm{\Sigma}_k\big)
	\label{eq:gauss_lik}
	\\
	&\bm{\Sigma}_k
	=
	\sum_{n=1}^{N_k}
	r_{k,n}\gamma_{k,n}
	\V{a}(\V{x}_{k,n})
	\V{a}(\V{x}_{k,n})^{\mathrm{H}}
	+
	\eta_k\M{I}_{C}.
	\label{eq:covariance}
\end{align}
The signal-dependent part of $\bm{\Sigma}_k$ has rank at most $\min(C,L_k)$, while $\eta_k>0$ makes the full covariance positive definite. The generally nonzero off-diagonal entries retain correlations among measurements within this block.
However, directly evaluating this Gaussian likelihood is generally intractable in the sensing regimes considered here, since the measurement dimension $C$ can reach several hundred thousand. For example, a typical radar setup with $128$ fast-time samples, $256$ chirps, and $16$ antennas yields $C = 524{,}288$. As illustrated in Table~\ref{tab:complexity_example}, storing the full covariance matrix for a single likelihood evaluation then requires approximately 2~TiB, which is infeasible on standard hardware. We therefore exploit the low-rank signal contribution together with the diagonal noise term and, when available, a Kronecker factorization of the array manifold. This makes likelihood evaluation tractable for typical problem sizes without forming or inverting a full $C \times C$ covariance matrix. 

\begin{table}[t]
	\centering
	\caption{Memory requirement for a single likelihood evaluation in a typical radar scenario with $C = 128 \times 256 \times 16 = 524{,}288$ complex values stored with 64~bits each.}
	\label{tab:complexity_example}
	\begin{tabular}{l l c}
		\toprule
		Item & Description & Memory Cost \\
		\midrule
		Input measurement $\V{z}_{k}$ & $C$ complex values & $\approx 4$ MiB \\
		Direct computation \eqref{eq:gauss_lik}
		& Full covariance matrix
		& $\approx 2.0$ TiB \\
		Efficient computation \cite{WeiRaoMey:SM27}
		& Decomposed components
		& $\approx 3.1$ KiB \\
		\bottomrule
	\end{tabular}
	\vspace{-3mm}
\end{table}

\subsection{Statistical Model of the Latent Field}
\label{sec:latent_field_prior}
We formalize the latent field as a positive-valued stochastic field indexed by
context. The context of an object comprises its position and possibly known sensing parameters and signal features. Let $(\Omega,\mathcal{F},P)$ be a probability space and let
$\mathcal{C}$ denote the context space. For learnable parameters $\bm{\theta}$,
the latent field is a collection of positive random variables
\begin{align}
	\Gamma_{\bm{\theta}}
	=
	\big\{\Gamma_{\bm{\theta}}(\,\cdot\,,\V{c})\big\}_{\V{c}\in\mathcal{C}}
	\qquad
	\Gamma_{\bm{\theta}}(\,\cdot\,,\V{c}):\Omega\longrightarrow\mathbb{R}_{>0}.
	\label{eq:latent_random_field}
\end{align}
For any fixed context $\V{c}\in\mathcal{C}$, $\Gamma_{\bm{\theta}}(\,\cdot\,,\V{c})$
is a positive random variable. Similarly, for any fixed outcome $\omega\in\Omega$,
the function $\Gamma_{\bm{\theta}}(\omega,\,\cdot\,):\mathcal{C}\rightarrow\mathbb{R}_{>0}$
is one realization of the latent field over the context space. For each fixed context $\V{c}$, the random variable $\Gamma_{\bm{\theta}}(\,\cdot\,,\V{c})$ has the PDF $f_{\bm{\theta}}(\gamma;\V{c})$.
For each pair $(k,n)$ of time step and potential object, we introduce an independent copy $\Gamma_{\bm{\theta}}^{(k,n)}$ of this field. The copies share the deterministic parameters $\bm{\theta}$ and the PDF $f_{\bm{\theta}}(\gamma;\V{c})$ at each context. If object $n$ exists at time $k$, its signal power is obtained by evaluating its copy at the context $\V{c}_{k,n}$, i.e.\vspace{0mm},
\begin{align}
	\gamma_{k,n}
	&=
	\Gamma_{\bm{\theta}}^{(k,n)}(\,\cdot\,,\V{c}_{k,n}).\label{eq:latent_field_evaluation}\\[-3mm]
	\nn
\end{align}
Thus, through the PDFs $f_{\bm{\theta}}(\gamma;\V{c})$, $\V{c}\in\mathcal{C}$, the latent field defines a family of context-dependent signal-power PDFs. Parameters shared across objects and time steps allow observations from multiple object tracks to update the latent field jointly.

By \eqref{eq:rho_gamma}, the signal power satisfies
\begin{align}
	\mathbb{E}\!\big[
		|\rho_{k,n}|^2
		\mathrel{\mid} \gamma_{k,n}
	\big]
	&=
	\gamma_{k,n}.
	\label{eq:latent_field_power} \\[-4.5mm]
	\nn
\end{align}
The latent field therefore models the statistics of the combined object and propagation effects on the signal power relative to the array response $\V{a}(\V{x}_{k,n})$, such as source strength, reflectivity, attenuation, and multipath-induced gain variations.

The context $\V{c}_{k,n}$ is treated as a fixed input during each inference and learning step. In our numerical studies in Section~\ref{sec:experiments}, the context consists of a positional encoding of the object position estimated from the preceding posterior, together with known sensing parameters and signal features. In our implementation, the context-dependent PDFs are represented by the conditional normalizing flow discussed in Section~\ref{sec:flow}.

Since $\gamma_{k,n}$ affects the measurement covariance in
\eqref{eq:covariance} only if $r_{k,n}=1$, we consider its PDF
conditioned on $r_{k,n}=1$, denoted\vspace{-2mm} by
\begin{align}
	f_{\bm{\theta}}(\gamma_{k,n}\mid r_{k,n}=1;\V{c}_{k,n}).
	\label{eq:latent_field_prior}
\end{align}
We combine the random variables $r_{k,n}$ and $\gamma_{k,n}$ as
$\bm{\phi}_{k,n}=(r_{k,n},\gamma_{k,n})$ to jointly represent object existence
and signal power. The PDF of $\bm{\phi}_{k,n}$ is given by
\begin{align}
	&f_{\bm{\theta}}(\bm{\phi}_{k,n})\nn\\
	&=
	\begin{cases}
		f(r_{k,n}=1)\,
		f_{\bm{\theta}}(\gamma_{k,n}\mid r_{k,n}=1;\V{c}_{k,n})
		& r_{k,n}=1\\[2pt]
		f(r_{k,n}=0)\,
		f_{\mathrm{D}}(\gamma_{k,n})
		& r_{k,n}=0
	\end{cases}
	\label{eq:phi-mixture}
\end{align}
where $f_{\mathrm{D}}(\gamma_{k,n})$ is a dummy PDF. Since the likelihood \eqref{eq:gauss_lik} does not depend on $\gamma_{k,n}$ if $r_{k,n}=0$, any normalized PDF that does not depend on $\bm{\theta}$ can be chosen for $f_{\mathrm{D}}(\gamma_{k,n})$ without affecting existence probabilities and state estimates \cite{LiaMey:J25} or the learning objective.

The
existence variable $r_{k,n}$ is modeled as independent of the field copies. It only
determines whether the corresponding copy models the signal power of object $n$ at time $k$. If
$r_{k,n}=1$, then $\gamma_{k,n}$ is given by
\eqref{eq:latent_field_evaluation}. If $r_{k,n}=0$, the object contributes no
power and $\gamma_{k,n}$ is distributed according to the dummy PDF $f_{\mathrm{D}}(\gamma_{k,n})$. Since
$\V{c}_{k,n}$ is fixed and $r_{k,n}$ is independent
of the field copies, the conditional PDF \eqref{eq:latent_field_prior} can be written as
\begin{align}
	f_{\bm{\theta}}(\gamma_{k,n}\mid r_{k,n}=1;\V{c}_{k,n})
	=
	f_{\bm{\theta}}(\gamma_{k,n};\V{c}_{k,n}).
	\label{eq:cond_equals_field}
\end{align}
We assume that, for fixed $\bm{\theta}$, the values of each field copy at distinct contexts are independent. The normalizing flow in Section~\ref{sec:flow} provides a construction of this field and its independent copies. Hence,
for any finite set of distinct contexts $\{\V{c}_j\}_{j=1}^{J}$\vspace{-3mm},
\begin{align}
	f_{\bm{\theta}}(\gamma_1,\dots,\gamma_J;\V{c}_1,\dots,\V{c}_J)
	=
	\prod_{j=1}^{J} f_{\bm{\theta}}(\gamma_j;\V{c}_j).
	\label{eq:field_fdd}
\end{align}
Hence, spatial structure, such as similar signal-power PDFs at nearby positions, is captured by the dependence of $f_{\bm{\theta}}(\gamma;\V{c})$ on the context $\V{c}$ through the shared parameters $\bm{\theta}$. By \eqref{eq:cond_equals_field} and \eqref{eq:field_fdd}, learning the latent field is
equivalent to learning the conditional PDF \eqref{eq:latent_field_prior} for all
contexts. The parameters $\bm{\theta}$ are updated during runtime as discussed in
Section~\ref{sec:Var}.

\subsection{Conditional Normalizing Flow}
\label{sec:flow}

We parameterize the conditional PDFs $f_{\bm{\theta}}(\gamma;\V{c})$, $\V{c}\in\mathcal{C}$, by a conditional normalizing flow~\cite{RezMoh:C15}. A flow provides an explicit expression for $f_{\bm{\theta}}(\gamma;\V{c})$, which is required by the learning objective in Section~\ref{sec:Var} and makes the learned latent field interpretable, e.g., through the mean and standard deviation of the signal power at each context. In addition, it can represent multimodal and heavy-tailed PDFs, such as those caused by multipath propagation. Since $\gamma_{k,n} > 0$, we first represent $\gamma_{k,n}$ by an unconstrained variable $\mu_{k,n}$, i.e.,
\begin{equation}
	\gamma_{k,n} = {G}(\mu_{k,n}) \qquad \mu_{k,n} \in \mathbb{R}
\end{equation}
where $G(\cdot):\mathbb{R}\to\mathbb{R}_{>0}$ is a smooth bijection. For notational simplicity, we suppress the indices $(k,n)$ in the following derivation. We then model the conditional PDF $f_{\bm{\theta}}(\mu;\V{c})$ using a conditional normalizing flow. Let $f_0(\varepsilon)$ denote a base PDF on $\mathbb{R}$ (e.g., $\varepsilon \sim \mathcal{N}(0,1)$). For a given context $\V{c}$, we define a sequence of $Q$ invertible, monotone transformations, i.e.,
\begin{equation}
	y_0 = \varepsilon\qquad
	y_\ell = F_{\theta_\ell}(y_{\ell-1};\, \V{c})\quad
	\ell = 1,\dots,Q \nn
\end{equation}
\begin{figure}[t]
	\centering
	\includegraphics[width=0.45\textwidth]{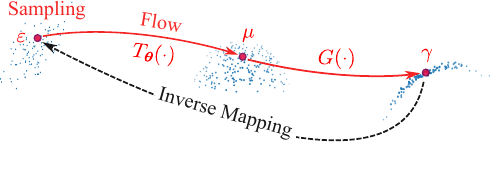}
	\vspace{-4mm}
	\caption{Computational steps of the normalizing flow.}
	\label{fig:flow}
	\vspace{-1.5mm}
\end{figure}%
where each $F_{\theta_\ell}(\cdot;\V{c}):\mathbb{R}\to\mathbb{R}$ is a differentiable bijection parameterized by $\theta_\ell$ and conditioned on $\V{c}$. In particular, $F_{\theta_\ell}(\cdot;\V{c})$ is a nonlinear, strictly monotone function whose coefficients are computed from the context $\V{c}$ by an encoder, i.e., a neural network with parameters $\theta_\ell$, such as a multilayer perceptron (MLP) or a CNN. Thus, the context determines the shape of each transformation, whereas the parameters $\bm{\theta}$ are shared across all contexts, objects, and time steps. The output of this sequence is given by
\[
\mu = y_{Q} =T_{\bm\theta}(\varepsilon;\V{c})
\vspace{1mm}
\]
where $\bm{\theta} = (\theta_1,\dots,\theta_{Q})$ collects the parameters of all transformations. By the change-of-variables formula \cite{PapPil:02}, the log-PDF of $\gamma$ given $\V{c}$ can be expressed\vspace{-1mm} as
\begin{align}
	\log f_{\bm\theta}(\gamma;\V{c})
	=&~
	\log f_0(\varepsilon)
	-
	\log
	\left|
	\frac{\mathrm{d}G(T_{\bm\theta}(\varepsilon;\V{c}))}{\mathrm{d}\varepsilon}
	\right|\notag\\
	= &~\log f_0(\varepsilon) -
	\sum_{\ell=1}^{Q}
	\log \left|
	\frac{\partial F_{\theta_\ell}(y_{\ell-1};\, \mathbf{c})}
	{\partial y_{\ell-1}}
	\right|\notag\\[1.5mm]
	&- \log \left|
	\frac{\mathrm{d} G(\mu)}{\mathrm{d} \mu}
	\right|.\label{eq:flow_gamma} 
\end{align}
In \eqref{eq:flow_gamma}, the base coordinate and the intermediate values are obtained from $\gamma$ by the inverse mappings, i.e., $\mu = G^{-1}(\gamma)$, $y_{Q} = \mu$, $y_{\ell-1} = F_{\theta_\ell}^{-1}(y_\ell;\V{c})$ for $\ell = Q,\dots,1$, and $\varepsilon = y_0 = T_{\bm{\theta}}^{-1}(\mu;\V{c})$. Hence, $\varepsilon$ and $y_{\ell-1}$ depend on $\bm{\theta}$ when \eqref{eq:flow_gamma} is differentiated with respect to $\bm{\theta}$.
The log-PDF in \eqref{eq:flow_gamma}, evaluated at the posterior signal-power particles and the corresponding contexts, supplies the terms of the learning objective derived in Section~\ref{sec:Var}.
The flow $T_{\bm{\theta}}(\,\cdot\,;\V{c})$, together with the transformation $G(\cdot)$, constructs the independent field copies introduced in Section~\ref{sec:latent_field_prior}. At a fixed context $\V{c}$,
\begin{align}
	\Gamma_{\bm{\theta}}^{(k,n)}(\,\cdot\,,\V{c})
	&=
	G\big(T_{\bm{\theta}}(\varepsilon_{k,n,\V{c}};\V{c})\big).
	\label{eq:flow_field}
\end{align}
The base random variables $\varepsilon_{k,n,\V{c}}$ are independent across distinct copy and context indices and each is drawn from the PDF $f_0(\varepsilon_{k,n,\V{c}})$. The generic variable $\varepsilon$ in \eqref{eq:flow_gamma} denotes the base coordinate for one such draw.
Hence, each field value $\Gamma_{\bm{\theta}}^{(k,n)}(\,\cdot\,,\V{c})$ has the PDF $f_{\bm{\theta}}(\gamma;\V{c})$ given by \eqref{eq:flow_gamma}. An overview of the flow steps is
shown in Fig.~\ref{fig:flow}. The forward mapping in~\eqref{eq:flow_field} draws
samples of $\gamma_{k,n}$ at the context of each object, which represent the signal-power prior PDF $f_{\bm{\theta}}(\gamma_{k,n};\V{c}_{k,n})$ used in Bayesian inference. The inverse mapping evaluates the
log-PDF~\eqref{eq:flow_gamma} at the fixed posterior signal-power particles for the parameter update in Section~\ref{sec:Var}.
Before the first time step, $\bm{\theta}$ is initialized such that the flow yields the uniform PDF on $(0,\gamma_{\max}]$ for all contexts, i.e., no prior knowledge of the signal power is assumed.

\section{Bayesian Inference and Latent-Field Learning}
\label{sec:inference_learning}

Using the model in Section~\ref{sec:method}, we now discuss how to infer object states and learn the latent field during runtime. We first describe Bayesian inference and then derive the learning objective and parameter updates.

\subsection{Bayesian Inference}
Let 
\(\V{h}_k = \big(\V{x}_{k},\bm{\phi}_{k},\eta_k\big)\), with $\bm{\phi}_k = \bm{\phi}_{k,1{:}N_k}$, collect all random variables at time \(k\). The joint prior PDF is denoted by \(f_{\bm{\theta}}(\mathbf h_k)\). Note that we adopt a sequential filtering perspective, where the prior is the result of prediction and update steps at previous time steps. That is, $f_{\bm{\theta}}(\mathbf h_k)$ is a shorthand for the predicted PDF $f_{\bm{\theta}}(\mathbf h_k \mid \V{z}_{1:k-1})$; for notational simplicity, we omit the conditioning on $\V{z}_{1:k-1}$. The posterior PDF at time step $k$ is given by Bayes' rule, i.e.,
\begin{align}
	f_{\bm{\theta}}(\V{h}_k\mid \V{z}_k)
	&=
	\frac{
	f(\V{z}_k\mid \V{h}_k)
	f_{\bm{\theta}}(\V{h}_k)
	}{
	f_{\bm{\theta}}(\V{z}_k)
	}.
	\label{eq:bayes}
\end{align}
Here,
$
f(\V{z}_k \mathrel{\mid} \V{h}_k)
=
\mathcal{CN}\!\big(\V{z}_k;\V{0},\bm{\Sigma}_k\big)
$
is the likelihood function. The denominator $f_{\bm{\theta}}(\V{z}_k)$ is the evidence, given by $f_{\bm{\theta}}(\V{z}_k) = \int f(\V{z}_k \mathrel{\mid} \V{h}_k) f_{\bm{\theta}}(\V{h}_k) \mathrm{d}\V{h}_k$.

\subsubsection{Posterior approximation} Direct computation of \eqref{eq:bayes} is generally intractable. We
therefore approximate the posterior PDF $f_{\bm{\theta}}(\mathbf h_k\mid \mathbf z_k)$ by a PDF $q(\mathbf h_k)$. This PDF serves as a formal auxiliary density; BP computes only its marginal posterior PDFs, which are represented by weighted particles. The resulting marginal posterior PDFs are obtained by marginalizing \(q(\mathbf h_k)\), i.e.,
\begin{align}
    q(u)
    =
    \int q(\mathbf h_k)
    \,\mathrm{d}\mathbf h_k^{\sim u}
    \qquad
    u\in\{\mathbf x_{k,n},\boldsymbol{\phi}_{k,n},\eta_k\}\label{eq:marg}
\end{align}
where $\mathbf h_k^{\sim u}$ denotes all components of $\mathbf h_k$ except $u$, and integration with respect to the binary existence variables in $\mathbf h_k$ is understood as summation over their values.

\subsubsection{Prediction} As is common, we model the kinematic states and the noise variance as first-order Markov processes with state-transition PDFs $f(\mathbf{x}_{k,n}\mid \mathbf{x}_{k-1,n})$ and $f(\eta_k\mid\eta_{k-1})$. Following \cite{LiaMey:J25}, we distinguish legacy potential objects $n\in\{1,\dots,N_{k-1}\}$, which were already considered at time step $k-1$, from new potential objects $n\in\{N_{k-1}+1,\dots,N_k\}$, which account for newly appearing objects. The predicted PDFs of the kinematic states of legacy potential objects and of the noise variance are given by
\begin{align}
	f(\mathbf{x}_{k,n}) =&~ \int f(\mathbf{x}_{k,n}\mid \mathbf{x}_{k-1,n})q(\mathbf{x}_{k-1,n}) \mathrm{d}\mathbf{x}_{k-1,n} \label{eq:pred_x}\\
    f(\eta_{k}) =&~ \int f(\eta_{k}\mid \eta_{k-1})q(\eta_{k-1}) \mathrm{d}\eta_{k-1}. \label{eq:pred_eta}
\end{align}
The prediction step is implemented using conventional particle-filtering operations \cite{AruMasGorCla:02}. As in \cite{LiaMey:J25}, $f(\eta_k\mid\eta_{k-1})$ is a Gamma PDF with mean $\eta_{k-1}$. The existence variable of a legacy potential object evolves according to $f(r_{k,n}=1\mid r_{k-1,n}=1)=p_{\mathrm{s}}$ and $f(r_{k,n}=1\mid r_{k-1,n}=0)=0$, where $p_{\mathrm{s}}$ is the probability of object survival. Thus, its predicted existence probability in \eqref{eq:phi-mixture} is $f(r_{k,n}=1)=p_{\mathrm{s}}\,q(r_{k-1,n}=1)$. For a new potential object, $f(r_{k,n}=1)=p_{\mathrm{B},n}$ is the birth probability, and $f(\mathbf x_{k,n})=f_{\mathrm{B},n}(\mathbf x_{k,n})$ is the birth PDF of its kinematic state. The birth PDFs used in our experiments are described in Section~\ref{sec:experiments}. For both legacy and new potential objects, the prior PDF of the signal power is provided by the latent field as described in Section~\ref{sec:latent_field_prior}.

\par Since the prediction step operates on the marginal posterior PDFs of time step $k-1$, the joint prior PDF at time step $k$ is represented by the product of the predicted PDFs \cite{LiaMey:J25}, i.e.,
\begin{align}
	f_{\bm{\theta}}(\mathbf h_k) = f(\eta_k) \prod_{n=1}^{N_k} f(\mathbf x_{k,n})\, f_{\bm{\theta}}(\boldsymbol{\phi}_{k,n})
	\label{eq:joint_prior}
\end{align}
where $f(\eta_k)$ is given by \eqref{eq:pred_eta}, $f(\mathbf x_{k,n})$ is given by \eqref{eq:pred_x} for legacy potential objects and by $f_{\mathrm{B},n}(\mathbf x_{k,n})$ for new potential objects, and $f_{\bm{\theta}}(\boldsymbol{\phi}_{k,n})$ is given by \eqref{eq:phi-mixture} with the predicted existence probability $f(r_{k,n}=1)$ discussed above. The parameters $\bm{\theta}$ enter \eqref{eq:joint_prior} only through the conditional PDFs $f_{\bm{\theta}}(\gamma_{k,n}\mid r_{k,n}=1;\V{c}_{k,n})$ in \eqref{eq:phi-mixture}.

\subsubsection{Update and state estimation} In BP, the marginal posterior PDFs in \eqref{eq:marg} are approximated by beliefs of the form
\begin{align*}
	q(\mathbf x_{k,n}) &\mathrel{\propto} f(\mathbf x_{k,n})\,\ell_{\mathbf x,n}(\mathbf x_{k,n})\\[1mm]
	q(\boldsymbol{\phi}_{k,n}) &\mathrel{\propto} f_{\bm{\theta}}(\boldsymbol{\phi}_{k,n})\,\ell_{\boldsymbol{\phi},n}(\boldsymbol{\phi}_{k,n})\\[1mm]
	q(\eta_k) &\mathrel{\propto} f(\eta_k)\,\ell_{\eta}(\eta_k)
\end{align*}
where \(\ell_{\mathbf x,n}(\mathbf x_{k,n})\), \(\ell_{\boldsymbol{\phi},n}(\boldsymbol{\phi}_{k,n})\), and
\(\ell_{\eta}(\eta_k)\) denote the marginal likelihood terms associated with the
kinematic state of object \(n\), its existence and signal power, and the noise
variance at time \(k\), respectively. These marginal likelihoods are
computed approximately using particle-based BP \cite{LiaMey:J25}. In particular, each marginal likelihood is approximated by a zero-mean complex Gaussian PDF via second-order moment matching, and the expected outer product $\mathbb{E}[\V{a}(\mathbf x_{k,n})\V{a}(\mathbf x_{k,n})^{\mathrm{H}}]$ is approximated by the rank-one matrix $\mathbb{E}[\V{a}(\mathbf x_{k,n})]\,\mathbb{E}[\V{a}(\mathbf x_{k,n})]^{\mathrm{H}}$ \cite[Secs.~II-C and III-A]{WeiRaoMey:SM27}; for Kronecker-structured array responses, the mean array response is further approximated by the Kronecker product of the means of its factors \cite[Sec.~III-D]{WeiRaoMey:SM27}. As in \cite{LiaMey:J25}, potential object $n$ is declared to exist at time step $k$ if $q(r_{k,n}=1)>T_{\mathrm{dec}}$, and potential objects with $q(r_{k,n}=1)<T_{\mathrm{pru}}$ are pruned before the next prediction step. The object state estimates of declared potential objects are obtained from the marginal posterior PDFs $q(\mathbf x_{k,n})$ using the minimum
mean-square error (MMSE) estimator.

\begin{figure}[t]
	\centering
	\includegraphics[scale=.75]{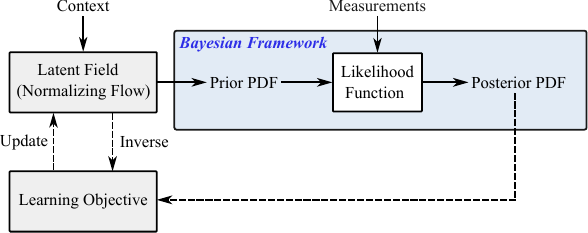}
	\vspace{1mm}
	\caption{The proposed inference and learning framework at one time step.}
	\label{fig:vb}
	\vspace{-3mm}
\end{figure}

\subsection{Learning Objective and Parameter Updates}
\label{sec:Var}

\subsubsection{Learning objective} We derive the learning objective from the standard ELBO. At time step \(k\), for any fixed PDF \(q(\mathbf h_k)\), the log-evidence satisfies
\begin{align}
    \log f_{\bm{\theta}}(\mathbf z_k)
    =
    \mathcal L_{k}(\bm{\theta})
    +
    \mathrm{KL}\!\left(
    q(\mathbf h_k)
    \,\|\, 
    f_{\bm{\theta}}(\mathbf h_k\mid \mathbf z_k)
    \right)
\end{align}
where
\begin{align}
    \mathcal L_{k}(\bm{\theta})
    =
    \mathbb E_{q(\mathbf h_k)}
    \left[
    \log f(\mathbf z_k\mid \mathbf h_k)
    +
    \log f_{\bm{\theta}}(\mathbf h_k)
    -
    \log q(\mathbf h_k)
    \right].
   \nn
\end{align}
Unlike conventional variational Bayes methods that restrict $q(\mathbf h_k)$ to a predefined parametric family, we do not specify $q(\mathbf h_k)$ explicitly; the learning objective below requires only its marginal posterior PDFs, which are represented by weighted particles provided by BP \cite{WeiRaoMey:SM27}. During the update of $\bm{\theta}$ at time step $k$, the predicted PDFs in \eqref{eq:joint_prior}, the contexts, and $q(\mathbf h_k)$ are held fixed, i.e., we do not differentiate through the computations of preceding time steps. By \eqref{eq:joint_prior}, the learnable parameters $\bm{\theta}$ appear only in the conditional
PDF~\eqref{eq:latent_field_prior}, which equals the signal-power PDF
$f_{\bm{\theta}}(\gamma_{k,n};\V{c}_{k,n})$ of the latent field by~\eqref{eq:cond_equals_field}. Hence, by \eqref{eq:phi-mixture}, the part of $\log f_{\bm{\theta}}(\mathbf h_k)$ that depends on $\bm{\theta}$ is $\sum_{n=1}^{N_k} r_{k,n}\log f_{\bm{\theta}}(\gamma_{k,n};\V{c}_{k,n})$. Since the likelihood $f(\mathbf z_k\mid\mathbf h_k)$ and $q(\mathbf h_k)$ do not depend on $\bm{\theta}$, and the expectation of $r_{k,n}\log f_{\bm{\theta}}(\gamma_{k,n};\V{c}_{k,n})$ with respect to $q(\mathbf h_k)$ depends only on the marginal posterior $q(\boldsymbol{\phi}_{k,n})$, maximizing the ELBO with respect to
$\bm{\theta}$ is equivalent to maximizing
\begin{align}
\mathcal J_k(\bm{\theta}) =&~
    \sum_{n=1}^{N_k}
    q(r_{k,n}=1) \label{eq:frame_objective}\\
   &\times \mathbb E_{q(\gamma_{k,n}\mid r_{k,n}=1)}
    \left[
    \log f_{\bm{\theta}}
    \big(\gamma_{k,n};\mathbf c_{k,n}\big)
    \right]\notag
\end{align}
where \(q(r_{k,n}=1)\) and \(q(\gamma_{k,n}\mid r_{k,n}=1)\) are obtained
from the marginal posterior \(q(\boldsymbol{\phi}_{k,n})\), and the log-PDF is given by \eqref{eq:flow_gamma}. At time step $k$, the accumulated objective is given by
\begin{equation}
\label{eq:objective}\mathcal J_{1:k}(\bm{\theta}) = \sum_{k'=1}^{k} \mathcal J_{k'}(\bm{\theta})
\end{equation}
which is evaluated across time and objects.

\subsubsection{Parameter updates} \looseness=-1 The posterior quantities of preceding time steps, i.e., the existence probabilities $q(r_{k',n}=1)$, the particles representing $q(\gamma_{k',n}\mid r_{k',n}=1)$, and the contexts $\V{c}_{k',n}$ with $k'<k$, are stored and held fixed; they are not recomputed with the updated parameters. At each time step $k$, $\bm{\theta}$ is initialized with its value from time step $k-1$ and updated by a few gradient-based optimization steps that increase $\mathcal J_{1:k}(\bm{\theta})$ (see Section~\ref{sec:experiments}). This procedure can be interpreted as an approximate, generalized variational expectation-maximization algorithm \cite{Bis:B06}, where the Bayesian inference step provides a particle-based approximate E-step and the normalizing flow update acts as the variational M-step. Fig.~\ref{fig:vb} summarizes the resulting inference/learning procedure at a single time step.

\section{Experiment and Evaluation}
\label{sec:experiments}
We first assess the proposed framework, referred to as \mbox{BP-TkBD} with normalizing flow (\mbox{BP-TkBD-NF}), in a simulated multisensor radar scenario, where the signal powers are generated by a known radar-equation model (Section~\ref{sec:exp_sim}). We then use two real-world datasets: a millimeter-wave radar dataset for meter-scale tracking in an urban environment (Section~\ref{sec:exp_radio}) and an acoustic dataset for kilometer-scale tracking in an underwater environment (Section~\ref{sec:exp_sonar}). For the acoustic setting, we additionally assess the design of the signal-power prior in a controlled simulation (Section~\ref{sec:latent_field_ablation}). In all experiments, new potential objects are introduced according to the birth model in \cite[Sec.~II-D]{LiaMey:J25}.

\looseness=-1 We report the generalized optimal sub-pattern assignment (GOSPA) metric \cite{RahGarSve:17} as a measure of performance. This metric has three main parameters $(\alpha, p, c)$: $\alpha$ controls the relative weighting of cardinality errors, $p$ specifies the metric order, and $c$ is the cutoff parameter for localization errors. We use the standard setting $(\alpha{=}2,\,p{=}1)$, and choose $c$ according to the spatial scale of the localization task. At time step $k$, let $\mathcal{X}_k$ and $\hat{\mathcal{X}}_k$ denote the sets of true and estimated object positions, respectively. For this setting, the GOSPA metric is given by
\begin{align}
	\mathrm{GOSPA}(\mathcal{X}_k,\hat{\mathcal{X}}_k)
	=&~\textstyle\sum\nolimits_{(\V{s},\hat{\V{s}})\in\mathcal{D}_k}\|\V{s}-\hat{\V{s}}\| + \tfrac{c}{2}\big(|\mathcal{X}_k|-|\mathcal{D}_k|\big)\notag\\
	&+ \tfrac{c}{2}\big(|\hat{\mathcal{X}}_k|-|\mathcal{D}_k|\big)
\end{align}
where the assignment $\mathcal{D}_k$, i.e., a set of pairs of true and estimated positions in which each position appears at most once, is chosen to minimize the right-hand side \cite{RahGarSve:17}. The three terms are the localization, missed-object, and false-object errors and correspond to the columns ``Loc,'' ``Miss,'' and ``False'' in Tables~\ref{tab:performance_uwcr} and~\ref{tab:performance_sonar}. The GOSPA is computed from 2-D object positions and averaged over time steps and scenes.

\subsection{Simulated Multisensor Radar Sensing}
\label{sec:exp_sim}
Because the true signal-power model is not available for the real-world datasets in Sections~\ref{sec:exp_radio} and~\ref{sec:exp_sonar}, the simulation makes it possible to compare the latent field learned during runtime directly with the power model that generated the data.

\subsubsection{Simulation setup}
The scenario comprises two frequency-modulated continuous-wave (FMCW) radar sensors $i\in\{1,2\}$ with low resolution, positioned at $[0~~0]^{\mathrm{T}}$ and $[50~~50]^{\mathrm{T}}$ in meters. Both sensors use a uniform linear array whose element spacing is half a wavelength and whose orientation is $\pi/4$. The measurements of each sensor constitute a separate measurement block, which has its own array manifold, signal powers $\gamma_{k,n}^{(i)}$, and noise variance $\eta_k^{(i)}$. Given these variables and the object states, the measurements of the two sensors are conditionally independent. Hence, the likelihood function factorizes into two Gaussian likelihood functions of the form \eqref{eq:gauss_lik} \cite{LiaMey:J25}, and the learning objective in \eqref{eq:frame_objective} contains an additional sum over the sensors. For each sensor, the array manifold in \eqref{eq:meas_model} takes the Kronecker\vspace{-.5mm} form
\begin{align}
	\V{a}(\V{x})
	&=
	\V{a}_f\otimes\V{a}_s\otimes\V{a}_\beta,
	\label{eq:radar_manifold}\\[.5mm]
	\V{v}(\xi,M)
	&=
	[1,e^{j\xi},\ldots,e^{j(M-1)\xi}]^{\mathrm{T}},
	\label{eq:radar_vandermonde}
\end{align}
where the sensor index $i$ is omitted and the order of the Kronecker factors corresponds to the order in which the entries of the measurement tensor are vectorized. The fast-time, slow-time, and angular factors are the Vandermonde vectors $\V{a}_f=\V{v}(\xi_f,M_f)$, $\V{a}_s=\V{v}(\xi_s,M_s)$, and $\V{a}_\beta=\V{v}(\xi_\beta,M_\beta)$. Their phase increments $\xi_f=4\pi\kappa T_{\mathrm{s}}R/c_0$, $\xi_s=4\pi T_{\mathrm{c}}\dot R/\lambda$, and $\xi_\beta=\pi\sin\beta$ depend on the range $R$, the radial velocity $\dot R$, and the angle of arrival $\beta$ of the object relative to the sensor, where $c_0$ and $\lambda$ denote the speed of light and the wavelength. The chirp rate is $\kappa=8$\,MHz/$\mu$s, the sampling interval is $T_{\mathrm{s}}=0.2\,\mu$s, and the pulse repetition interval is $T_{\mathrm{c}}=40\,\mu$s. Each sensor acquires $M_f=20$ samples in fast time and $M_s=10$ samples in slow time with an array of $M_\beta=10$ elements, which results in $C=2{,}000$ measurement entries.

\par To generate the data, each existing object is assigned a complex amplitude $\rho_{k,n}^{(i)}$ whose phase is drawn uniformly from $[0,2\pi)$ and whose squared magnitude follows the radar equation
\begin{align}
	\big|\rho_{k,n}^{(i)}\big|^2 = \frac{c_0^2}{f_c^2(4\pi)^3\big(R_{k,n}^{(i)}\big)^4}
	\label{eq:radar_equation}
\end{align}
with carrier frequency $f_c=77$\,GHz and range $R_{k,n}^{(i)}$ between object $n$ and sensor $i$ at time step $k$. This is the radar equation for unit transmit power, unit antenna gains, and a unit radar cross section. The amplitude magnitudes are therefore determined by the range instead of being drawn from the conditionally Gaussian model in \eqref{eq:rho_gamma}. We set the noise variance such that an object located at the center of the region of interest (ROI) $[0,50]\times[0,50]$\,m has a signal-to-noise ratio per measurement entry of $10\log_{10}\big(|\rho_{k,n}^{(i)}|^2/\eta_k^{(i)}\big)=-3$\,dB.

\par Over $80$ time steps, five objects with 2-D position and velocity states move inside the ROI according to a constant-velocity model \cite[Sec.~6.3.2]{ShaKirLi:B02}. The initial states of objects 1 and 2 are $[10~~25~~0.5~~0]^{\mathrm{T}}$ and $[25~~10~~0~~0.5]^{\mathrm{T}}$, where the position precedes the velocity, and these objects are present from time step $2$ to $72$ and from $5$ to $75$, respectively. Objects 3, 4, and 5 start from random initial states and are present from time step $10$ to $70$, $18$ to $78$, and $25$ to $75$, respectively. Both sensors share a single latent field, i.e., a single flow with parameters $\bm{\theta}$, whose context $\V{c}_{k,n}^{(i)}$ consists of the estimated range and angle of the object with respect to sensor $i$. All results are averaged across $100$ Monte Carlo runs.

\subsubsection{Baselines}
Three reference methods are considered. \mbox{MP-BP} and \mbox{SBL-BP} are detect-then-track methods that extract point detections by matching pursuit (MP) \cite{MalZha:J93} or SBL \cite{WipRao:J07,GerMecChrXenNan:J16}, respectively, and process them with the multisensor BP tracking method in \cite{MeyBraWilHla:J17} with $30{,}000$ kinematic-state particles. \mbox{BP-TkBD-RW} uses the same raw measurements and the same efficient likelihood evaluation as \mbox{BP-TkBD-NF}, but its signal powers evolve according to a prescribed random-walk model, as in \mbox{BP-TkBD} \cite{LiaMey:J25}. For both \mbox{BP-TkBD-RW} and \mbox{BP-TkBD-NF}, the kinematic states, signal powers, and noise variances are represented by $10{,}000$, $500$, and $500$ particles, respectively. The GOSPA cutoff is set to $c=5$\,m.

\begin{figure}[t]
	\centering
	\includegraphics[width=\columnwidth]{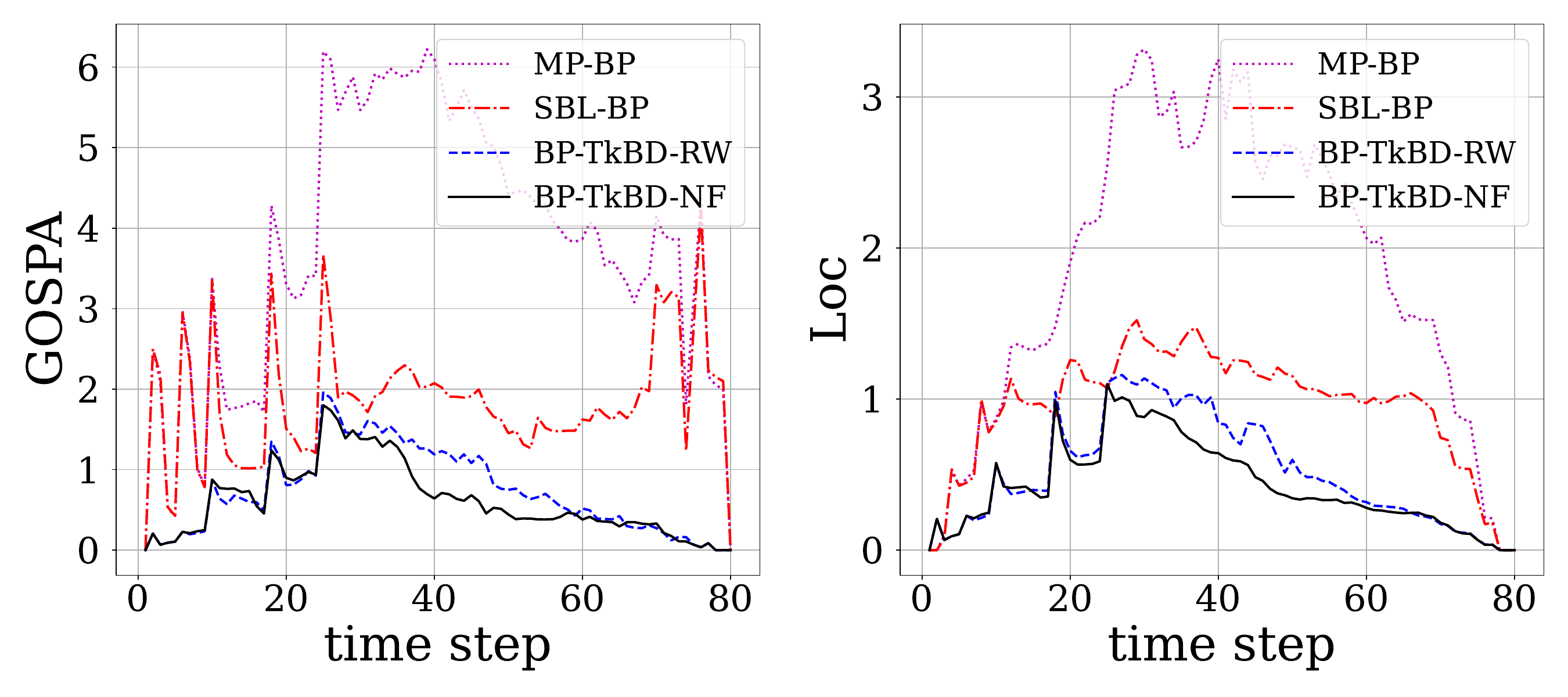}
	\caption{Total GOSPA error (left) and localization error (right) as functions of time in the simulated multisensor radar scenario, averaged over $100$ runs. Both errors are given in meters\vspace{-2mm}.}
	\label{fig:sim_gospa}
\end{figure}

\begin{figure*}[t]
	\centering
	\subfloat[Radar equation (left) and mean field (right)]{\includegraphics[width=0.62\textwidth]{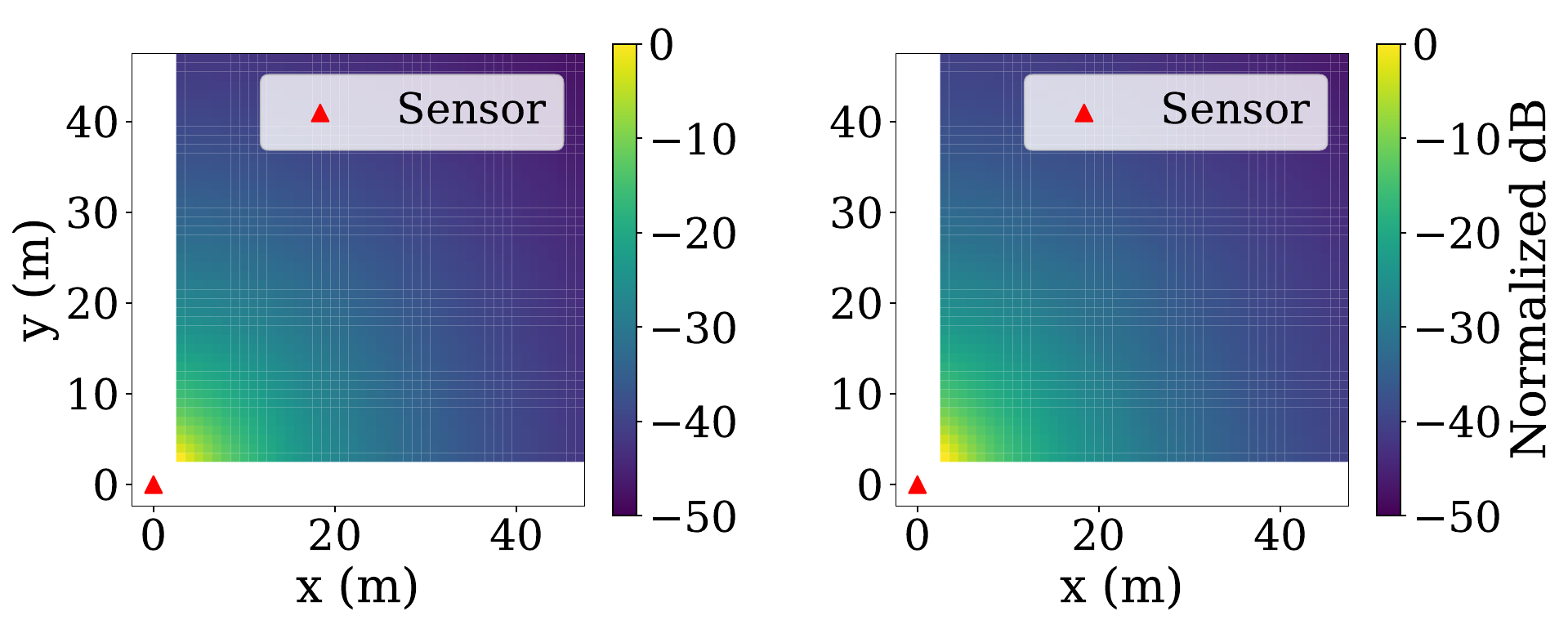}\label{fig:sim_intensity_maps}}\hfill
	\subfloat[Difference between the maps in (a)]{\includegraphics[width=0.31\textwidth]{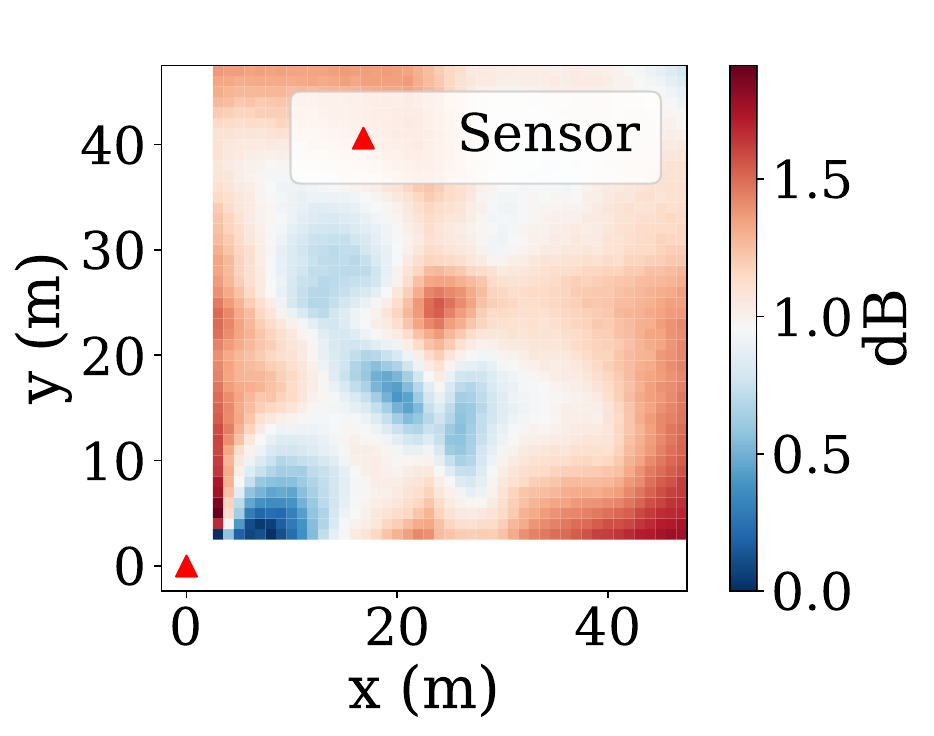}\label{fig:sim_intensity_diff}}
	\caption{Signal power of sensor 1 in the simulated multisensor radar scenario. (a) Normalized power in dB according to the radar equation \eqref{eq:radar_equation} (left) and according to the mean field $m_{\bm{\theta}}(\V{c})$ in \eqref{eq:mean_field} (right). (b) Pointwise deviation of the learned map from the radar-equation map\vspace{-2mm}.}
	\label{fig:sim_intensity}
\end{figure*}

\subsubsection{Results}
Fig.~\ref{fig:sim_gospa} depicts the time evolution of the GOSPA error and its localization component. The detect-then-track baselines yield the largest errors. For \mbox{SBL-BP}, the error exhibits peaks at the time steps where objects appear or disappear, which can be attributed to missed and false detections and the resulting data-association errors. The two TkBD methods achieve considerably lower errors. For approximately the first $30$ time steps, during which the flow has been updated with only limited posterior information, the proposed method and \mbox{BP-TkBD-RW} perform similarly. Afterward, the proposed method yields lower GOSPA and localization errors than \mbox{BP-TkBD-RW}, and the gap is largest between approximately $k=35$ and $k=55$. At the end of the sequence, the errors of the two methods are again similar. This suggests that the latent field describes the range-dependent signal powers more accurately than the random-walk model of \mbox{BP-TkBD-RW}.

\par For visualization, we use the mean field, i.e., the mean signal power of the latent field at context $\V{c}$, given by
\begin{align}
	m_{\bm{\theta}}(\V{c})
	\mathrel{=}
	\mathbb{E}\big[\Gamma_{\bm{\theta}}(\,\cdot\,,\V{c})\big]
	\mathrel{=}
	\int_{0}^{\infty}\gamma\, f_{\bm{\theta}}(\gamma;\V{c})\,\mathrm{d}\gamma
	\label{eq:mean_field}
\end{align}
which we approximate by averaging samples drawn from $f_{\bm{\theta}}(\gamma;\V{c})$. For sensor 1, the mean field is evaluated at the cells of a regular grid with a cell size of $0.5\times0.5$\,m$^2$, using $512$ samples per cell. Fig.~\ref{fig:sim_intensity} juxtaposes this map with the power predicted by the radar equation \eqref{eq:radar_equation}. Over the ROI, both maps decrease by about $50$\,dB from the region near the sensor to the far corner, while their pointwise difference stays below about $2$\,dB. The latent field, which is learned during runtime and without labeled data, thus accurately captures the range dependence of the signal power used to generate the data.

\par Tracking results at selected time steps, which show that the proposed method resolves closely spaced objects despite the low sensor resolution, are provided in \cite[Sec.~IV]{WeiRaoMey:SM27}.

\subsection{Active Radio Sensing}
\label{sec:exp_radio}

\begin{figure*}[t]
	\centering
	\subfloat[Camera reference]{\includegraphics[width=0.31\textwidth,height=3.8cm,keepaspectratio]{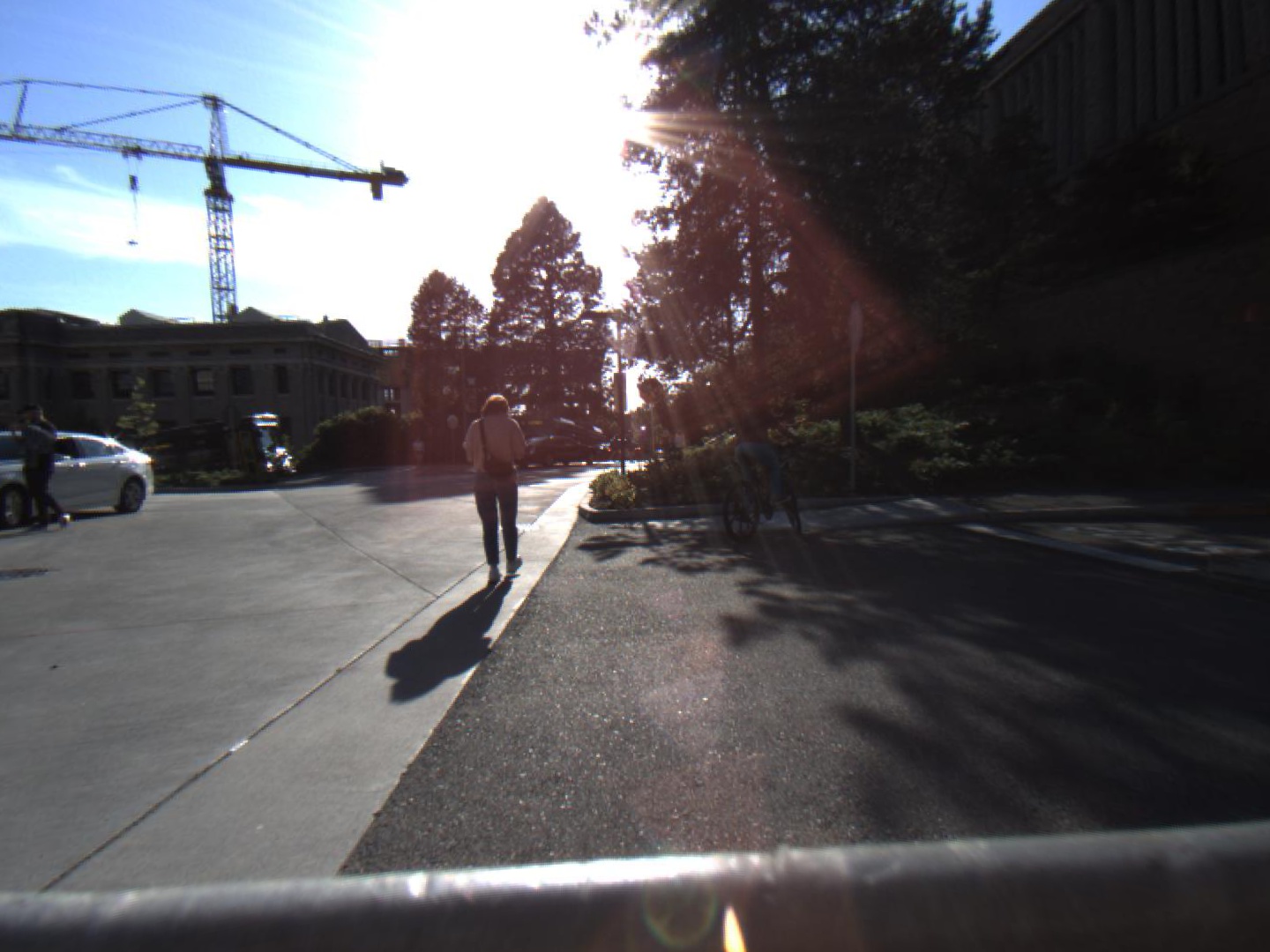}\label{fig:radar_camera}}\hfill
	\subfloat[Observed power]{\includegraphics[width=0.33\textwidth]{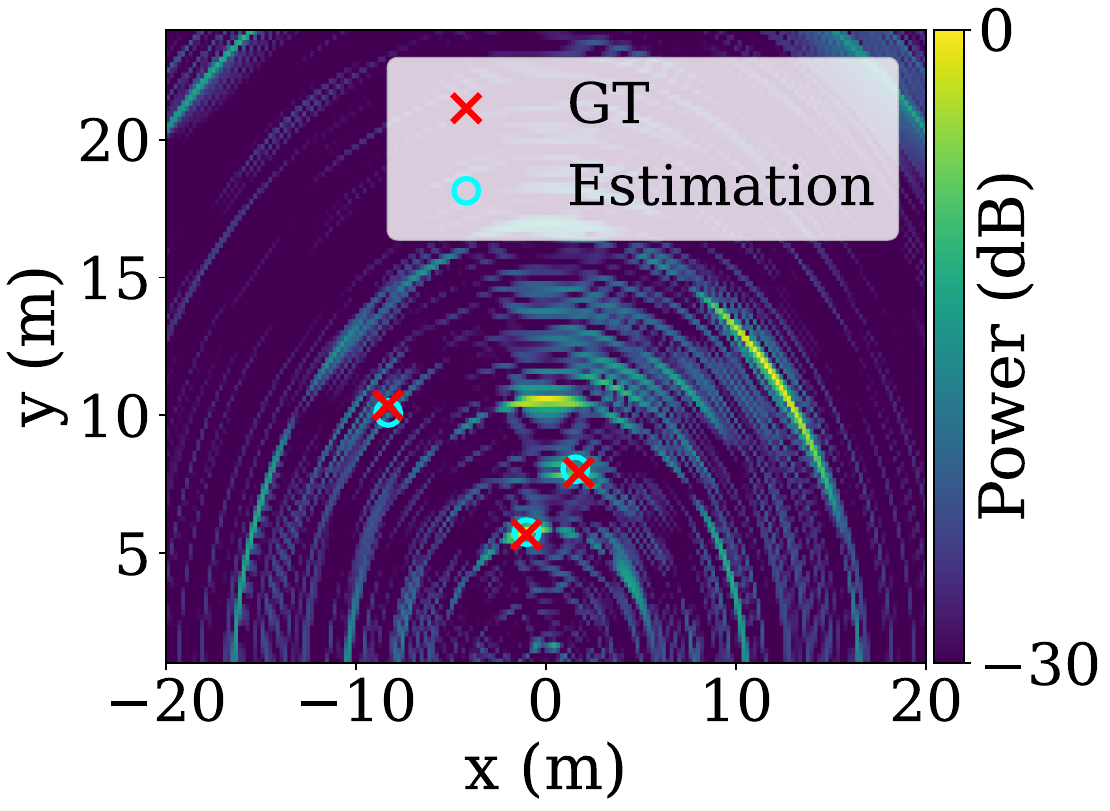}\label{fig:radar_spectrum}}\hfill
	\subfloat[Est. positions and velocities]{\includegraphics[width=0.31\textwidth]{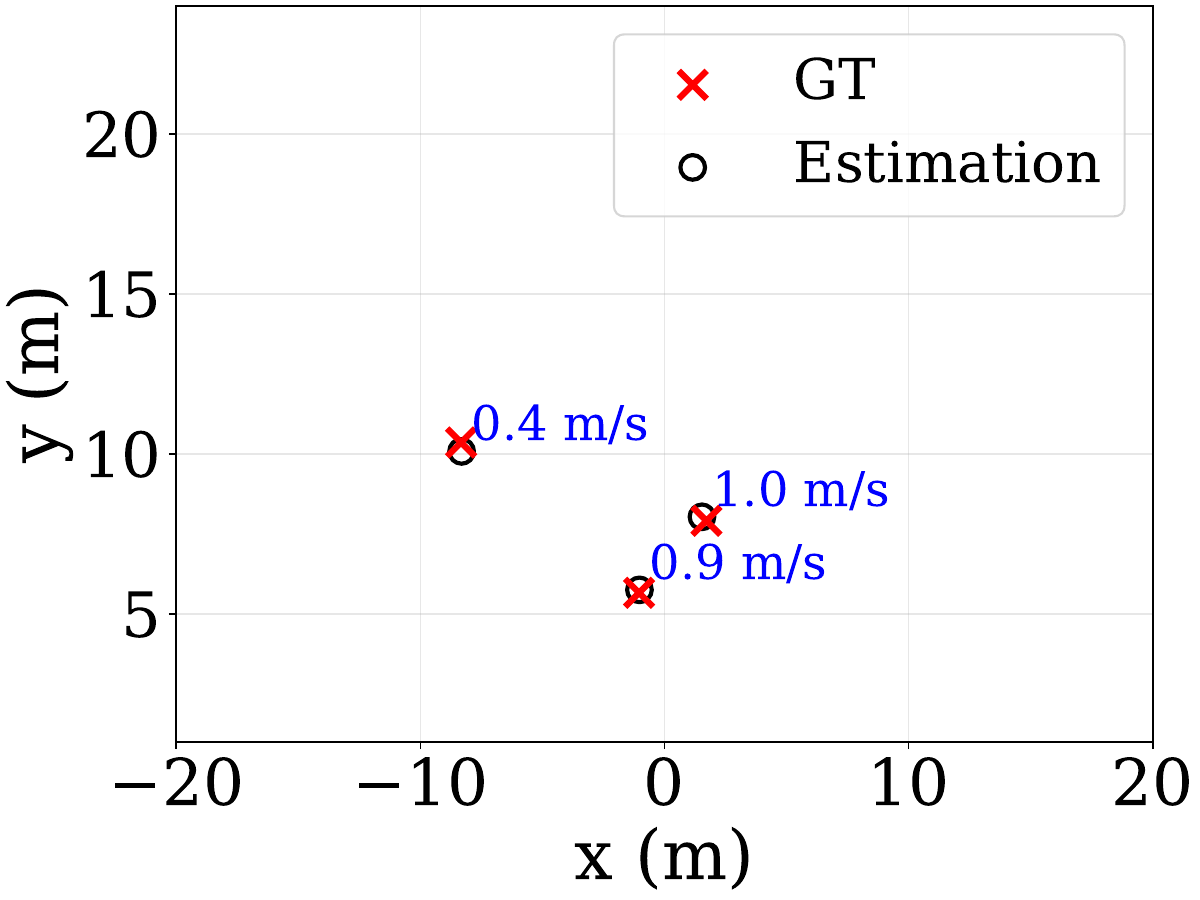}\label{fig:radar_xy}}
	\caption{Example scene from the UWCR dataset. (a) Synchronized camera view in an urban environment. (b) Observed spatial power computed from the received measurement. (c) Estimated positions and velocities of all moving objects. In (b) and (c), ``GT'' denotes the ground truth\vspace{-2mm}.}
	\label{fig:radar_results}
\end{figure*}

\begin{figure}[t]
	\centering
	\subfloat[Activity map]{\includegraphics[width=0.48\columnwidth]{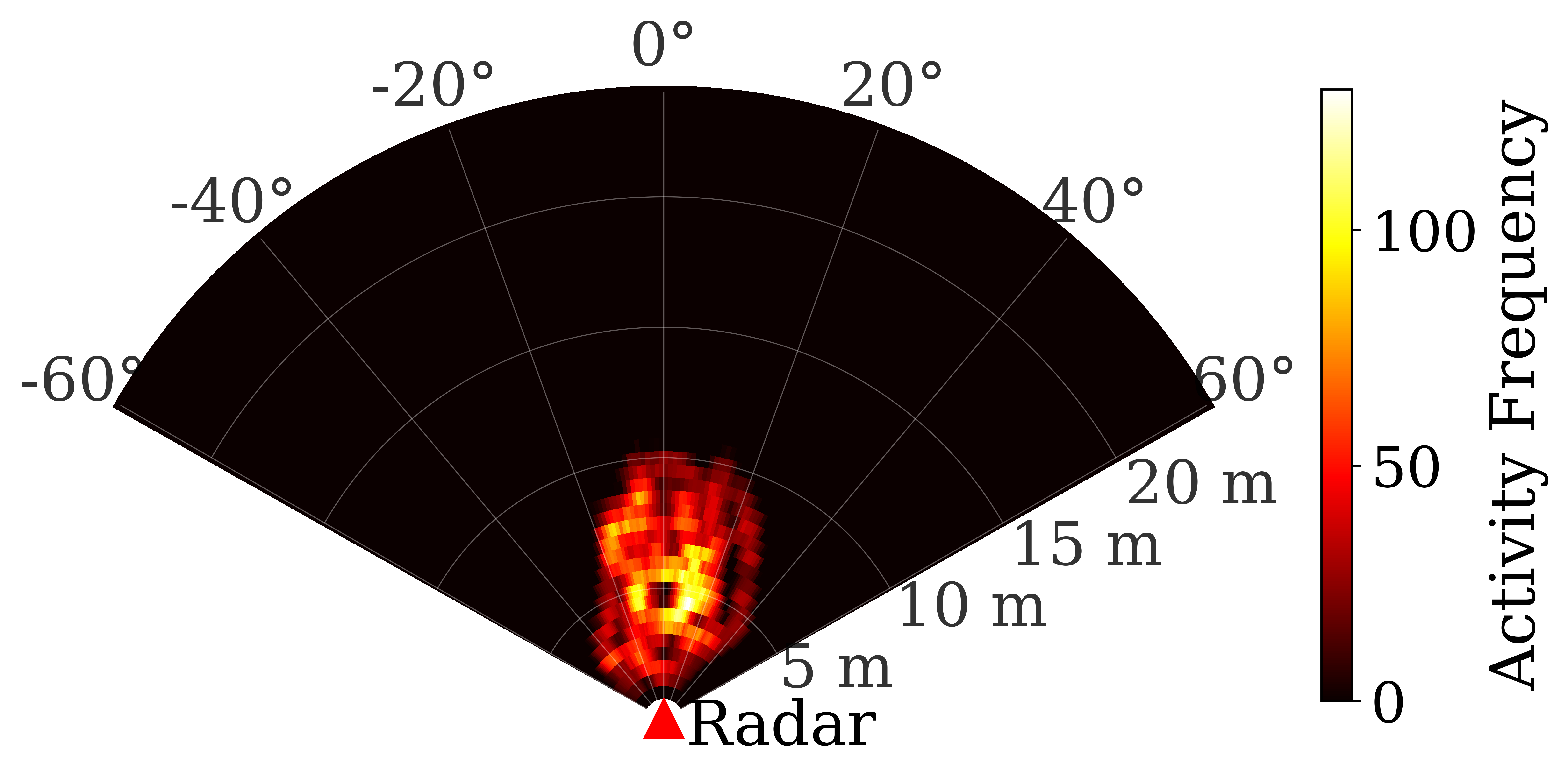}\label{fig:radar_coverage}}\hfill
	\subfloat[Mean field of the learned latent field]{\includegraphics[width=0.48\columnwidth]{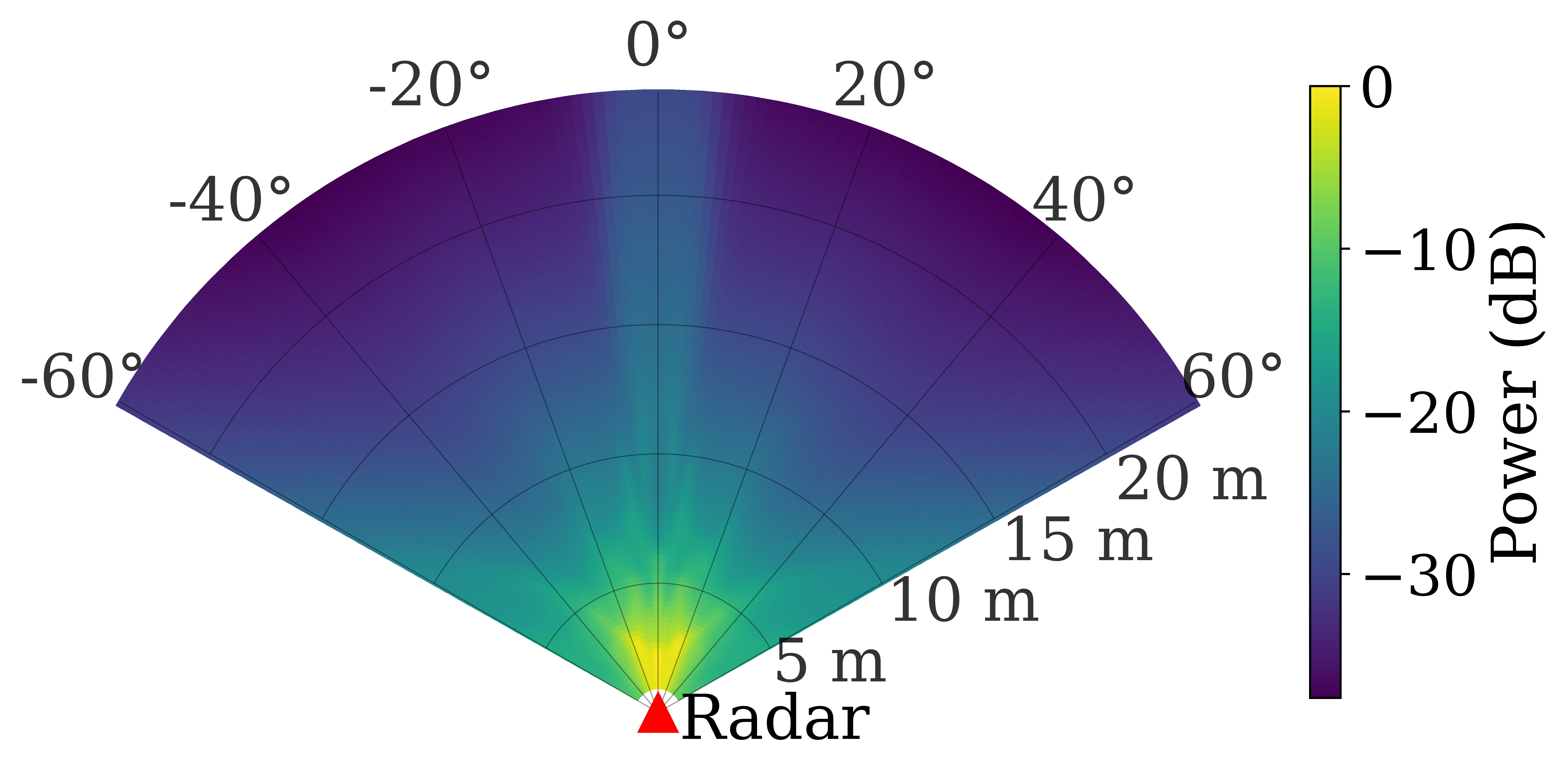}\label{fig:radar_intensity}}
	\caption{Visualization of the activity map and the learned latent field in a parking-lot environment\vspace{-4mm}.}
	\label{fig:coverage_intensity}
\end{figure}

We use the UWCR dataset \cite{GaoLuoXinRoyLiu:22}, which provides raw analog-to-digital converter (ADC) data from a 77\,GHz FMCW signal. A Texas Instruments AWR1843 board equipped with 2 transmit and 4 receive antennas serves as the radar sensor. At each time step, the ADC data of one radar frame is a four-dimensional complex-valued tensor of size $128 \times 255 \times 2 \times 4$ (fast-time samples $\times$ chirps $\times$ transmitters $\times$ receivers), yielding a measurement dimension of $C = 261{,}120$.

\subsubsection{Implementation details}
We first concatenate the two transmit paths and four receive paths into eight virtual channels and reorder each frame as a $128\times8\times255$ tensor (fast time $\times$ virtual array $\times$ slow time). The proposed method processes this complex-valued tensor directly: apart from scaling by a fixed normalization constant for numerical stability, it does not construct an intermediate range--Doppler (RD) or RA intensity map. A 3-D FFT is used only to initialize new potential objects: it identifies coarse search regions from peak responses without CFAR. A Gaussian birth PDF $f_{\mathrm{B},n}(\mathbf x_{k,n})$ is centered at the peak in each search region, and the posterior PDF of the new potential object is subsequently refined by sequential Bayesian inference on the raw ADC samples.

For this FMCW configuration, we instantiate the physics-based array manifold in \eqref{eq:meas_model} using the Kronecker structure in \eqref{eq:radar_manifold}, with the Kronecker factors ordered according to the vectorization of the $128\times8\times255$ tensor. The fast-time, slow-time, and virtual-array factors all have the Vandermonde form in \eqref{eq:radar_vandermonde}, with phase increments determined by range, radial velocity, and azimuth, respectively. These phase increments are given by the expressions in Section~\ref{sec:exp_sim}, evaluated with the chirp parameters of the dataset; as in Section~\ref{sec:exp_sim}, the virtual array has half-wavelength element spacing, and $\beta$ denotes the azimuth. This factorization retains the phase response prescribed by the radar geometry while the latent field models the context-dependent signal-power PDFs. The unit-magnitude Vandermonde factors give $\|\V{a}(\V{x})\|_2^2=C$, so an existing object's expected squared signal norm is $C\gamma_{k,n}$ in the measurement scaling used for inference.

The normalizing-flow context contains positional encodings of the range and azimuth obtained by MMSE estimation from the previous posterior, together with estimated radar-cross-section (RCS) statistics of each object. The flow is initialized at the beginning of each sequence and trained during runtime, without a separate training set or offline pretraining. We use Adam with a learning rate of $10^{-4}$ and perform five gradient steps per time step. Forward Bayesian inference uses $10{,}000$ particles for the object state $\V{x}$, $3{,}000$ particles for the signal power $\gamma$, and $500$ particles for the noise variance $\eta$. Additional details about the dataset are provided in \cite{GaoLuoXinRoyLiu:22}.

\begin{table*}[thp]
	\centering
	\caption{Performance comparison based on the UWCR dataset. $\downarrow$ indicates lower is better. GOSPA values are given in meters, runtimes are given in milliseconds per frame, and bold values indicate the best result in each column. For detect-then-track methods, the runtimes of the detector and the tracking method are listed separately. Performance is averaged over 50 Monte Carlo runs. \mbox{BP-TkBD-RW} uses the same efficient likelihood evaluation as \mbox{BP-TkBD-NF} but replaces the latent field with a random-walk signal-power model.}
	\label{tab:performance_uwcr}
			\begin{tabular}{l c c c c c c c}
			\toprule
			\textbf{Method} & \textbf{Methodology} & \textbf{Key Representation} & \textbf{GOSPA $\downarrow$} & \textbf{Loc $\downarrow$} & \textbf{Miss $\downarrow$} & \textbf{False $\downarrow$} & \textbf{Runtime $\downarrow$} \\
			\midrule
			3-D FFT & Detector & RAD Cube & 1.187 & 0.726 & 0.342 & 0.119 & \textbf{3.15} \\
			MUSIC \cite{Sch:J86}& Detector & RD Map + CFAR Dets. & 1.142 & 0.722 & 0.312 & 0.107 & 4.81 \\
			MVDR \cite{Cap:J69}& Detector & RD Map + CFAR Dets. & 1.125 &0.707 & 0.326 & 0.093 & 4.57 \\
			RODNet \cite{WanJiaGaoHwaXinLiu:C21} & Detector & Complex RA maps & 1.077 & 0.698 & 0.100 & 0.279 & 420.28 \\
			RAMP-CNN \cite{GaoXinRoyLiu:J21} & Detector & RA/RV/VA maps & 0.802 & 0.708 & 0.050 & 0.044 & 414.25 \\
			\midrule
			MVDR + BP \cite{MeyKroWilLauHlaBraWin:J18} & Det. + Track & Point Cloud & 0.871 & 0.759 & 0.048 &0.065 & 4.57 + 8.86 \\
			MVDR + PMBM \cite{GarWilGraSve:J18} & Det. + Track & Point Cloud & 0.877 & 0.753 & 0.089 & 0.035 & 4.57 + 8.63 \\
			MVDR + PHD \cite{Mah:J03} & Det. + Track & Point Cloud & 1.105 & 0.696 & 0.366 & 0.044 & 4.57 + 0.27 \\
			RAMP-CNN + BP \cite{MeyKroWilLauHlaBraWin:J18} & Det. + Track & Point Cloud & 0.765 & 0.690 & 0.040 & 0.035 & 414.25 + 9.18 \\
			\midrule
			\textbf{BP-TkBD-RW} & TkBD & Raw Complex (ADC) & 0.577 & 0.542 & 0.017 & 0.018 & 29.09 \\
			\textbf{BP-TkBD-NF} & TkBD & Raw Complex (ADC) & \textbf{0.552} & \textbf{0.526} & \textbf{0.013} & \textbf{0.013} & 69.86 \\
			\bottomrule
		\end{tabular}
\end{table*}

\subsubsection{Baselines}
We use the GOSPA metric with a cutoff of $c=5$\,m, averaged across all scenes and $50$ Monte Carlo runs. The methods in Table~\ref{tab:performance_uwcr} comprise detector-only methods, detect-then-track methods, and the proposed raw-measurement method. The classical detectors apply moving target indication (MTI) \cite{Ric:B14}, windowed range and Doppler FFTs, non-coherent accumulation, 2-D CFAR, angle estimation, and density-based spatial clustering of applications with noise (DBSCAN) \cite{EstKriSan:C96}. They differ in the angle estimator, which is a 3-D FFT that forms a range--angle--Doppler (RAD) cube, MUSIC \cite{Sch:J86}, or the minimum variance distortionless response (MVDR) beamformer \cite{Cap:J69}. MUSIC and MVDR use spatial smoothing and are applied only to the array snapshots at CFAR-selected RD cells, which yields the representation ``RD Map + CFAR Dets.'' in Table~\ref{tab:performance_uwcr}. The learning-based detectors RODNet \cite{WanJiaGaoHwaXinLiu:C21}, which we retrain on the UWCR dataset, and RAMP-CNN \cite{GaoXinRoyLiu:J21}, for which we use the authors' checkpoint, process FFT-derived RA maps and, in the case of RAMP-CNN, additional RV and VA maps. For both, we retain only the positions of moving objects and obtain point detections by clustering. The tracking methods BP \cite{MeyKroWilLauHlaBraWin:J18}, the probability hypothesis density (PHD) filter \cite{Mah:J03}, and the Poisson multi-Bernoulli mixture (PMBM) filter \cite{GarWilGraSve:J18} process these point detections, whereas the proposed method updates the multiobject posterior directly from the raw complex ADC samples after coarse birth initialization. Since the methods also differ in their measurement models and priors, this comparison does not isolate the effect of avoiding intermediate measurement compression.

\subsubsection{Results}
Table~\ref{tab:performance_uwcr} compares the proposed method with different detectors and tracking methods based on the UWCR dataset. The detectors rely on compressed representations, and the tracking methods operate on the resulting detection sets. These representations provide efficient summaries of the measurements but may discard information useful for downstream localization. Our method directly takes the complex ADC data as input without generating a compressed representation in a preprocessing step, apart from the coarse birth initialization. As shown in Table~\ref{tab:performance_uwcr}, our method achieves the lowest GOSPA. These results indicate that combining inference based on the raw data with a learned latent field is beneficial for this multiobject tracking task.

\par Table~\ref{tab:performance_uwcr} also reports the average runtime per frame, computed with single-precision arithmetic; for the learning-based detectors, it comprises network inference but excludes offline training and separately executed preprocessing. All runtimes are measured on the same workstation, which is equipped with an AMD Ryzen Threadripper PRO 5955WX CPU with 16 cores (32 threads), 128 GB of RAM, and an NVIDIA GeForce RTX 4090 GPU with 24 GB of memory, using Python 3.8 with PyTorch 2.0.1 and CUDA 11.8, running on Ubuntu 22.04. The proposed method requires $69.86$\,ms per frame, of which $21.45$\,ms, $45.06$\,ms, and $3.34$\,ms are spent on forward Bayesian inference, the flow update, and prior generation and initialization, respectively. This runtime exceeds those of the classical detectors and of the MVDR-based detect-then-track methods but is substantially lower than those of the learning-based detectors. Thus, the low-rank and Kronecker-structured likelihood evaluation makes sequential inference based on approximately $2.6\times10^5$ complex ADC samples per frame computationally feasible.

\par Fig.~\ref{fig:radar_results} shows a synchronized camera view of an example scene, the observed spatial power computed by beamforming, and the positions and velocities of the moving objects estimated by the proposed method. Although the strongest responses in the observed power are not located at the object positions, the estimated positions agree closely with the ground truth. A particle representation of the posterior PDF, which is sharply concentrated because the likelihood function of the proposed signal model is highly informative, is provided in \cite[Sec.~V]{WeiRaoMey:SM27}.

\par Fig.~\ref{fig:coverage_intensity} visualizes the latent field learned during runtime by the normalizing flow in a parking-lot environment. The activity map in Fig.~\ref{fig:coverage_intensity}(a) aggregates the spatial regions occupied by localized objects over all time steps, and therefore indicates the area from which measurements are collected. Fig.~\ref{fig:coverage_intensity}(b) shows the mean field \eqref{eq:mean_field} evaluated on a dense spatial grid. The learned field exhibits a dominant mainlobe around $0^\circ$ and weaker sidelobes near $\pm20^\circ$ and $\pm60^\circ$, which may reflect the radiation pattern of the radar antennas. Moreover, the power decreases with increasing range. Hence, the learned structure is not merely a copy of the activity map: while the activity map reflects where objects have been observed, the learned latent field describes how the mean signal power varies across space. In regions that are not covered by the activity map, e.g., at ranges below 5\,m and near $\pm60^\circ$, the latent field is an extrapolation determined by the context parameterization.
For this visualization, we intentionally omit object-specific features and use the polynomial positional context
\begin{align}
	\V{c}
	=
	\big[
	\bar R,\bar R^2,\ldots,\bar R^{U},\,
	\{\sin^2(u\bar\beta),\cos(u\bar\beta)\}_{u=1}^{U}
	\big]^{\mathrm{T}},
	\label{eq:radar_visualization_context}
\end{align}
where $\bar R$ and $\bar\beta$ denote normalized range and azimuth, respectively, and $U$ is the order of the encoding. This context encourages the flow to represent spatial propagation effects and enforces symmetry about $0^\circ$; hence, the symmetry in Fig.~\ref{fig:coverage_intensity}(b) results from this choice of context and is not learned from the data.

\par The evolution of the signal-power PDF through the successive flow transformations is illustrated in \cite[Sec.~V]{WeiRaoMey:SM27}.

\par Replacing the 3-D FFT peak that serves as the mean of each Gaussian birth PDF by detections from MVDR or RAMP-CNN changes the GOSPA by less than $0.005$\,m \cite[Sec.~V]{WeiRaoMey:SM27}. Hence, the birth PDF needs only to identify a plausible search region.

\subsection{Passive Acoustic Sensing}
\label{sec:exp_sonar}
\begin{figure}[t]
	\centering
	\subfloat[Sensing environment]{\includegraphics[width=0.6\columnwidth]{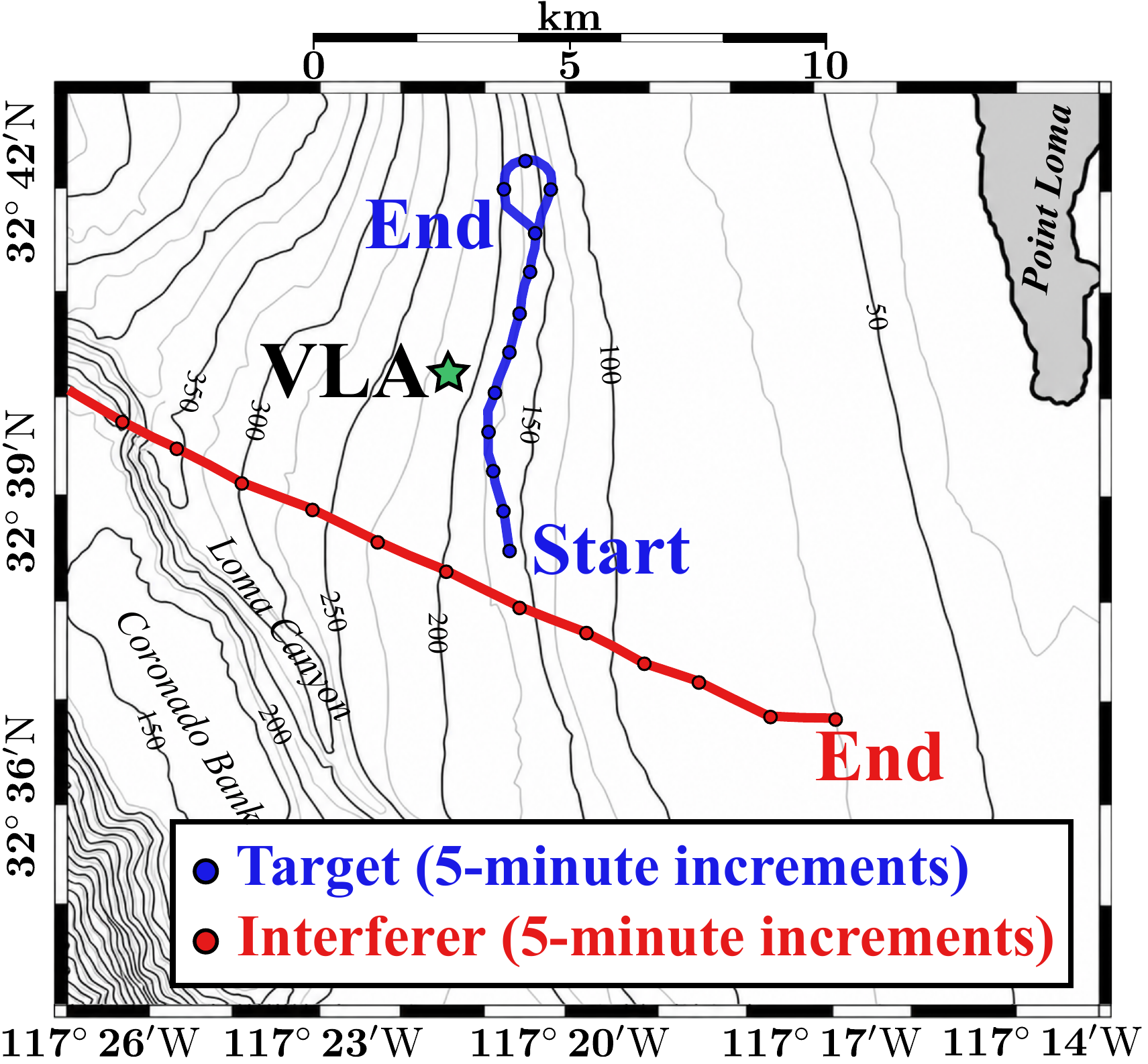}\label{fig:sonar_s59}}\\[4mm]
	\subfloat[Estimated trajectory over 5 minutes]{\includegraphics[scale=.0205]{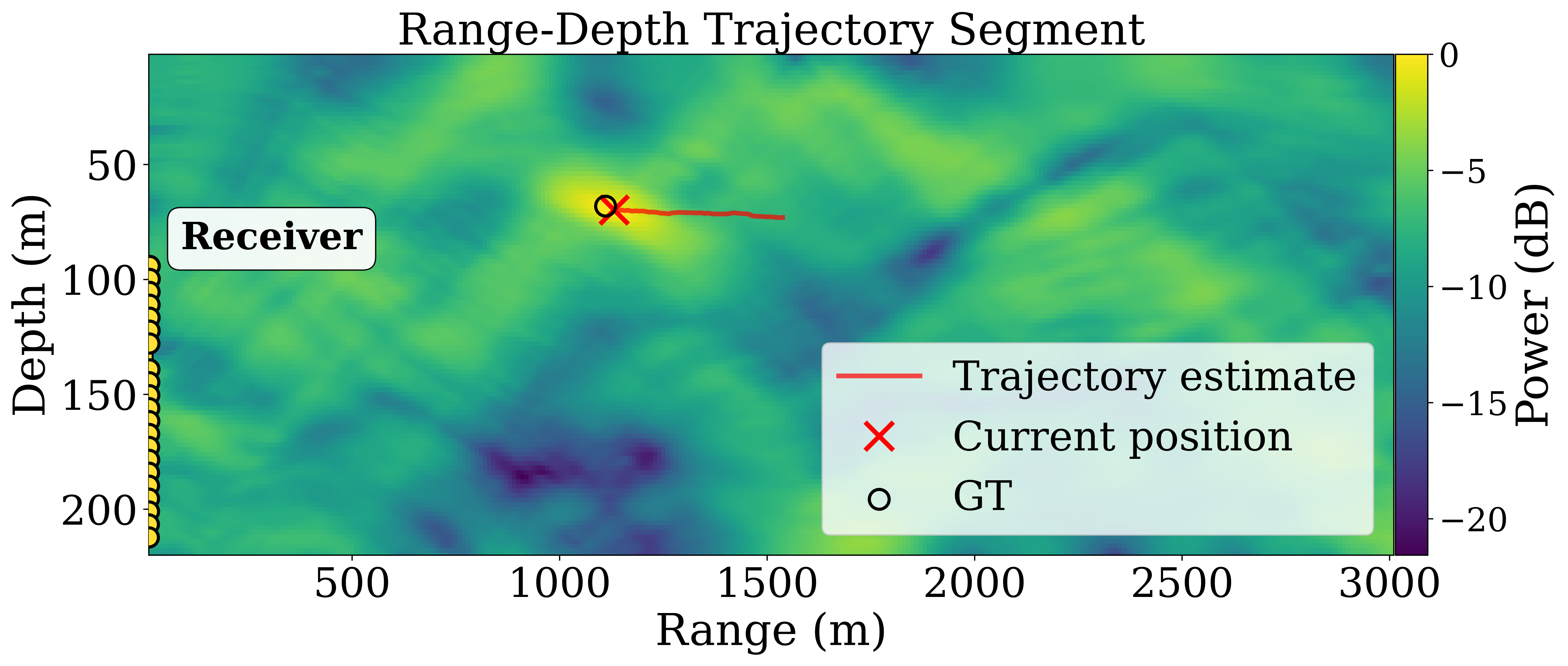}\label{fig:sonar_est}}
	\caption{Event S59 of the SWellEx-96 dataset. (a) Sensing environment. The blue trajectory corresponds to the source to be tracked, which is labeled ``Target'' in the legend, and the red trajectory corresponds to an interfering source. (b) Trajectory estimated by the proposed method. The background depicts the spatial power of the acoustic pressure field obtained by matched-field processing \cite{BagKupMik:J93}\vspace{-2mm}.}
	\label{fig:sonar_panel}
\end{figure}

\begin{table*}[thbp]
	\centering
	\caption{Performance comparison based on the SWellEx-96 dataset. $\downarrow$ indicates lower is better. GOSPA values are given in meters, and bold values indicate the best result in each column\vspace{1mm}.}
	\label{tab:performance_sonar}
			\begin{tabular}{l c c c c c c}
			\toprule
			\textbf{Method} & \textbf{Methodology} & \textbf{Key Representation} & \textbf{GOSPA $\downarrow$} & \textbf{Loc $\downarrow$} & \textbf{Miss $\downarrow$} & \textbf{False $\downarrow$} \\
			\midrule
			MP \cite{MalZha:J93} & Detector & Raw Complex & 35.973 & 30.322 & \textbf{2.825} & 2.825 \\
			SBL \cite{GerMecChrXenNan:J16} & Detector & Raw Complex & 36.312 & 30.148 & 3.253 & 2.911 \\
			\midrule
			MP-BP \cite{MeyKroWilLauHlaBraWin:J18}& Det. + Track & Point Cloud & 32.377 & 29.123 & 3.082 & 0.172 \\
			SBL-BP \cite{MeyKroWilLauHlaBraWin:J18} & Det. + Track & Point Cloud & 32.624 & 29.285 & 3.082 & 0.257 \\
			\midrule
			\textbf{BP-TkBD-NF} & TkBD & Raw Complex & \textbf{30.832 $\pm$ 0.09} & \textbf{27.835} & 2.997 & \textbf{0.000} \\
			\bottomrule
		\end{tabular}
		\vspace{-1mm}
\end{table*}

\begin{figure*}[t]
	\centering
	\includegraphics[width=0.75\textwidth]{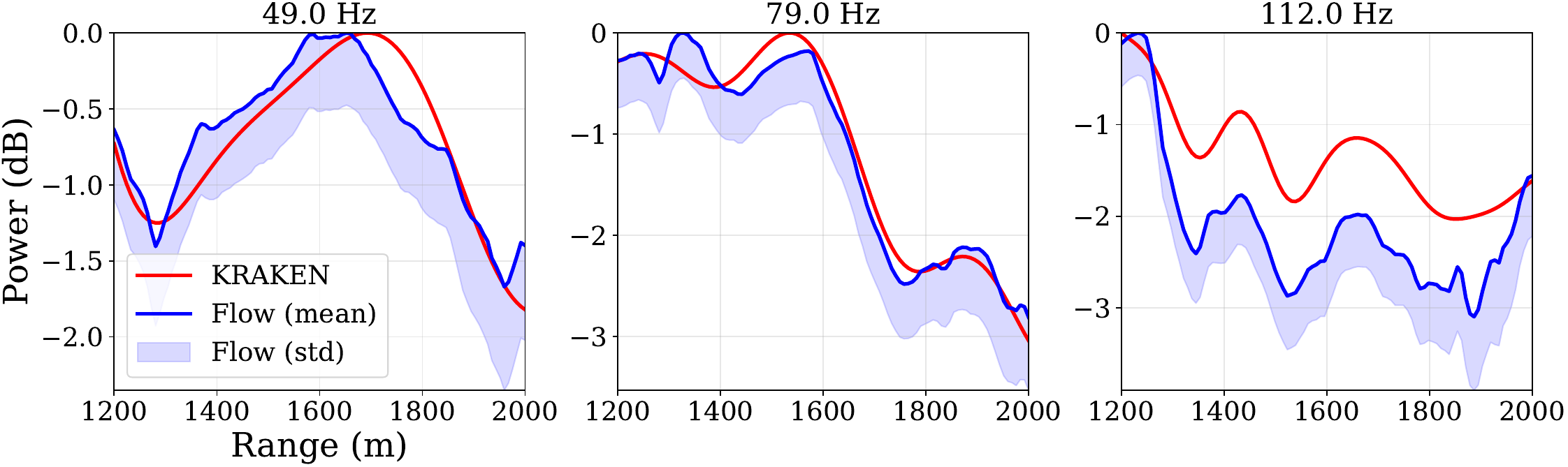}
	\vspace{-2mm}
	\caption{Normalized power as a function of range at a source depth of 70~m for the three lowest tones at 49, 79, and 112\,Hz, according to KRAKEN and to the learned latent field. Each curve is normalized to a maximum of 0~dB. For the latent field, the mean field $m_{\bm{\theta}}(\V{c})$ and, for readability, only the lower half of the $\pm1$ standard-deviation band (shaded) are shown\vspace{-2mm}.}
	\label{fig:sonar_curves}
\end{figure*}
For underwater acoustic multiobject tracking, we use the SWellEx-96 dataset \cite{BooAbaSchHod:J00}. In particular, we use data from Events S5 and S59 recorded on May 11, 1996, from 00:05 to 00:25 and on May 13, 1996, from 12:00 to 12:20, respectively. The experiment was conducted approximately 12\,km southwest of Point Loma, San Diego, California, in water approximately 216.5\,m deep. A 21-element vertical line array (VLA) with an aperture of 120\,m serves as the receiver. Fig.~\ref{fig:sonar_panel} shows the layout of Event S59 and the corresponding estimation results: the R/V~Sproul towed an underwater acoustic source (blue) northward along an isobath while a surface-ship interferer (red) simultaneously traversed the area. We use seven tonal signals transmitted by the underwater source, with frequencies from 49\,Hz to 388\,Hz, which are recorded by the 21 hydrophones.

\subsubsection{Implementation details}
The physical measurements are the raw time series from the 21 VLA hydrophones. We apply the same front end to all methods: short-time Fourier analysis extracts complex coefficients at the seven transmitted tones, and the measurement of each time step consists of three consecutive snapshots. The hydrophone signals are sampled at 1500\,Hz, and the short-time Fourier analysis uses a Hamming window with an FFT length of $8{,}192$ samples, i.e., approximately $5.46$\,s, and an overlap of $75\%$. Consecutive snapshots are therefore $2{,}048$ samples apart, and the time between consecutive time steps is $4.096$\,s. Each tone defines a dictionary, and each tone/snapshot block contains 21 complex hydrophone coefficients. We use the observation model for multiple dictionaries and snapshots in \cite[Sec.~II-A]{LiaMey:J25}: kinematic states and existence variables are shared across blocks, signal powers and noise variances are dictionary-specific and shared across snapshots, and the likelihood factorizes across blocks given these variables. The proposed method performs inference jointly across these tones and snapshots without first reducing them to point detections.

For an underwater source state $\V{x}=[R,\dot R,z,\dot z]^{\mathrm{T}}$, where $R$, $\dot R$, $z$, and $\dot z$ denote range, range rate, depth, and depth rate, respectively, as in \cite{LiaMey:J25}, the array manifold depends only on $R$ and $z$ and must account for multipath propagation and dispersion in the shallow-water waveguide; a far-field plane-wave approximation is therefore inadequate. For one tone, with its frequency index suppressed, we define the 21-component unit-norm replica vector $\V{a}(\V{x}) \rmv=\rmv \V{p}(\V{x}) / \|\V{p}(\V{x})\|_2 \rmv\in\rmv \mathbb{C}^{21}$, where $\V{p}(\V{x})$ is the complex pressure vector with entries $p_i(R,z) \approx
	\sum_{m=1}^{M}
	\Psi_m(z)\Psi_m(z_i)
	\frac{e^{j k_m R}}{\sqrt{k_m R}}$ for each hydrophone $i \in \{1,\dots,21 \}$. Here, $\Psi_m(\cdot)$ is the depth function of mode $m$, $k_m$ is its horizontal wavenumber, and $z_i$ is the depth of hydrophone $i$ \cite{JenKupPorSchTol:B11}. The unit-norm replica preserves the relative complex response across hydrophones and satisfies $\|\V{a}(\V{x})\|_2^2=1$. Thus, an existing object's coefficient power equals its expected squared signal norm within one tone/snapshot block, in the chosen measurement scaling. This formulation follows the matched-field processing paradigm \cite{BagKupMik:J93}, in which the received array data are coherently compared with modeled wavefields for candidate source positions. Note that the interferer is not tracked, since its contribution to the measurements is suppressed by the matched-field processing approach, which only considers field values at the seven tones transmitted by the acoustic source~\cite{BagKupMik:J93}.

The flow context contains the tone frequency and a positional encoding of the MMSE range and depth estimates from the preceding posterior. This choice corresponds to the flow with spatial context only in the ablation study of Section~\ref{sec:latent_field_ablation}. For example, the normalized range $\bar R$ is encoded as
\begin{align*}
	[
	\sin(2^0\pi\bar R),\cos(2^0\pi\bar R),
	\ldots,
	\sin(2^{U-1}\pi\bar R),\cos(2^{U-1}\pi\bar R)
	]^{\mathrm{T}}
\end{align*}
with an analogous encoding for depth, where $U$ is the number of encoding frequencies. We train the normalizing flow during runtime using Adam with a learning rate of $10^{-4}$ and perform five gradient steps per time step. The source follows a constant-velocity motion model in range and depth with zero-mean Gaussian driving noise of covariance $\mathrm{diag}(10^{-3},10^{-5})$, as described in \cite{LiaMey:J25}. Sequential inference uses $10{,}000$ particles for $\V{x}$, $3{,}000$ particles for $\gamma$, and $1{,}000$ particles for $\eta$.

\subsubsection{Baselines}
We use the GOSPA metric with a cutoff of $c=100$\,m. MP \cite{MalZha:J93} represents the measurements of each time step as a sparse combination of range--depth replica vectors. It greedily selects the dictionary atom most aligned with the residual energy averaged across tones and snapshots, estimates its coefficient, and removes its contribution. SBL \cite{GerMecChrXenNan:J16} uses the same dictionary but estimates a sparse probabilistic representation by evidence maximization, shrinking the variances of inactive grid cells toward zero. The detect-then-track methods \mbox{MP-BP} and \mbox{SBL-BP} pass the resulting point detections to BP \cite{MeyKroWilLauHlaBraWin:J18}. In contrast, the proposed method retains the joint complex measurements across all seven tones and three snapshots and conditions its normalizing flow on the estimated source location and tone frequency.

This comparison therefore parallels the radio experiment: the baseline detectors and tracking methods operate on an intermediate sparse representation, whereas the proposed method performs the posterior update before such compression. The acoustic setting additionally tests whether the learned latent field can recover frequency-dependent propagation structure in a dispersive multipath environment.
\subsubsection{Results}
The results in Table~\ref{tab:performance_sonar} lead to conclusions similar to those of the radio experiment in Section~\ref{sec:exp_radio}. Detector-only methods exhibit the largest errors. Applying a tracking method to the detections improves performance by exploiting temporal and spatial structure, but the overall accuracy remains limited by the quality of the upstream point estimates. In contrast, the proposed method achieves the lowest GOSPA as well as the lowest localization and false-object errors by jointly exploiting raw complex measurements and a learned latent field.

\subsubsection{Validation of the latent field model by comparison with KRAKEN}
We compare the learned latent field with the power predicted by the KRAKEN normal-mode program \cite{Por:91}. Fig.~\ref{fig:sonar_curves} shows the normalized power of both fields as a function of range for the three lowest tones at 49, 79, and 112\,Hz. KRAKEN computes a deterministic propagation prediction from an idealized waveguide model specified by the sound-speed profile, seabed parameters, and array geometry. We use this prediction as a physics-based reference rather than ground truth: the flow is trained only through the variational objective and is not supervised by KRAKEN amplitudes.

For a quantitative comparison, we evaluate the mean field $m_{\bm{\theta}}(\V{c})$ and the KRAKEN-predicted power as a function of range for the fixed estimated source depth of 70~m. Note that in this mildly range-dependent environment, the estimated source position is deeper than the true source position because of the mirage effect \cite{DspMurHodBooSch:J99}. The learned and KRAKEN-predicted fields follow the same range-dependent trend at low frequencies. The corresponding curves for the tones at 148, 201, 283, and 388\,Hz, which are provided in \cite[Sec.~VI]{WeiRaoMey:SM27}, show a growing gap at higher frequencies.
 At these frequencies and ranges, volume absorption in seawater is negligible, whereas bottom attenuation and boundary scattering increase with frequency and can be underrepresented by an idealized normal-mode calculation. Moreover, the modes at higher frequencies are more sensitive to small mismatches in the sound-speed profile, seabed properties, and array geometry. Consequently, KRAKEN itself may also deviate from the unknown true range dependence of the signal power at these frequencies. In addition, the latent field may be less accurate at higher frequencies, because the modal interference pattern varies on shorter range scales, which the chosen positional encoding may not resolve.

\subsubsection{Ablation study of the signal-power prior}
\label{sec:latent_field_ablation}
In a controlled acoustic simulation in which KRAKEN generates the measurements, we compare flow-based and parametric signal-power priors while keeping the measurement model, inference procedure, and object trajectories fixed. A flow whose context contains summary statistics of the preceding posterior achieves the lowest GOSPA of $21.44$, compared with $22.13$ for a flow with spatial context only, $23.57$ for a flow with a shifted base PDF, $27.16$ for a context-conditioned Gamma prior, and $31.92$ for a Gamma random walk. These results indicate that both spatial conditioning and a flexible density model are important, and that temporal information is most effective when it is supplied through the flow context. The simulation setup, the compared priors, and a detailed discussion are provided in \cite[Sec.~VII]{WeiRaoMey:SM27}\vspace{-4mm}.

\section{Conclusion}
\label{sec:conclusion}
We proposed a Bayesian framework for multiobject tracking that directly exploits raw complex-valued measurements. Our approach integrates a physics-based array manifold with a learnable latent field that is parameterized by a normalizing flow and optimized during runtime via variational inference. In simulations with a known radar-equation power model, the latent field learned during runtime closely reproduced this model. On real active radio and passive acoustic datasets, the proposed framework achieved lower GOSPA errors than detector-only and detect-then-track baselines. For the acoustic data, the learned latent field followed the same range-dependent trend as KRAKEN predictions at low frequencies. These results indicate that performing inference on raw complex-valued signals while learning a latent field that captures the effects of the objects and the propagation environment offers a scalable new paradigm for wireless sensing. This paradigm can enable safer robot navigation, improved situational awareness in search and rescue, and more effective environmental monitoring\vspace{0mm}.

\renewcommand{\baselinestretch}{0.97}
\bibliographystyle{IEEEtran}
\bibliography{IEEEabrv,StringDefinitions,SALBooks,SALPapers,temp}

\begin{thebibliography}{10}
\providecommand{\url}[1]{#1}
\csname url@samestyle\endcsname
\providecommand{\newblock}{\relax}
\providecommand{\bibinfo}[2]{#2}
\providecommand{\BIBentrySTDinterwordspacing}{\spaceskip=0pt\relax}
\providecommand{\BIBentryALTinterwordstretchfactor}{4}
\providecommand{\BIBentryALTinterwordspacing}{\spaceskip=\fontdimen2\font plus
\BIBentryALTinterwordstretchfactor\fontdimen3\font minus
  \fontdimen4\font\relax}
\providecommand{\BIBforeignlanguage}[2]{{%
\expandafter\ifx\csname l@#1\endcsname\relax
\typeout{** WARNING: IEEEtran.bst: No hyphenation pattern has been}%
\typeout{** loaded for the language `#1'. Using the pattern for}%
\typeout{** the default language instead.}%
\else
\language=\csname l@#1\endcsname
\fi
#2}}
\providecommand{\BIBdecl}{\relax}
\BIBdecl

\bibitem{PatAshKypHerMosCor:05}
N.~Patwari, J.~N. Ash, S.~Kyperountas, A.~O. Hero, III, R.~L. Moses, and N.~S.
  Correal, ``Locating the nodes: Cooperative localization in wireless sensor
  networks,'' \emph{{IEEE} Signal Process. Mag.}, vol.~22, no.~4, pp. 54--69,
  Jul. 2005.

\bibitem{WymLieWin:09}
H.~Wymeersch, J.~Lien, and M.~Z. Win, ``Cooperative localization in wireless
  networks,'' \emph{Proc. {IEEE}}, vol.~97, no.~2, pp. 427--450, Feb. 2009.

\bibitem{Kay:B93}
S.~M. Kay, \emph{{Fundamentals of Statistical Signal Processing: Estimation
  Theory}}.\hskip 1em plus 0.5em minus 0.4em\relax Upper Saddle River, NJ, USA:
  Prentice-Hall, 1993.

\bibitem{KriVib:96}
H.~Krim and M.~Viberg, ``Two decades of array signal processing research: {T}he
  parametric approach,'' \emph{{IEEE} Signal Process. Mag.}, vol.~13, no.~4,
  pp. 67--94, Jul. 1996.

\bibitem{Cap:J69}
J.~Capon, ``High-resolution frequency-wavenumber spectrum analysis,''
  \emph{Proc. {IEEE}}, vol.~57, no.~8, pp. 1408--1418, Aug. 1969.

\bibitem{GaoXinRoyLiu:J21}
X.~Gao, G.~Xing, S.~Roy, and H.~Liu, ``{RAMP-CNN}: A novel neural network for
  enhanced automotive radar object recognition,'' \emph{IEEE Sensors J.},
  vol.~21, no.~4, pp. 5119--5132, Feb. 2021.

\bibitem{WanJiaGaoHwaXinLiu:C21}
Y.~Wang, Z.~Jiang, X.~Gao, J.-N. Hwang, G.~Xing, and H.~Liu, ``{RODN}et: Radar
  object detection using cross-modal supervision,'' in \emph{Proc. WACV-21},
  Waikoloa, HI, USA, Jan. 2021, pp. 504--513.

\bibitem{Mah:B07}
R.~P.~S. Mahler, \emph{{Statistical Multisource-Multitarget Information
  Fusion}}.\hskip 1em plus 0.5em minus 0.4em\relax Norwood, MA, USA: Artech
  House, 2007.

\bibitem{BarWilTia:B11}
Y.~Bar-Shalom, P.~K. Willett, and X.~Tian, \emph{{Tracking and Data Fusion: A
  Handbook of Algorithms}}.\hskip 1em plus 0.5em minus 0.4em\relax Storrs, CT,
  USA: YBS Publishing, 2011.

\bibitem{RutGorMas:J05}
M.~G. Rutten, N.~J. Gordon, and S.~Maskell, ``Recursive track-before-detect
  with target amplitude fluctuations,'' \emph{IEE Proc. Radar Sonar Navig.},
  vol. 152, no.~5, pp. 345--352, Oct. 2005.

\bibitem{RisVoVoFar:J13}
B.~Ristic, B.-T. Vo, B.-N. Vo, and A.~Farina, ``A tutorial on {Bernoulli}
  filters: {Theory}, implementation and applications,'' \emph{{IEEE} Trans.
  Signal Process.}, vol.~61, no.~13, pp. 3406--3430, Jul. 2013.

\bibitem{FisTsaTapNurShl:C24}
E.~Fishel, N.~Tsarov, T.~Tapiro, I.~Nuri, and N.~Shlezinger, ``Learn to
  track-before-detect via neural dynamic programming,'' in \emph{Proc. IEEE
  ICASSP-24}, Seoul, South Korea, Apr. 2024, pp. 9586--9590.

\bibitem{LiaMey:J25}
\BIBentryALTinterwordspacing
M.~Liang and F.~Meyer, ``Sparse signal reconstruction for
  track-before-detect,'' \emph{{IEEE} Trans. Signal Process.}, 2026, in press.
  [Online]. Available: \url{https://arxiv.org/abs/2506.18177v4}
\BIBentrySTDinterwordspacing

\bibitem{KscFreLoe:01}
F.~R. Kschischang, B.~J. Frey, and H.-A. Loeliger, ``Factor graphs and the
  sum-product algorithm,'' \emph{{IEEE} Trans. Inf. Theory}, vol.~47, no.~2,
  pp. 498--519, Feb. 2001.

\bibitem{DouFreGor:01}
A.~Doucet, N.~{de Freitas}, and N.~Gordon, Eds., \emph{Sequential {M}onte
  {C}arlo Methods in Practice}.\hskip 1em plus 0.5em minus 0.4em\relax New
  York, NY, USA: Springer, 2001.

\bibitem{RezMoh:C15}
D.~J. Rezende and S.~Mohamed, ``Variational inference with normalizing flows,''
  in \emph{Proc. ICML-15}, Lille, France, Jul. 2015, pp. 1530--1538.

\bibitem{Tur:J60}
G.~L. Turin, ``An introduction to matched filters,'' \emph{IRE Trans. Inf.
  Theory}, vol.~6, no.~3, pp. 311--329, Jun. 1960.

\bibitem{VanBuc:88}
B.~D. Van~Veen and K.~M. Buckley, ``Beamforming: {A} versatile approach to
  spatial filtering,'' \emph{{IEEE} {ASSP} Mag.}, vol.~5, no.~2, pp. 4--24,
  Apr. 1988.

\bibitem{Sch:J86}
R.~O. Schmidt, ``Multiple emitter location and signal parameter estimation,''
  \emph{IEEE Trans. Antennas Propag.}, vol.~34, no.~3, pp. 276--280, Mar. 1986.

\bibitem{Tip:J01}
M.~E. Tipping, ``Sparse {B}ayesian learning and the relevance vector machine,''
  \emph{J. Mach. Learn. Res.}, vol.~1, pp. 211--244, Jun. 2001.

\bibitem{GerMecChrXenNan:J16}
P.~Gerstoft, C.~F. Mecklenbräuker, A.~Xenaki, and S.~Nannuru, ``Multisnapshot
  sparse {B}ayesian learning for {DOA},'' \emph{{IEEE} Signal Process. Lett.},
  vol.~23, no.~10, pp. 1469--1473, Oct. 2016.

\bibitem{Roh:J83}
H.~Rohling, ``Radar {CFAR} thresholding in clutter and multiple target
  situations,'' \emph{{IEEE} Trans. Aerosp. Electron. Syst.}, vol. AES-19,
  no.~4, pp. 608--621, Jul. 1983.

\bibitem{WanGaoMaoPan:J17}
X.~Wang, L.~Gao, S.~Mao, and S.~Pandey, ``{CSI}-based fingerprinting for indoor
  localization: A deep learning approach,'' \emph{{IEEE} Trans. Veh. Technol.},
  vol.~66, no.~1, pp. 763--776, Jan. 2017.

\bibitem{AyyAruWu:C20}
R.~Ayyalasomayajula, A.~Arun, C.~Wu, S.~Sharma, A.~R. Sethi, D.~Vasisht, and
  D.~Bharadia, ``Deep learning based wireless localization for indoor
  navigation,'' in \emph{Proc. ACM MobiCom-20}, London, U.K., Sep. 2020, pp.
  214--227.

\bibitem{BarLi:95}
Y.~Bar-Shalom and X.-R. Li, \emph{{Multitarget-Multisensor Tracking: Principles
  and Techniques}}.\hskip 1em plus 0.5em minus 0.4em\relax Storrs, CT, USA: YBS
  Publishing, 1995.

\bibitem{ForBarSch:83}
T.~E. Fortmann, Y.~Bar-Shalom, and M.~Scheffe, ``Sonar tracking of multiple
  targets using joint probabilistic data association,'' \emph{IEEE J. Ocean.
  Eng.}, vol.~8, no.~3, pp. 173--184, Jul. 1983.

\bibitem{GarWilGraSve:J18}
{\'A}.~F. Garc{\'\i}a-Fern{\'a}ndez, J.~L. Williams, K.~Granstr{\"o}m, and
  L.~Svensson, ``Poisson multi-{Bernoulli} mixture filter: Direct derivation
  and implementation,'' \emph{{IEEE} Trans. Aerosp. Electron. Syst.}, vol.~54,
  no.~4, pp. 1883--1901, Aug. 2018.

\bibitem{Mah:J03}
R.~P.~S. Mahler, ``Multitarget {Bayes} filtering via first-order multitarget
  moments,'' \emph{{IEEE} Trans. Aerosp. Electron. Syst.}, vol.~39, no.~4, pp.
  1152--1178, Oct. 2003.

\bibitem{MeyKroWilLauHlaBraWin:J18}
F.~Meyer, T.~Kropfreiter, J.~L. Williams, R.~A. Lau, F.~Hlawatsch, P.~Braca,
  and M.~Z. Win, ``Message passing algorithms for scalable multitarget
  tracking,'' \emph{Proc. {IEEE}}, vol. 106, no.~2, pp. 221--259, Feb. 2018.

\bibitem{LiaMey:J23}
M.~Liang and F.~Meyer, ``Neural enhanced belief propagation for multiobject
  tracking,'' \emph{{IEEE} Trans. Signal Process.}, vol.~72, pp. 15--30, 2024.

\bibitem{WeiLiaMey:J26}
S.~Wei, M.~Liang, and F.~Meyer, ``{B}ayesian multiobject tracking with
  neural-enhanced motion and measurement models,'' \emph{{IEEE} Trans. Signal
  Process.}, vol.~74, pp. 622--638, Jan. 2026.

\bibitem{MeyWil:J21}
F.~Meyer and J.~L. Williams, ``Scalable detection and tracking of geometric
  extended objects,'' \emph{{IEEE} Trans. Signal Process.}, vol.~69, pp.
  6283--6298, Oct. 2021.

\bibitem{FeiPinZis:C17}
C.~Feichtenhofer, A.~Pinz, and A.~Zisserman, ``Detect to track and track to
  detect,'' in \emph{Proc. ICCV-17}, Venice, Italy, Oct. 2017, pp. 3057--3065.

\bibitem{BraLau:C20}
G.~Bras{\'o} and L.~Leal-Taix{\'e}, ``Learning a neural solver for multiple
  object tracking,'' in \emph{Proc. CVPR-20}, Seattle, WA, USA, Jun. 2020, pp.
  6246--6256.

\bibitem{ZenDonZha:C21}
F.~Zeng, B.~Dong, Y.~Zhang, T.~Wang, X.~Zhang, and Y.~Wei, ``{MOTR}: End-to-end
  multiple-object tracking with transformer,'' in \emph{Proc. ECCV-22}, Tel
  Aviv, Israel, Oct. 2022, pp. 659--675.

\bibitem{WesModLeiPed:26}
A.~M. Westerkam, J.~M{\"o}derl, E.~Leitinger, and T.~Pedersen, ``A variational
  message passing framework for multi-sensor multi-object tracking using raw
  radar signals,'' \emph{arXiv preprint arXiv:2604.13884}, 2026.

\bibitem{StrGraWal:J02}
R.~L. Streit, M.~L. Graham, and M.~J. Walsh, ``Multitarget tracking of
  distributed targets using histogram-{PMHT},'' \emph{Digit. Signal Process.},
  vol.~12, no. 2--3, pp. 394--404, 2002.

\bibitem{DavGae:B18}
S.~J. Davey and H.~X. Gaetjens, \emph{{Track-Before-Detect Using Expectation
  Maximisation}}.\hskip 1em plus 0.5em minus 0.4em\relax Singapore: Springer,
  2018.

\bibitem{VoVoPhaSut:10}
B.-N. Vo, B.-T. Vo, N.-T. Pham, and D.~Suter, ``Joint detection and estimation
  of multiple objects from image observations,'' \emph{{IEEE} Trans. Signal
  Process.}, vol.~58, no.~10, pp. 5129--5141, Oct. 2010.

\bibitem{KroWilMey:21}
T.~Kropfreiter, J.~L. Williams, and F.~Meyer, ``A scalable track-before-detect
  method with {P}oisson/multi-{B}ernoulli model,'' in \emph{Proc. FUSION-21},
  Sun City, South Africa, Nov. 2021, pp. 1--8.

\bibitem{FerHahFox:C06}
B.~Ferris, D.~H{\"a}hnel, and D.~Fox, ``Gaussian processes for signal
  strength-based location estimation,'' in \emph{Proc. RSS-06}, Philadelphia,
  PA, USA, Aug. 2006.

\bibitem{FerFoxLaw:C07}
B.~Ferris, D.~Fox, and N.~Lawrence, ``{WiFi-SLAM} using {Gaussian} process
  latent variable models,'' in \emph{Proc. IJCAI-07}, Hyderabad, India, Jan.
  2007, pp. 2480--2485.

\bibitem{WeiRaoMey:C26}
S.~Wei, B.~Rao, and F.~Meyer, ``Normalizing flow for neural-enhanced
  track-before-detect,'' in \emph{Proc. FUSION-26}, Trondheim, Norway, Jun.
  2026.

\bibitem{WeiRaoMey:SM27}
\BIBentryALTinterwordspacing
------, ``Learning latent fields in multiobject tracking: {S}upplementary
  material,'' 2027. [Online]. Available:
  \url{https://fmeyer.ucsd.edu/TSP-27-SM.pdf}
\BIBentrySTDinterwordspacing

\bibitem{PapPil:02}
A.~Papoulis and S.~U. Pillai, \emph{Probability, Random Variables and
  Stochastic Processes}, 4th~ed.\hskip 1em plus 0.5em minus 0.4em\relax New
  York, NY, USA: McGraw-Hill, 2002.

\bibitem{AruMasGorCla:02}
M.~S. Arulampalam, S.~Maskell, N.~Gordon, and T.~Clapp, ``{A tutorial on
  particle filters for online nonlinear/non-Gaussian {B}ayesian tracking},''
  \emph{{IEEE} Trans. Signal Process.}, vol.~50, no.~2, pp. 174--188, Feb.
  2002.

\bibitem{Bis:B06}
C.~M. Bishop, \emph{Pattern Recognition and Machine Learning}.\hskip 1em plus
  0.5em minus 0.4em\relax New York, NY, USA: Springer, 2006.

\bibitem{RahGarSve:17}
A.~S. Rahmathullah, {\'A}.~F. Garc{\'\i}a-Fern{\'a}ndez, and L.~Svensson,
  ``Generalized optimal sub-pattern assignment metric,'' in \emph{Proc.
  FUSION-17}, Xi'an, China, Jul. 2017, pp. 1--8.

\bibitem{ShaKirLi:B02}
Y.~Bar-Shalom, X.-R. Li, and T.~Kirubarajan, \emph{{Estimation with
  Applications to Tracking and Navigation: Theory, Algorithms and
  Software}}.\hskip 1em plus 0.5em minus 0.4em\relax New York, NY, USA: Wiley,
  2002.

\bibitem{MalZha:J93}
S.~G. Mallat and Z.~Zhang, ``Matching pursuits with time-frequency
  dictionaries,'' \emph{{IEEE} Trans. Signal Process.}, vol.~41, no.~12, pp.
  3397--3415, Dec. 1993.

\bibitem{WipRao:J07}
D.~P. {Wipf} and B.~D. {Rao}, ``{An empirical Bayesian strategy for solving the
  simultaneous sparse approximation problem},'' \emph{{IEEE} Trans. Signal
  Process.}, vol.~55, no.~7, pp. 3704--3716, Jul. 2007.

\bibitem{MeyBraWilHla:J17}
F.~Meyer, P.~Braca, P.~Willett, and F.~Hlawatsch, ``{A scalable algorithm for
  tracking an unknown number of targets using multiple sensors},'' \emph{{IEEE}
  Trans. Signal Process.}, vol.~65, no.~13, pp. 3478--3493, Jul. 2017.

\bibitem{GaoLuoXinRoyLiu:22}
\BIBentryALTinterwordspacing
X.~Gao, Y.~Luo, G.~Xing, S.~Roy, and H.~Liu, ``Raw {ADC} data of 77{GHz}
  {MMWave} radar for automotive object detection,'' IEEE DataPort, 2022.
  [Online]. Available:
  \url{https://ieee-dataport.org/documents/raw-adc-data-77ghz-mmwave-radar-automotive-object-detection}
\BIBentrySTDinterwordspacing

\bibitem{Ric:B14}
M.~A. Richards, \emph{Fundamentals of Radar Signal Processing}, 2nd~ed.\hskip
  1em plus 0.5em minus 0.4em\relax New York, NY, USA: McGraw-Hill, 2014.

\bibitem{EstKriSan:C96}
M.~Ester, H.-P. Kriegel, J.~Sander, and X.~Xu, ``A density-based algorithm for
  discovering clusters in large spatial databases with noise,'' in \emph{Proc.
  KDD-96}, Portland, OR, USA, Aug. 1996, pp. 226--231.

\bibitem{BagKupMik:J93}
A.~B. {Baggeroer}, W.~A. {Kuperman}, and P.~N. {Mikhalevsky}, ``An overview of
  matched field methods in ocean acoustics,'' \emph{IEEE J. Ocean. Eng.},
  vol.~18, no.~4, pp. 401--424, Oct. 1993.

\bibitem{BooAbaSchHod:J00}
N.~O. Booth, A.~T. Abawi, P.~W. Schey, and W.~S. Hodgkiss, ``Detectability of
  low-level broad-band signals using adaptive matched-field processing with
  vertical aperture arrays,'' \emph{{IEEE} J. Ocean. Eng.}, vol.~25, no.~3, pp.
  296--313, Jul. 2000.

\bibitem{JenKupPorSchTol:B11}
F.~B. Jensen, W.~A. Kuperman, M.~B. Porter, and H.~Schmidt, \emph{Computational
  Ocean Acoustics}, 2nd~ed.\hskip 1em plus 0.5em minus 0.4em\relax New York,
  NY, USA: Springer, 2011.

\bibitem{Por:91}
M.~B. Porter, ``The {KRAKEN} normal mode program,'' SACLANT Undersea Research
  Centre, La Spezia, Italy, Memorandum SM-245, 1991.

\bibitem{DspMurHodBooSch:J99}
G.~L. D’Spain, J.~J. Murray, W.~S. Hodgkiss, N.~O. Booth, and P.~W. Schey,
  ``Mirages in shallow water matched field processing,'' \emph{J. Acoust. Soc.
  Am.}, vol. 105, no.~6, pp. 3245--3265, Jun. 1999.

\end{thebibliography}


\begin{thebibliography}{1}
\providecommand{\url}[1]{#1}
\csname url@samestyle\endcsname
\providecommand{\newblock}{\relax}
\providecommand{\bibinfo}[2]{#2}
\providecommand{\BIBentrySTDinterwordspacing}{\spaceskip=0pt\relax}
\providecommand{\BIBentryALTinterwordstretchfactor}{4}
\providecommand{\BIBentryALTinterwordspacing}{\spaceskip=\fontdimen2\font plus
\BIBentryALTinterwordstretchfactor\fontdimen3\font minus
  \fontdimen4\font\relax}
\providecommand{\BIBforeignlanguage}[2]{{%
\expandafter\ifx\csname l@#1\endcsname\relax
\typeout{** WARNING: IEEEtran.bst: No hyphenation pattern has been}%
\typeout{** loaded for the language `#1'. Using the pattern for}%
\typeout{** the default language instead.}%
\else
\language=\csname l@#1\endcsname
\fi
#2}}
\providecommand{\BIBdecl}{\relax}
\BIBdecl

\bibitem{WeiRaoMey:SM27}
S.~Wei, B.~Rao, and F.~Meyer, ``Learning latent fields in multiobject
  tracking,'' 2027, in peer review.

\bibitem{LiaMey:J25}
\BIBentryALTinterwordspacing
M.~Liang and F.~Meyer, ``Sparse signal reconstruction for
  track-before-detect,'' \emph{{IEEE} Trans. Signal Process.}, 2026, in press.
  [Online]. Available: \url{https://arxiv.org/abs/2506.18177v4}
\BIBentrySTDinterwordspacing

\bibitem{KscFreLoe:01}
F.~R. Kschischang, B.~J. Frey, and H.-A. Loeliger, ``Factor graphs and the
  sum-product algorithm,'' \emph{{IEEE} Trans. Inf. Theory}, vol.~47, no.~2,
  pp. 498--519, Feb. 2001.

\bibitem{Por:91}
M.~B. Porter, ``The {KRAKEN} normal mode program,'' SACLANT Undersea Research
  Centre, La Spezia, Italy, Memorandum SM-245, 1991.

\bibitem{swellex:Env96}
\BIBentryALTinterwordspacing
``{SWellEx-96} experiment: Environment,'' accessed Oct. 2026. [Online].
  Available: \url{https://swellex96.ucsd.edu/environment.htm}
\BIBentrySTDinterwordspacing

\end{thebibliography}

\end{document}


\title{\huge Learning Latent Fields in Multiobject Tracking: Supplementary Material}
\author{\large{Shaoxiu Wei, Bhaskar Rao, and Florian Meyer}}
\maketitle
\frenchspacing
\vspace{-4mm}

This manuscript provides additional derivations and results for the paper~\cite{WeiRaoMey:SM27}. Equations, figures, and tables labeled with a section prefix (e.g., (I.1), Fig.~I.1, Table~I.1) refer to this supplementary material; all other references refer to~\cite{WeiRaoMey:SM27}.

The remainder of this manuscript is organized as follows. Section~\ref{app:vi} derives the evidence lower bound (ELBO) used to learn the latent field during runtime without labeled data and summarizes the processing at one time step. Section~\ref{app:bp} derives the forward Bayesian inference, including the sum-product message passing, the Gaussian approximation of the marginal likelihoods, and their Monte Carlo implementation with weighted particles. Section~\ref{app:ec} develops a low-rank and Kronecker-structured formulation that makes the inference computationally feasible for high-dimensional raw measurements and analyzes its storage and computational complexity. Section~\ref{app:sim} presents additional results for the simulated multisensor radar scenario. Section~\ref{app:radio} provides additional results for the active radio sensing experiment, including the posterior PDF at one time step, the evolution of the learned signal-power PDF through the flow transformations, and the sensitivity to birth initialization. Section~\ref{app:sonar} compares the learned latent field with KRAKEN predictions for the passive acoustic sensing experiment. Section~\ref{app:ablation} presents the ablation study of the signal-power prior in a controlled acoustic environment.

\numberwithin{equation}{section}
\numberwithin{figure}{section}
\numberwithin{table}{section}


\section{Derivations of the Variational Learning for the Latent Field}
\label{app:vi}
In this section we derive the learning objective used by the proposed method.
\subsection{System Model and Notation}
\label{A:model}

We consider a single raw measurement vector $\mathbf z_k\in\mathbb{C}^C$ for a given time step $k$, $N_k$ potential objects, and a scalar random variable $\eta_k\in\mathbb{R}_{>0}$ to denote the noise variance.
For each object $n\in\{1,\dots,N_k\}$ we define
\[
\mathbf x_{k,n} \in \mathbb{R}^{d},\qquad
{\boldsymbol{\phi}}_{k,n} = (r_{k,n},\,\gamma_{k,n}),
\]
where $\mathbf x_{k,n}$ is a random variable denoting the kinematic state (e.g., $d=4$ for 2-D position and velocity), $r_{k,n}\in\{0,1\}$ is the binary existence variable, and $\gamma_{k,n}>0$ is a random variable denoting the signal power of object $n$, i.e., the conditional mean squared magnitude of its complex amplitude.

\subsubsection{Likelihood}
Given the complex amplitude $\rho_{k,n}\sim\mathcal{CN}(0,\gamma_{k,n})$ and noise $\V{w}_k\sim\mathcal{CN}(\V{0},\eta_k\M{I}_C)$, the measurement at time $k$ is taken as
\begin{equation}
	\V{z}_{k} = \sum_{n=1}^{N_k} r_{k,n}\,\rho_{k,n}\,\V{a}(\V{x}_{k,n}) + \V{w}_k.
\end{equation}
Conditional on the existence variables $\V{r}_k=r_{k,1:N_k}$, the kinematic states $\V{x}_k=\V{x}_{k,1:N_k}$, the signal powers $\bm{\gamma}_k=\gamma_{k,1:N_k}$, and $\eta_k$, the amplitudes $\rho_{k,n}$ are mutually independent and independent of the noise $\V{w}_k$, and all of them are proper complex Gaussian. Marginalizing the amplitudes yields a zero-mean complex Gaussian likelihood:
\begin{align}
	f(\V{z}_{k}\mid \V{r}_{k},\V{x}_{k},\bm{\gamma}_{k},\eta_k)
	&= \mathcal{CN}\!\big(\V{z}_{k};\,\V{0},\,\bm{\Sigma}_k\big),\\
	\bm{\Sigma}_k &= \sum_{n=1}^{N_k} r_{k,n}\gamma_{k,n}\, \V{a}(\V{x}_{k,n})\V{a}(\V{x}_{k,n})^{\mathrm{H}} + \eta_k\, \M{I}_{C},
\end{align}
where $\V{a}(\cdot)\in\mathbb{C}^{C}$ denotes the physics-based array manifold, which is in general a nonlinear function of the kinematic state.

\subsubsection{Prior PDF}
At each time step $k$, every random variable has a prior density as
\begin{equation}
	f(\V{x}_{k,n}),\qquad
	f_{\bm{\theta}}({\boldsymbol{\phi}}_{k,n}),\qquad
	f(\eta_k),
\end{equation}
where the prior $f_{\bm{\theta}}({\boldsymbol{\phi}}_{k,n})$ is denoted as
\begin{align}
	f_{\bm{\theta}}(\bm{\phi}_{k,n}) =
	\begin{cases}
		f(r_{k,n}=1)\,
		f_{\bm{\theta}}(\gamma_{k,n}\mid r_{k,n}=1;\V{c}_{k,n}),
		& r_{k,n}=1,\\[2pt]
		f(r_{k,n}=0)\,
		f_{\mathrm{D}}(\gamma_{k,n}),
		& r_{k,n}=0,
	\end{cases}
	\label{eq:phi-mixture}
\end{align}
where $f_{\mathrm{D}}(\gamma_{k,n})$ is a \emph{dummy} PDF assigned when $r_{k,n}=0$, since the measurement model does not depend on $\gamma_{k,n}$ in this case and it can be assigned an arbitrary reference PDF without affecting inference.
The learnable component is the conditional prior $f_{\bm{\theta}}(\gamma_{k,n}\mid r_{k,n}\!=\!1; \V{c}_{k,n})$, which equals the signal-power PDF $f_{\bm{\theta}}(\gamma_{k,n};\V{c}_{k,n})$ of the latent field (cf.~\eqref{M-eq:cond_equals_field} of the main paper). The latent field is the family of context-dependent signal-power PDFs introduced in Section~\ref{M-sec:latent_field_prior} of the main paper, and its parameters $\bm{\theta}$ are shared across objects and time. The context $\V{c}_{k,n}$ of object $n$ at time step $k$ comprises its position and possibly known sensing parameters and signal features. It is treated as a deterministic quantity, not a random variable.

The predicted PDFs $f(r_{k,n})$, $f(\V{x}_{k,n})$, and $f(\eta_k)$ are not learned. As is common, they are obtained from first-order Markov models with the transition PDFs $f(r_{k,n}\mid r_{k-1,n})$, $f(\V{x}_{k,n}\mid \V{x}_{k-1,n})$, and $f(\eta_k\mid\eta_{k-1})$, as described in Section~\ref{M-sec:inference_learning} of the main paper. For example,
\begin{align*}
	f(\V{x}_{k,n}) = \int f(\V{x}_{k,n}\mid \V{x}_{k-1,n})\,q(\V{x}_{k-1,n})\, \mathrm{d}\V{x}_{k-1,n}
\end{align*}
where $q(\V{x}_{k-1,n})$ is the marginal posterior PDF of time step $k-1$.
However, the variational framework developed below does not depend on these specific choices. The concrete form of $f_{\bm{\theta}}(\cdot)$ -- either a nonparametric or parametric statistical family -- is specified below.
\subsubsection{Posterior PDF}
Let 
\(\V{h}_k = \big(\V{x}_{k,1{:}N_k},
\boldsymbol{\phi}_{k,1{:}N_k},\eta_k\big)\)
collect all hidden variables at time \(k\). We adopt a sequential filtering
perspective, where the predictive prior before observing \(\mathbf z_k\) is
\(f_{\bm\theta}(\mathbf h_k\mid \mathbf z_{1:k-1})\). For notational simplicity,
we denote this predictive prior by \(f_{{\bm\theta}}(\mathbf h_k)\). The joint density is then written as
\begin{align}
\label{A:joint_pdf}
    f_{\bm\theta}(\mathbf z_k,\mathbf h_k)
    =
    f(\mathbf z_k\mid \mathbf h_k)\,
    f_{\bm\theta}(\mathbf h_k),
\end{align}
where $ f(\mathbf z_k\mid \mathbf h_k) = \mathcal{CN}\!\big(\mathbf z_k;\mathbf 0,\bm{\Sigma}_k\big)$ and $f_{\bm\theta}(\mathbf h_k)  = f(\eta_k)\prod_n
    f(\mathbf x_{k,n})
    f_{\bm\theta}(\boldsymbol{\phi}_{k,n})
    .$
The dependence on previous measurements is absorbed into the predictive
prior through posterior propagation. The evidence at time $k$ is obtained as
\begin{align}
    f_{\bm\theta}(\mathbf z_k)
    =
    \int
    f(\mathbf z_k\mid \mathbf h_k)
    f_{\bm\theta}(\mathbf h_k)
    d\mathbf h_k,
\end{align}
where \(f_{\bm\theta}(\mathbf z_k)\) abbreviates
\(f_{\bm\theta}(\mathbf z_k\mid \mathbf z_{1:k-1})\). Therefore, the posterior PDF at time step \(k\) is given by Bayes' rule:
\begin{align}
    \label{A:bayes}
    f_{\bm\theta}(\mathbf h_k\mid \mathbf z_k)
    =
    \frac{
    f_{\bm\theta}(\mathbf z_k,\mathbf h_k)
    }{
    f_{\bm\theta}(\mathbf z_k)
    } .
\end{align}
Direct computation of \eqref{A:bayes} is generally intractable. We
therefore approximate the posterior PDF $f_{\bm\theta}(\mathbf h_k\mid \mathbf z_k)$ by a PDF $q(\mathbf h_k)$. This PDF serves as a formal auxiliary density for the derivation; belief propagation (BP) computes only its marginal posterior PDFs, which are represented by weighted particles. The resulting marginal
posterior PDFs are obtained by marginalizing \(q(\mathbf h_k)\):
\begin{align}
    q(\mathbf x_{k,n})
    &=
    \int q(\mathbf h_k)\,
    d\boldsymbol{\phi}_{k,1:N_k}\, d\eta_k
    \prod_{n'\neq n} d\mathbf x_{k,n'}
    \propto
    f(\mathbf x_{k,n})\,\ell_{\mathbf x,n}(\mathbf x_{k,n}),
    \\
    q(\boldsymbol{\phi}_{k,n})
    &=
    \int q(\mathbf h_k)\,
    d\mathbf x_{k,1:N_k}\, d\eta_k
    \prod_{n'\neq n} d\boldsymbol{\phi}_{k,n'}
    \propto
    f_{\bm\theta}(\boldsymbol{\phi}_{k,n})\,\ell_{\boldsymbol{\phi},n}(\boldsymbol{\phi}_{k,n}),
    \\
    q(\eta_k)
    &=
    \int q(\mathbf h_k)\,
    d\mathbf x_{k,1:N_k}\, d\boldsymbol{\phi}_{k,1:N_k}
    \propto
    f(\eta_k)\,\ell_{\eta}(\eta_k),
\end{align}
where each marginal is obtained by integrating out all hidden variables in $\mathbf h_k$ except the one of interest; for the binary existence variables $r_{k,n'}$ contained in $\boldsymbol{\phi}_{k,n'}$, the integration is understood as a summation over $\{0,1\}$.
Here \(\ell_{\mathbf x,n}(\mathbf x_{k,n})\), \(\ell_{\boldsymbol{\phi},n}(\boldsymbol{\phi}_{k,n})\), and
\(\ell_{\eta}(\eta_k)\) denote the marginal likelihood terms associated with the
kinematic state of object \(n\), its existence and signal power, and the noise
variance at time \(k\), respectively. The proportionality relations hold exactly for the factor graph of a single measurement block, which is a tree (cf.~Section~\ref{B:likelihood}). The Gaussian approximation in Section~\ref{B:gaussian_approx} and, for multiple measurement blocks, the loops of the factor graph make them approximate, i.e., the marginal posterior PDFs are then approximated by the beliefs of BP.
\subsection{Variational Inference}
\label{A:vi}
We derive the learning objective from the
standard evidence decomposition. At time step $k$, for any fixed distribution
$q(\mathbf h_k)$, the one-step evidence satisfies
\begin{align}
    \log f_{\bm{\theta}}(\mathbf z_k)
    =
    \mathcal L_k(\bm{\theta})
    +
    \mathrm{KL}\!\left(
    q(\mathbf h_k)
    \,\|\,
    f_{\bm{\theta}}(\mathbf h_k\mid \mathbf z_k)
    \right),
    \label{A:evidence-decomp}
\end{align}
where the corresponding ELBO is
\begin{align}
    \mathcal L_k(\bm{\theta})
    =
    \mathbb E_{q(\mathbf h_k)}
    \left[
    \log f_{\bm{\theta}}(\mathbf z_k, \mathbf h_k)
    -
    \log q(\mathbf h_k)
    \right].
    \label{A:local-elbo}
\end{align}
Since the KL divergence is nonnegative, $\mathcal L_k(\bm{\theta})$ is a lower
bound on the evidence $\log f_{\bm{\theta}}(\mathbf z_k)$. Using the factorization in \eqref{A:joint_pdf}, the ELBO can be written as
\begin{align}
    \mathcal L_k(\bm{\theta})
    &=
    \underbrace{
    \mathbb E_{q(\mathbf h_k)}
    \big[
    \log f(\mathbf z_k\mid \mathbf h_k)
    \big]
    }_{\text{likelihood}}
    +
    \underbrace{
    \mathbb E_{q(\mathbf h_k)}
    \big[
    \log f_{\bm{\theta}}(\boldsymbol{\phi}_{k,1:N_k})
    \big]
    }_{\text{signal-power prior}}
    \notag\\
    &\quad
    +
    \underbrace{
    \mathbb E_{q(\mathbf h_k)}
    \big[
    \log f(\mathbf x_{k,1:N_k})
    +
    \log f(\eta_k)
    \big]
    }_{\text{other priors}}
    -
    \underbrace{
    \mathbb E_{q(\mathbf h_k)}
    \big[
    \log q(\mathbf h_k)
    \big]
    }_{\text{entropy}}.
    \label{A:elbo-decomp}
\end{align}
When $q(\mathbf h_k)$ is fixed during the flow update, the likelihood term,
the kinematic prior, the noise prior, and the entropy term are independent
of $\bm{\theta}$. Therefore, the $\bm{\theta}$-dependent part of
$\mathcal L_k(\theta)$ comes only from the prior $f_{\bm{\theta}}(\boldsymbol{\phi}_{k,1:N_k})$. Thus, we only need to evaluate the expected log-prior
$\mathbb E_{q(\mathbf h_k)}\big[\log f_{\bm{\theta}}(\boldsymbol{\phi}_{k,1:N_k})\big]$, which we expand step by step in the following.

\par First, the prior of the existence and signal-power states factorizes across objects, i.e., $f_{\bm{\theta}}(\boldsymbol{\phi}_{k,1:N_k}) = \prod_{n=1}^{N_k} f_{\bm{\theta}}(\boldsymbol{\phi}_{k,n})$. The logarithm turns this product into a sum, and by the linearity of expectation we obtain
\begin{align}
    \mathbb E_{q(\mathbf h_k)}
    \big[
    \log f_{\bm{\theta}}(\boldsymbol{\phi}_{k,1:N_k})
    \big]
    =
    \mathbb E_{q(\mathbf h_k)}
    \bigg[
    \sum_{n=1}^{N_k}
    \log f_{\bm{\theta}}(\boldsymbol{\phi}_{k,n})
    \bigg]
    =
    \sum_{n=1}^{N_k}
    \mathbb E_{q(\mathbf h_k)}
    \big[
    \log f_{\bm{\theta}}(\boldsymbol{\phi}_{k,n})
    \big].
    \label{A:phi-sum}
\end{align}

\par Second, each summand in \eqref{A:phi-sum} depends on $\mathbf h_k$ only through $\boldsymbol{\phi}_{k,n}$. Integrating out all remaining hidden variables, i.e., $\mathbf x_{k,1:N_k}$, $\boldsymbol{\phi}_{k,n'}$ with $n'\neq n$, and $\eta_k$, yields the marginal posterior PDF $q(\boldsymbol{\phi}_{k,n})$, so that
\begin{align}
    \mathbb E_{q(\mathbf h_k)}
    \big[
    \log f_{\bm{\theta}}(\boldsymbol{\phi}_{k,n})
    \big]
    =
    \int q(\mathbf h_k)\,
    \log f_{\bm{\theta}}(\boldsymbol{\phi}_{k,n})\,
    d\mathbf h_k
    =
    \int q(\boldsymbol{\phi}_{k,n})\,
    \log f_{\bm{\theta}}(\boldsymbol{\phi}_{k,n})\,
    d\boldsymbol{\phi}_{k,n}
    =
    \mathbb E_{q(\boldsymbol{\phi}_{k,n})}
    \big[
    \log f_{\bm{\theta}}(\boldsymbol{\phi}_{k,n})
    \big].
    \label{A:phi-marginal}
\end{align}

\par Third, since $\boldsymbol{\phi}_{k,n} = (r_{k,n},\gamma_{k,n})$ consists of a binary existence variable and a continuous signal power, the expectation in \eqref{A:phi-marginal} is a sum over $r\in\{0,1\}$ and an integral over $\gamma_{k,n}$. Factorizing the marginal posterior as $q(r_{k,n}=r,\gamma_{k,n}) = q(r_{k,n}=r)\, q(\gamma_{k,n}\mid r_{k,n}=r)$, we have
\begin{align}
    \mathbb E_{q(\boldsymbol{\phi}_{k,n})}
    \big[
    \log f_{\bm{\theta}}(\boldsymbol{\phi}_{k,n})
    \big]
    &=
    \sum_{r\in\{0,1\}}
    \int
    q(r_{k,n}=r,\gamma_{k,n})\,
    \log f_{\bm{\theta}}(r_{k,n}=r,\gamma_{k,n})\,
    d\gamma_{k,n}
    \notag\\
    &=
    \sum_{r\in\{0,1\}}
    q(r_{k,n}=r)
    \int
    q(\gamma_{k,n}\mid r_{k,n}=r)\,
    \log f_{\bm{\theta}}(r_{k,n}=r,\gamma_{k,n})\,
    d\gamma_{k,n}
    \notag\\
    &=
    \sum_{r\in\{0,1\}}
    q(r_{k,n}=r)\,
    \mathbb E_{q(\gamma_{k,n}\mid r_{k,n}=r)}
    \big[
    \log f_{\bm{\theta}}(r_{k,n}=r,\gamma_{k,n})
    \big].
    \label{A:phi-r-sum}
\end{align}

\par Fourth, according to the prior in \eqref{eq:phi-mixture}, the logarithm of the prior evaluated for the two existence hypotheses reads
\begin{align}
    \log f_{\bm{\theta}}(r_{k,n}=1,\gamma_{k,n})
    &=
    \log f(r_{k,n}=1)
    +
    \log f_{\bm{\theta}}
    \big(
    \gamma_{k,n}\mid r_{k,n}=1;\mathbf c_{k,n}
    \big),
    \label{A:logprior-r1}
    \\
    \log f_{\bm{\theta}}(r_{k,n}=0,\gamma_{k,n})
    &=
    \log f(r_{k,n}=0)
    +
    \log f_{\mathrm{D}}(\gamma_{k,n}).
    \label{A:logprior-r0}
\end{align}

\par Finally, substituting \eqref{A:phi-marginal}, \eqref{A:phi-r-sum}, \eqref{A:logprior-r1}, and \eqref{A:logprior-r0} into \eqref{A:phi-sum}, the expected log-prior is obtained as
\begin{align}
    &\mathbb E_{q(\mathbf h_k)}
    \big[
    \log f_{\bm{\theta}}(\boldsymbol{\phi}_{k,1:N_k})
    \big]
    \notag\\
    &=
    \sum_{n=1}^{N_k}
    \bigg[
    q(r_{k,n}=1)
    \mathbb E_{q(\gamma_{k,n}\mid r_{k,n}=1)}
    \Big[
    \log f(r_{k,n}=1)
    +
    \log f_{\bm{\theta}}
    \big(
    \gamma_{k,n}\mid r_{k,n}=1;\mathbf c_{k,n}
    \big)
    \Big]
    \notag\\
    &\qquad\quad
    +
    q(r_{k,n}=0)
    \mathbb E_{q(\gamma_{k,n}\mid r_{k,n}=0)}
    \Big[
    \log f(r_{k,n}=0)
    +
    \log f_{\mathrm{D}}(\gamma_{k,n})
    \Big]
    \bigg].
    \label{A:phi-expanded}
\end{align}
Since $\log f(r_{k,n}=r)$ does not depend on $\gamma_{k,n}$ and $\int q(\gamma_{k,n}\mid r_{k,n}=r)\, d\gamma_{k,n} = 1$, the existence terms can be taken out of the expectations. Hence, \eqref{A:phi-expanded} can be separated into a $\bm{\theta}$-dependent part and a $\bm{\theta}$-independent part, i.e.,
\begin{align}
    \mathbb E_{q(\mathbf h_k)}
    \big[
    \log f_{\bm{\theta}}(\boldsymbol{\phi}_{k,1:N_k})
    \big]
    &=
    \sum_{n=1}^{N_k}
    q(r_{k,n}=1)\,
    \mathbb E_{q(\gamma_{k,n}\mid r_{k,n}=1)}
    \Big[
    \log f_{\bm{\theta}}
    \big(
    \gamma_{k,n}\mid r_{k,n}=1;\mathbf c_{k,n}
    \big)
    \Big]
    \notag\\
    &\quad
    +
    \underbrace{
    \sum_{n=1}^{N_k}
    \Big(
    \sum_{r\in\{0,1\}}
    q(r_{k,n}=r)
    \log f(r_{k,n}=r)
    +
    q(r_{k,n}=0)\,
    \mathbb E_{q(\gamma_{k,n}\mid r_{k,n}=0)}
    \big[
    \log f_{\mathrm{D}}(\gamma_{k,n})
    \big]
    \Big)
    }_{\text{independent of }\bm{\theta}}.
    \label{A:phi-split}
\end{align}
Here $q(r_{k,n}=1)$ and
$q(\gamma_{k,n}\mid r_{k,n}=1)$ are obtained from the marginal posterior PDF
$q(\boldsymbol{\phi}_{k,n})$. Since $f(r_{k,n})$ and the dummy density $f_{\mathrm{D}}(\gamma_{k,n})$ are independent of $\bm{\theta}$, the last term in \eqref{A:phi-split} is a constant. Therefore, maximizing the ELBO with respect to $\bm{\theta}$ is
equivalent to maximizing its $\bm{\theta}$-dependent part:
\begin{align}
    \mathcal J_k(\bm{\theta})
    =
    \sum_n
    q(r_{k,n}=1)
    \mathbb E_{q(\gamma_{k,n}\mid r_{k,n}=1)}
    \left[
    \log f_{\bm{\theta}}
    \big(
    \gamma_{k,n}\mid r_{k,n}=1;\mathbf c_{k,n}
    \big)
    \right].
    \label{A:objective-local}
\end{align}
At time step $k$, the accumulated objective is given by
\begin{align}
    \mathcal J_{1:k}(\bm{\theta})
    = \sum_{k'=1}^{k}
    \mathcal J_{k'}(\bm{\theta})
    \label{A:objective}
\end{align}
\par We now describe how the expectation in \(\mathcal J_k(\bm{\theta})\) is evaluated
in practice. Since the posterior $q$ is represented nonparametrically via Monte Carlo method (Section~\ref{B:mc}),
the conditional posterior of \(\gamma_{k,n}\) given \(r_{k,n}=1\) is 
represented by a weighted particle set:
\begin{align}
    q(\gamma_{k,n}\mid r_{k,n}=1)
    \approx
    \left\{
    \big(\omega_{\gamma,k,n,p},\gamma_{k,n,p}\big)
    \right\}_{p=1}^{P},
    \label{A:particle_posterior}
\end{align}
where the weights satisfy
$\sum_{p=1}^{P}\omega_{\gamma,k,n,p}=1$. Substituting
\eqref{A:particle_posterior} into \eqref{A:objective-local} gives the
Monte Carlo objective as:
\begin{align}
    \mathcal J_{1:k}(\bm{\theta})
    =
    \sum_{k'=1}^{k}\sum_{n}
    q(r_{k',n}=1)
    \sum_{p=1}^{P}
    \omega_{\gamma,k',n,p}
    \log f_{\bm{\theta}}
    \big(
    \gamma_{k',n,p}\mid r_{k',n}=1;\mathbf c_{k',n}
    \big).
    \label{A:mc_objective}
\end{align}
During the parameter update, the inferred particles and their weights are
treated as fixed outputs of the forward inference. Therefore, the
gradient with respect to $\bm{\theta}$ acts only on the flow density:
\begin{align}
    \nabla_{\bm{\theta}}\mathcal J_{1:k}(\bm{\theta})
    =
    \sum_{k'=1}^{k}\sum_{n}
    q(r_{k',n}=1)
    \sum_{p=1}^{P}
    \omega_{\gamma,k',n,p}
    \nabla_{\bm{\theta}}
    \log f_{\bm{\theta}}
    \big(
    \gamma_{k',n,p}\mid r_{k',n}=1;\mathbf c_{k',n}
    \big).
    \label{A:gradient}
\end{align}
The posterior power particles are samples from the prior PDF obtained with the parameters before the update, reweighted by the likelihood. Hence, maximizing \eqref{A:mc_objective} amounts to a weighted maximum-likelihood fit whose support is restricted to that of the preceding prior, and its accuracy depends on the effective sample size $1/\sum_{p=1}^{P}\omega_{\gamma,k',n,p}^2$ of the power particles.

In the following sections, we discuss several choices of this conditional PDF.

\subsection{Nonparametric Prior PDF via Conditional Normalizing Flow}

We model the conditional prior $f_{\bm{\theta}}(\gamma_{k,n}\mid r_{k,n}\!=\!1;\V{c}_{k,n})$, i.e., the signal-power PDF of the latent field, using a conditional normalizing flow as in Section~\ref{M-sec:flow} of the main paper, which provides a flexible family of PDFs over $\gamma_{k,n}$.
Since $\gamma_{k,n} > 0$, we first map to an unconstrained domain:
\begin{equation}
	\gamma_{k,n} = {G}(\mu_{k,n}), \qquad  \mu_{k,n} \in \mathbb{R},
\end{equation}
where $G(\cdot):\mathbb{R}\to\mathbb{R}_{>0}$ is a smooth bijection. For notational simplicity, we suppress the indices $(k,n)$ in the following derivation. We then model the conditional PDF $f_{\bm{\theta}}(\mu;\V{c})$ using a conditional normalizing flow.
Let $f_0(\varepsilon)$ be a base PDF on $\mathbb{R}$ (e.g., $\varepsilon \sim \mathcal{N}(0,1)$).
For a given context $\V{c}$ as input, define a composition of $Q$ invertible, monotone transformations:
\begin{equation}
	y_0 = \varepsilon,\qquad
	y_\ell = F_{\theta_\ell}(y_{\ell-1};\, \V{c}),\quad
	\ell = 1,\dots,Q,
\end{equation}
where each $F_{\theta_\ell}(\cdot;\,\V{c}):\mathbb{R}\to\mathbb{R}$ is a differentiable bijection parameterized by $\theta_\ell$ and conditioned on $\V{c}$. In particular, $F_{\theta_\ell}(\cdot;\V{c})$ is a nonlinear, strictly monotone function whose coefficients are computed from the context $\V{c}$ by an encoder, i.e., a neural network with parameters $\theta_\ell$.
The output is $
\mu = y_{Q} = T_{\bm{\theta}}(\varepsilon\mathpunct{;}\, \V{c}),
$
with $\bm{\theta} = (\theta_1,\dots,\theta_{Q})$.
By the change-of-variables formula, the log-PDF of $\mu$ given $\V{c}$ is
\begin{equation}
	\log f_{\bm{\theta}}(\mu;\V{c})
	= \log f_0(\varepsilon) -
	\sum_{\ell=1}^{Q}
	\log \left|
	\frac{\partial F_{\theta_\ell}(y_{\ell-1};\, \V{c})}%
	{\partial y_{\ell-1}}
	\right|,
	\label{A:flow_logpdf}
\end{equation}
where $\varepsilon$ and $\{y_{\ell}\}$ are recovered by inverting the flow for the given $(\mu,\, \V{c})$.
Transforming back to the original signal-power variable via $\gamma = G(\mu)$, we obtain
\begin{equation}
	\log f_{\bm{\theta}}(\gamma\mid r\!=\!1;\, \V{c})
	=
	\log f_{\bm{\theta}}(\mu;\, \V{c})
	- \log \left|
	\frac{d G(\mu)}%
	{d \mu}
	\right|.
	\label{A:flow_gamma}
\end{equation}
Substituting~\eqref{A:flow_gamma} into the Monte Carlo objective~\eqref{A:mc_objective} yields the objective that is maximized to train the flow parameters $\bm{\theta}$. Before the first time step, $\bm{\theta}$ is initialized such that the flow yields the uniform PDF on $(0,\gamma_{\max}]$ for all contexts.

\subsubsection{Training and Inference Pipeline}
\label{A:pipeline}
The training and inference procedures are coupled during multiobject tracking. The complete procedure for a given context $\V{c}$ consists of four steps:
\begin{enumerate}
	\item \textbf{Prior sampling.}
	For each power particle index $p = 1,\dots,P$:
	\begin{itemize}
		\item Draw a base sample $\varepsilon_p \sim f_0(\varepsilon)$.
		\item Propagate through the flow:
		$\mu_p = T_{\bm{\theta}}(\varepsilon_p\mathpunct{;}\, \V{c})$.
		\item Transform to signal-power space:
		$\gamma_p = G(\mu_p) > 0$.
	\end{itemize}
	This yields prior particles
	$\{\gamma_p\}_{p=1}^{P} \sim f_{\bm{\theta}}(\gamma\mid r\!=\!1;\,\V{c})$ with uniform weights.
	
	\item \textbf{Posterior reweighting.}
	\begin{itemize}
		\item After the new measurement is received, each prior particle $\gamma_p$ is reweighted by importance sampling, which yields the posterior weights $\omega_{\gamma, p}$.
		\item The resulting particle set
		$q(\gamma\mid r\!=\!1) \approx
		\{(\omega_{\gamma, p},\,\gamma_p)\}_{p=1}^{P}$
		approximates the marginal posterior.
	\end{itemize}
	
	\item \textbf{Density evaluation.}
	For each posterior particle $\gamma_p$:
	\begin{itemize}
		\item Invert the transformation: $\mu_p = G^{-1}(\gamma_p)$.
		\item Invert the flow to recover the base sample $\varepsilon_p$
		and the intermediate values $\{y_\ell\}$.
		\item Compute $\log f_{\bm{\theta}}(\gamma_p\mid r\!=\!1;\,\V{c})$ via
		\eqref{A:flow_gamma} by accumulating the Jacobian terms.
	\end{itemize}
	
	\item \textbf{Parameter update.}
	\begin{itemize}
		\item Evaluate the gradient
		$\nabla_{\bm{\theta}}\mathcal J_{1:k}(\bm{\theta})$ approximately using the posterior weights
		$\omega_p$ and the log-PDFs from step~3, which are buffered across time steps and scenes. A window is applied if available memory becomes limited.
	\end{itemize}
\end{enumerate}

\subsection{Parametric Prior PDF}
\label{A:parametric}
When a parametric family is considered, for example, we model the conditional prior as a Gamma PDF
\begin{equation}
	f_{\bm{\theta}}(\gamma_{k,n}\mid r_{k,n}\!=\!1;\,\V{c}_{k,n})
	=
	\mathrm{Gamma}\!\left(\gamma_{k,n}\,;\, \alpha_{\bm{\theta}}(\V{c}_{k,n}),\,\beta_{\bm{\theta}}(\V{c}_{k,n})\right),
	\label{A:gamma_direct}
\end{equation}
where the shape $\alpha_{\bm{\theta}}(\cdot)>0$ and rate $\beta_{\bm{\theta}}(\cdot)>0$ are produced by a neural network using the context $\V{c}_{k,n}$ as input.
Substituting into the Monte Carlo objective~\eqref{A:mc_objective} and expanding the Gamma log-PDF gives
\begin{align}
	\mathcal J_{1:k}(\bm{\theta})  &= \sum_{k'=1}^{k}\sum_{n}\;q(r_{k',n}\!=\!1) \sum_p\omega_{\gamma,k',n,p}\,\log\mathrm{Gamma}\!\left(\gamma_{k',n,p};\,\alpha_{\bm{\theta}}(\V{c}_{k',n}),\,\beta_{\bm{\theta}}(\V{c}_{k',n})\right)\label{A:gamma_loss}\\
	&= \sum_{k'=1}^{k}\sum_{n}\;q(r_{k',n}\!=\!1)\Big[
	\alpha_{\bm{\theta}}(\V{c})\log\beta_{\bm{\theta}}(\V{c}) - \log\Gamma\!\big(\alpha_{\bm{\theta}}(\V{c})\big)
	+ \big(\alpha_{\bm{\theta}}(\V{c})-1\big)\,\overline{\log\gamma}_{k',n}
	- \beta_{\bm{\theta}}(\V{c})\,\hat{\gamma}_{k',n}\Big],\notag	
\end{align}
where $\V{c}=\V{c}_{k',n}$, and the sufficient statistics are
\begin{equation}
	\overline{\log\gamma}_{k',n}=\sum_p \omega_{\gamma,k',n,p}\log \gamma_{k',n,p},\qquad
	\hat{\gamma}_{k',n} =\sum_p \omega_{\gamma,k',n,p}\,\gamma_{k',n,p}.
\end{equation}
\subsection{Processing at One Time Step}
\label{A:summary}
Algorithm~\ref{alg:time_step} summarizes the processing at time step $k$, which combines the prediction step described in Section~\ref{M-sec:inference_learning} of the main paper, the Bayesian inference in Section~\ref{app:bp}, its efficient implementation in Section~\ref{app:ec}, and the parameter update derived in this section.
\begin{algorithm}[htbp]
	\caption{Processing at time step $k$}
	\label{alg:time_step}
	\KwIn{Particle representations of $q(\V{x}_{k-1,n})$ and $q(\eta_{k-1})$, existence probabilities $q(r_{k-1,n}\!=\!1)$, flow parameters $\bm{\theta}$, and measurement $\V{z}_k$}
	\KwOut{State estimates, particle representations of the marginal posterior PDFs at time step $k$, and updated parameters $\bm{\theta}$}
	\textbf{Prediction:} propagate the particles of the kinematic states of legacy potential objects and of the noise variance through the transition PDFs, compute $f(r_{k,n}\!=\!1)=p_{\mathrm{s}}\,q(r_{k-1,n}\!=\!1)$, and introduce new potential objects with birth probabilities $p_{\mathrm{B},n}$ and birth PDFs $f_{\mathrm{B},n}(\V{x}_{k,n})$\;
	\textbf{Context:} compute the contexts $\V{c}_{k,n}$ of all potential objects\;
	\textbf{Prior sampling:} draw the power particles $\gamma_{k,n,p}$ from $f_{\bm{\theta}}(\gamma;\V{c}_{k,n})$ using the flow (Section~\ref{A:pipeline})\;
	\textbf{Bayesian inference:} compute the marginal likelihoods with the formulation in Section~\ref{app:ec}, and update the particle weights \eqref{B:weight_update} and the existence probabilities \eqref{B:existence_update}; for multiple measurement blocks, iterate the messages (Section~\ref{B:mc})\;
	\textbf{Estimation:} declare the potential objects with $q(r_{k,n}\!=\!1)>T_{\mathrm{dec}}$ to exist and compute the MMSE estimates of their kinematic states\;
	\textbf{Learning:} store $q(r_{k,n}\!=\!1)$, the weighted power particles, and $\V{c}_{k,n}$, and update $\bm{\theta}$ by a few gradient steps that increase \eqref{A:mc_objective}\;
	\textbf{Pruning and resampling:} remove the potential objects with $q(r_{k,n}\!=\!1)<T_{\mathrm{pru}}$ and resample the particles of the kinematic states and of the noise variance\;
\end{algorithm}

\newpage

\section{Derivations of Forward Bayesian Inference}
\label{app:bp}

This section details the forward Bayesian inference of our proposed method. The key derivations proceed in three main steps:
\begin{enumerate}
    \item We use the sum-product algorithm to compute the marginal likelihood messages in a factor graph (Section~\ref{B:likelihood}).
    \item We show that these marginal likelihoods are zero-mean continuous Gaussian mixtures. To ensure analytical tractability, we approximate each likelihood as a single zero-mean Gaussian PDF via second-order moment matching (Section~\ref{B:gaussian_approx}).
    \item Because the integrals required to compute the marginal likelihoods remain intractable for nonparametric priors, we employ Monte Carlo methods to complete the Bayesian inference procedure, where all prior and posterior PDFs are represented by weighted particles (Section~\ref{B:mc}).
\end{enumerate}

\subsection{Model and Notation}
\label{B:model}

We consider a single time step with a single raw measurement vector $\mathbf z\in\mathbb{C}^C$, $N$ potential objects, and a scalar noise parameter $\eta\in\mathbb{R}_{>0}$.
For each object $n\in\{1,\dots,N\}$ we define
\[
\mathbf x_n \in \mathbb{R}^{d},\qquad
{\boldsymbol{\phi}}_n = (r_n,\,\gamma_n),
\]
where $\mathbf x_n$ is the kinematic state (e.g., $d=4$ for 2-D position and velocity), $r_n\in\{0,1\}$ is the existence variable of object $n$, and $\gamma_n>0$ is the signal power of object $n$. The time index $k$ is omitted in this section.

The likelihood is
\begin{equation}
	f_{\V{z}} \equiv f(\mathbf z \mid \mathbf x_{1:N},{\boldsymbol{\phi}}_{1:N},\eta)	=
	\mathcal{CN}\!\big(\mathbf z;\mathbf 0,\bm{\Sigma}\big).
\end{equation}
with covariance
\begin{equation}
	\bm{\Sigma}
	=
	\eta \mathbf I_C
	+
	\sum_{n=1}^{N}
	r_n\gamma_n\, \mathbf a(\mathbf x_n)\mathbf a(\mathbf x_n)^{\mathrm{H}}.
	\label{B:cov_full}
\end{equation}
Here, $\mathbf a(\mathbf x_n)$ is the array response of object $n$. We use the following simplified notations to denote the prior PDFs
\begin{equation}
f_{\V{x},n}\equiv f(\V{x}_{n}),\qquad f_{{\boldsymbol{\phi}},n}\equiv f({\boldsymbol{\phi}}_{n}),\qquad f_\eta\equiv f(\eta),
\end{equation}
The marginal posterior PDFs are computed as
\begin{align}
	q(\mathbf x_n\mid \mathbf z)
	&\propto
	f(\mathbf x_n)\cdot\ell_{\V{x},n}(\mathbf x_n)
	\label{B:local_post_x}
	\\
	q({\boldsymbol{\phi}}_n\mid \mathbf z)
	&\propto
	f({\boldsymbol{\phi}}_n)\cdot\ell_{{\boldsymbol{\phi}},n}({\boldsymbol{\phi}}_n)
	\label{B:local_post_phi}
	\\
	q(\eta\mid \mathbf z)
	&\propto
	f(\eta)\cdot\ell_\eta(\eta)
	\label{B:local_post_eta}
\end{align}
where $\ell_{\V{x},n}(\mathbf x_n)$, $\ell_{{\boldsymbol{\phi}},n}({\boldsymbol{\phi}}_n)$, and $\ell_\eta(\eta)$ denote the marginal likelihoods for the kinematic state of object $n$, for its existence variable and signal power, and for the noise variance, respectively.

\subsection{Likelihood Computation}
\label{B:likelihood}
\begin{figure}[!t]
	\centering
	\includegraphics[width=3in]{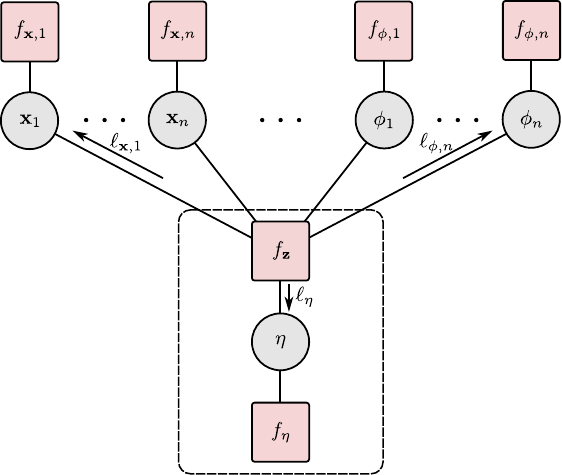}
	\caption{Factor graph of the statistical model at one time step for a single measurement block. Circles denote the variable nodes, i.e., the kinematic states $\V{x}_n$, the existence variables and signal powers $\boldsymbol{\phi}_n$, and the noise variance $\eta$. Squares denote the factor nodes, i.e., the prior PDFs and the likelihood function $f_{\V{z}}$. Arrows indicate the marginal likelihood messages. The complete factor graph for multiple sensors, dictionaries, and snapshots can be found in \cite[Sec.~II-E]{LiaMey:J25}.}
	\label{B:FG}
\end{figure}
The marginal likelihoods in \eqref{B:local_post_x}--\eqref{B:local_post_eta} can be computed by applying the sum-product algorithm~\cite{KscFreLoe:01}, i.e., BP, to the factor graph shown in Fig.~\ref{B:FG}. Because the measurement is a coherent superposition of the contributions of all objects, the marginal likelihood for one object is obtained by marginalizing the contributions of all other objects.
Table~\ref{B:tab_bp} summarizes the key quantities used in Sections~\ref{B:model}--\ref{B:mc}.

\begin{table}[htbp]
	\centering
	\small
	\caption{Key quantities for the model (Sections~\ref{B:model}--\ref{B:mc}).}
	\label{B:tab_bp}
	\begin{tabular}{@{}lll@{}}
		\toprule
		Symbol & Description & Dimension \\
		\midrule
		$\V{z}$ & Measurement vector & $\mathbb{C}^{C}$ \\
		$\V{x}_n$ & Kinematic state of object $n$ & $\mathbb{R}^{d}$ \\
		${\boldsymbol{\phi}}_n = (r_n,\gamma_n)$ & Existence variable and signal power & $\{0,1\}\times\mathbb{R}_{>0}$ \\
		$\eta$ & Background noise variance & $\mathbb{R}_{>0}$ \\
		$\V{a}(\V{x}_n)$ & Array response & $\mathbb{C}^{C}$ \\
		$\bar\gamma_n$ & Effective mean signal power $f(r_n\!=\!1)\,\mathbb{E}[\gamma_n\!\mid\! r_n\!=\!1]$ & scalar \\
		$\M{R}_n$ & Mean outer product of the array response $\mathbb{E}[\V{a}\V{a}^{\mathrm{H}}]$ & $C \times C$ \\
		$\bar\eta$ & Mean noise variance $\mathbb{E}[\eta]$ & scalar \\
		$\M{C}_{{\boldsymbol{\phi}},n},\,\M{C}_{\V{x},n},\,\M{C}_\eta$ & Gaussian PDF covariances & $C \times C$ \\
		\bottomrule
	\end{tabular}
\end{table}

From the factor graph, the marginal likelihood to the random variable ${\boldsymbol{\phi}}_n$ can be computed by the sum-product algorithm:
	\begin{align}
		\ell_{{\boldsymbol{\phi}},n}({\boldsymbol{\phi}}_n)
		&=
		\int\cdots\int
		f(\mathbf z \mid \mathbf x_{1:N},{\boldsymbol{\phi}}_{1:N},\eta)\, f(\eta)\,d\eta
		\nonumber\\
		&\quad\times
		\prod_{m\neq n}
		f({\boldsymbol{\phi}}_m)\, d{\boldsymbol{\phi}}_m
		\prod_{m=1}^{N}
		f(\mathbf x_m)\, d\mathbf x_m,
		\label{B:lphi_exact}
	\end{align}
where we marginalize all other random variables out except for ${\boldsymbol{\phi}}_n$ itself. Similarly, the marginal likelihood to $\mathbf x_n$ is obtained by
	\begin{align}
		\ell_{\V{x},n}(\V{x}_n)
		&=
		\int\cdots\int
		f(\mathbf z \mid \mathbf x_{1:N},{\boldsymbol{\phi}}_{1:N},\eta)\, f(\eta)\, d\eta
		\nonumber\\
		&\quad\times
		\prod_{m=1}^{N}
		f({\boldsymbol{\phi}}_m)\, d{\boldsymbol{\phi}}_m
		\prod_{m\neq n}
		f(\mathbf x_m)\, d\mathbf x_m.
		\label{B:lx_exact}
	\end{align}
The marginal likelihood to noise variable $\eta$ is obtained by
	\begin{align}
		\ell_\eta(\eta)
		&=
		\int\cdots\int
		f(\mathbf z \mid \mathbf x_{1:N},{\boldsymbol{\phi}}_{1:N},\eta)~\prod_{m=1}^{N}
		f({\boldsymbol{\phi}}_m)\, d{\boldsymbol{\phi}}_m
		\prod_{m=1}^{N}
		f(\mathbf x_m)\, d\mathbf x_m.
		\label{B:leta_exact}
	\end{align}

\subsection{Gaussian Approximation of Likelihoods}
\label{B:gaussian_approx}
As shown in the derivations from \eqref{B:lphi_exact} to \eqref{B:leta_exact}, the marginal likelihood messages take the form of continuous Gaussian mixtures. We first show that the first moments (mean) of these likelihoods are exactly zero. For example, considering $\ell_{{\boldsymbol{\phi}},n}$:
\begin{align}
	\mathbb{E}_{\ell_{{\boldsymbol{\phi}},n}}[\V{z}]
	&= \int \V{z} \, \ell_{{\boldsymbol{\phi}},n} d\V{z} \\
	&= \int \V{z} \left[ \int\cdots\int f(\V{z} \mid \V{x}_{1:N},{\boldsymbol{\phi}}_{1:N},\eta) f(\eta)d\eta \prod_{m\ne n}f({\boldsymbol{\phi}}_m)d{\boldsymbol{\phi}}_m \prod_{m=1}^{N}f(\V{x}_m)d\V{x}_m \right] d\V{z} \notag\\
	&= \int\cdots\int \underbrace{\left[ \int \V{z} f(\V{z} \mid \V{x}_{1:N},{\boldsymbol{\phi}}_{1:N},\eta) d\V{z} \right]}_{=\, \mathbb{E}[\V{z} \mid \V{x}_{1:N},{\boldsymbol{\phi}}_{1:N},\eta] \,=\, \V{0}} f(\eta)d\eta \prod_{m\ne n}f({\boldsymbol{\phi}}_m)d{\boldsymbol{\phi}}_m \prod_{m=1}^{N}f(\V{x}_m)d\V{x}_m \notag\\
	&= \V{0},\notag
\end{align}
However, a closed-form evaluation of these mixtures is intractable due to the high-dimensional integration over the continuous latent variables. To ensure analytical tractability, we seek to approximate them with simpler distributions. Therefore, we apply second-order moment matching to approximate these likelihoods as single zero-mean Gaussian PDFs:
\begin{align}
	\ell_{{\boldsymbol{\phi}},n}({\boldsymbol{\phi}}_n)
	&\approx
	\mathcal{CN}\!\big(\mathbf z;\mathbf 0,\mathbf C_{{\boldsymbol{\phi}},n}\big),
	\label{B:lphi_gaussian}
	\\
	\ell_{\V{x},n}(\V{x}_n)
	&\approx
	\mathcal{CN}\!\big(\mathbf z;\mathbf 0,\mathbf C_{x,n}\big),
	\label{B:lx_gaussian}
	\\
	\ell_\eta(\eta)
	&\approx
	\mathcal{CN}\!\big(\mathbf z;\mathbf 0,\mathbf C_{\eta}\big).
	\label{B:leta_gaussian}
\end{align}
Now we derive how to compute the covariance of each Gaussian in the following sections.
\subsubsection{Covariance Computation}
\label{B:cov_comp}

We now derive the covariance matrices
$\mathbf C_{{\boldsymbol{\phi}},n}$, $\mathbf C_{x,n}$, and $\mathbf C_\eta$
in \eqref{B:lphi_gaussian}--\eqref{B:leta_gaussian}. Since the outgoing likelihoods are obtained by marginalizing over the \emph{incoming} densities $f_{\V{x},n}$, $f_{\boldsymbol{\phi},n}$, and $f_{\eta}$, the required moments are defined with respect to these quantities:
\begin{align}
	\bar\gamma_n
	&=
	\mathbb E[r_n\gamma_n]
	=
	\sum_{r_n\in\{0,1\}} r_n\, f(r_n)\int \gamma_n\, f(\gamma_n\mid r_n)\, d\gamma_n
	=
	f(r_n\!=\!1)\,\mathbb{E}[\gamma_n\mid r_n\!=\!1],
	\\
	\M{R}_n
	&=
	\mathbb E\left[\mathbf a(\mathbf x_n)\mathbf a(\mathbf x_n)^{\mathrm{H}}\right]
	=
	\int
	\mathbf a(\mathbf x_n)\mathbf a(\mathbf x_n)^{\mathrm{H}}\,
	f(\mathbf x_n)\, d\mathbf x_n,
	\\
	\bar\eta
	&=
	\mathbb E[\eta]
	=
	\int \eta\, f(\eta)\, d\eta.
\end{align}

\paragraph{Covariance $\mathbf C_{{\boldsymbol{\phi}},n}$}
When computing the marginal likelihood to ${\boldsymbol{\phi}}_n$, the variable ${\boldsymbol{\phi}}_n$ remains while all other variables are averaged out. Conditioned on all hidden variables, the measurement $\mathbf z$ follows a zero-mean complex Gaussian PDF with covariance matrix $\sum_{m=1}^{N} r_m\gamma_m\, \mathbf a(\mathbf x_m)\mathbf a(\mathbf x_m)^{\mathrm{H}}+\eta \mathbf I_C$. Thus, the second moment of $\mathbf z$ is obtained by averaging this conditional covariance over the incoming densities of all variables except ${\boldsymbol{\phi}}_n$. Hence,
\begin{align}
	\mathbf C_{{\boldsymbol{\phi}},n}
	=&~\int \mathbf z\mathbf z^{\mathrm{H}}\, \ell_{{\boldsymbol{\phi}},n}\, d\mathbf z\notag\\
	=&~\int\cdots \int\left(
	\sum_{m=1}^{N}
	r_m\gamma_m\, \mathbf a(\mathbf x_m)\mathbf a(\mathbf x_m)^{\mathrm{H}}+\eta \mathbf I_C
	\right)f(\eta)\,d\eta
	\prod_{m\neq n}
	f({\boldsymbol{\phi}}_m)\, d{\boldsymbol{\phi}}_m
	\prod_{m=1}^{N}
	f(\mathbf x_m)\, d\mathbf x_m
	\notag\\
	=&~
	r_n\gamma_n
	\int \mathbf a(\mathbf x_n)\mathbf a(\mathbf x_n)^{\mathrm{H}}\, f(\mathbf x_n)\, d\mathbf x_n
	+
	\bigg(\int \eta\, f(\eta)\, d\eta\bigg)\mathbf I_C
	\notag\\
	&~+
	\sum_{m\neq n}
	\bigg(\int r_m\gamma_m\, f({\boldsymbol{\phi}}_m)\, d{\boldsymbol{\phi}}_m\bigg)
	\int \mathbf a(\mathbf x_m)\mathbf a(\mathbf x_m)^{\mathrm{H}}\, f(\mathbf x_m)\, d\mathbf x_m
	\notag\\
	=&~
	r_n\gamma_n\,\M{R}_n
	+
	\sum_{m\neq n}\bar\gamma_m\,\M{R}_m
	+
	\bar\eta\,\mathbf I_C,
\end{align}
where the third and fourth lines follow from the linearity of the integral and the fact that the densities of the variables that do not appear in a term integrate to one, and the last line follows from the definitions of $\bar\gamma_m$, $\M{R}_m$, and $\bar\eta$ above.

\paragraph{Covariance $\mathbf C_{\V{x},n}$}

When computing the marginal likelihood to $\mathbf x_n$, the state $\mathbf x_n$ remains while all other variables are averaged out. Thus,
\begin{equation}
	\mathbf C_{\V{x},n}
	=
	\bar\gamma_n\, \mathbf a(\mathbf x_n)\mathbf a(\mathbf x_n)^{\mathrm{H}}
	+
	\sum_{m\neq n}\bar\gamma_m\,\M{R}_m
	+
	\bar\eta\,\mathbf I_C.
\end{equation}
Note that the moment matching combines the hypotheses $r_m=0$ and $r_m=1$ of each object into the effective mean power $\bar\gamma_m=f(r_m\!=\!1)\,\mathbb{E}[\gamma_m\mid r_m\!=\!1]$, and that $\mathbf C_{\V{x},n}$ also uses $\bar\gamma_n$ for object $n$ itself. Hence, for a potential object with a small existence probability, the message $\ell_{\V{x},n}(\V{x}_n)$ is less informative than the exact Gaussian mixture in \eqref{B:lx_exact}, which can delay the detection of newly appearing objects. This approximation is analogous to the Gaussian moment matching in \cite{LiaMey:J25}.

\paragraph{Covariance $\mathbf C_{\eta}$}
When computing the marginal likelihood to $\eta$, the noise variable remains while all object contributions are averaged out. Therefore,
\begin{equation}
	\mathbf C_{\eta}
	=
	\sum_{m=1}^{N}\bar\gamma_m\,\M{R}_m
	+
	\eta\,\mathbf I_C.
\end{equation}

\subsection{Monte Carlo Implementation}
\label{B:mc}
The integrals in Section~\ref{B:cov_comp} involve expectations over continuous priors that generally lack closed-form solutions.
We therefore represent each prior by a weighted set of $P$ Monte Carlo samples (particles) and replace integrals with weighted sums.

\subsubsection{Particle Representation of Prior PDFs}
\begin{itemize}
	\item \textbf{Object state.}
	The prior of the $n$-th object state is approximated by a weighted particle set
	\[
	f(\V{x}_n) \approx \left\{ \big(\omega_{\V{x},n,p},\V{x}_{n,p}\big) \right\}_{p=1}^{P},
	\]
	where $\V{x}_{n,p}$ denotes the $p$-th particle and $\omega_{\V{x},n,p}$ is its associated weight, with $\sum_p \omega_{\V{x},n,p} = 1$.
	
	\item \textbf{Signal power and existence.}
	The prior of the signal power state ${\boldsymbol{\phi}}_n =(\gamma_n,r_n)$ is denoted as,
\begin{align}
	f(\bm{\phi}_{n}) =
	\begin{cases}
		f(r_{n}=1)\,
		f_{\bm{\theta}}(\gamma_{n}\mid r_{n}=1),
		& r_{n}=1,\\[2pt]
		f(r_{n}=0)\,
		f_{\mathrm{D}}(\gamma_{n}),
		& r_{n}=0,
	\end{cases}
	\label{eq:phi-mixture-bp}
\end{align}
	where $f_{\mathrm{D}}(\gamma_n)$ is a \emph{dummy} density assigned when $r_n=0$, since the measurement model does not depend on $\gamma_n$ in this case and it can be chosen arbitrarily without affecting inference.
	The conditional power prior is approximated by particles
	\[
	f(\gamma_n \mid r_n=1) \approx \left\{ \big(\omega_{\gamma,n,p},\gamma_{n, p}\big) \right\}_{p=1}^{P}.
	\]
	The binary variable $r_n$ follows a Bernoulli distribution, i.e., $f(r_n\!=\!1) = 1 - f(r_n\!=\!0)$.
	\item \textbf{Background noise.}
	The prior of the background noise parameter $\eta$ is approximated by
	\[
	f(\eta) \approx \left\{ \big(\omega_{\eta,p},\eta_p\big) \right\}_{p=1}^{P}.
	\]
\end{itemize}
In practice, the number of particles used for different random variables need not be the same; here we use the same notation $P$ for simplicity.

\subsubsection{Per-Particle Covariance Matrices}
\label{B:mc_covariance}
Substituting the particle representations into the covariance expressions from Section~\ref{B:cov_comp}, each expectation becomes a weighted sum over the corresponding particles.
For a given particle of the target variable, the moments of all other variables are computed with respect to their priors, yielding
\begin{align}
	\mathbf C_{{\boldsymbol{\phi}},n,p}
	=&~
	r_n\gamma_{n,p}\,\M{R}_n
	+
	\sum_{m\neq n}\bar\gamma_m\,\M{R}_m
	+
	\bar\eta\,{\M I}_{C},\\
	\mathbf C_{\V{x},n,p}
	=&~
	\bar\gamma_n\, \V{a}(\mathbf x_{n,p})\V{a}(\V{x}_{n,p})^{\mathrm{H}}
	+
	\sum_{m\neq n}\bar\gamma_m\,\M{R}_m
	+
	\bar\eta\,\mathbf I_C,\\
	\mathbf C_{\eta,p}
	=&~
	\sum_{n=1}^{N}\bar\gamma_n\,\M{R}_n
	+
	\eta_p\,\mathbf I_C,
\end{align}
where the expected statistics are computed from the prior particles:
\begin{align}
	\bar\gamma_n
	&= f(r_n\!=\!1)\sum_p\omega_{\gamma,n,p}\,\gamma_{n,p},
	\\
	\M{R}_n
	&=
	\sum_p
	\omega_{\V{x}, n,p}\,\mathbf a(\mathbf x_{n,p})\mathbf a(\mathbf x_{n,p})^{\mathrm{H}},
	\\
	\bar\eta
	&=
	\sum_p\omega_{\eta,p}\,\eta_p.
\end{align}

\subsubsection{Weight Update}
Each per-particle covariance is used to compute the Gaussian likelihood.
The posterior weight of each particle is obtained by multiplying its prior weight with the corresponding likelihood value and renormalizing. This update corresponds to importance sampling with proposal given by the prior PDF, where the importance weights are proportional to the Gaussian likelihood.
For example, the weight update for the $p$-th power particle of object $n$ is
\begin{equation}
	\omega_{\gamma,n,p}^{\mathrm{post}}
	=
	\frac{
		\omega_{\gamma,n,p}\;\mathcal{CN}\!\big(\V{z};\,\V{0},\,\M{C}_{{\boldsymbol{\phi}},n,p}\big)
	}{
		\sum_{p'=1}^{P}\omega_{\gamma,n,p'}\;\mathcal{CN}\!\big(\V{z};\,\V{0},\,\M{C}_{{\boldsymbol{\phi}},n,p'}\big)
	},
	\label{B:weight_update}
\end{equation}
and similarly for $\omega_{\V{x},n,p}^{\mathrm{post}}$ and $\omega_{\eta,p}^{\mathrm{post}}$.
The resulting weighted particle sets form a nonparametric approximation of the marginal posterior PDFs \eqref{B:local_post_x}--\eqref{B:local_post_eta}.
After the weight update, a resampling step is applied to mitigate particle degeneracy before propagating the particles of the kinematic states and of the noise variance to the next time step. The prior particles of the signal powers are not propagated; they are drawn anew from the flow at each time step (Section~\ref{A:pipeline}).

\paragraph{Existence probability}
The posterior existence probability follows from \eqref{B:local_post_phi} by summing over $r_n$ and integrating over $\gamma_n$.
For $r_n=0$, the covariance $\M{C}_{{\boldsymbol{\phi}},n,p}$ reduces to $\sum_{m\neq n}\bar\gamma_m\M{R}_m+\bar\eta\,\M{I}_C$, which does not depend on $\gamma_n$, and the dummy density $f_{\mathrm{D}}(\gamma_n)$ integrates to one.
For $r_n=1$, the integral over $\gamma_n$ is replaced by the weighted sum over the power particles. Hence,
\begin{align}
	q(r_n\!=\!1\mid\V{z})
	=
	\frac{f(r_n\!=\!1)\sum_{p}\omega_{\gamma,n,p}\,\mathcal{CN}\big(\V{z};\V{0},\M{C}_{{\boldsymbol{\phi}},n,p}\big)}
	{f(r_n\!=\!1)\sum_{p}\omega_{\gamma,n,p}\,\mathcal{CN}\big(\V{z};\V{0},\M{C}_{{\boldsymbol{\phi}},n,p}\big)
	+f(r_n\!=\!0)\,\mathcal{CN}\big(\V{z};\V{0},\sum_{m\neq n}\bar\gamma_m\M{R}_m+\bar\eta\,\M{I}_C\big)},
	\label{B:existence_update}
\end{align}
where $\M{C}_{{\boldsymbol{\phi}},n,p}$ is evaluated at $r_n=1$.
Moreover, $q(\gamma_n\mid r_n\!=\!1,\V{z})$ is given by the weights in \eqref{B:weight_update}, and $q(\gamma_n\mid r_n\!=\!0,\V{z})=f_{\mathrm{D}}(\gamma_n)$.

\paragraph{Remark}
When multiple snapshots, sensors, or dictionaries are available, the measurement consists of several measurement blocks that are conditionally independent given the states, and the factor graph contains one likelihood factor per block. Since the kinematic states and existence variables are shared across blocks, the factor graph contains loops. In this case, the message passing described above is performed for each block, where the prior PDFs of the shared variables are replaced by the products of the prior PDFs and the messages from all other blocks, and the messages are iterated as in \cite{LiaMey:J25}. The per-particle weight update rule remains unchanged.

\newpage
\section{Derivations of Efficient Computation}
\label{app:ec}
This section derives an efficient computation for the Bayesian inference with high dimensional measurement in our model. All derivations in this section build upon the results of Section~\ref{app:bp}. We proceed in three stages:
\begin{enumerate}

	\item \textbf{Low-rank formulation} (Sections~\ref{B:lowrank}--\ref{B:messages}). A rank-one approximation of the covariance, combined with the Woodbury lemma and the Sherman--Morrison formula, reduces the computation complexity of Gaussian PDFs from $O(C^3 PN)$ to $O(CPN^2)$ (Given $N$ objects, $P$ Monte Carlo samples and $C$-dimensional measurement).
	
	\item \textbf{Kronecker Decomposition} (Section~\ref{B:kronecker}). When the manifold admits a Kronecker product structure, all $C$-dimensional operations could be replaced by operations on the smaller factors, reducing total storage complexity from $O(CPN)$ to $O(C_{\max}NP)$.
	
	\item \textbf{Example} (Section~\ref{B:example}). A practical radar sensing example -- 3-D tensor analog-to-digital-converter (ADC) data, is used to quantify the specific reductions in computational and storage complexity.
\end{enumerate}
\subsection{Efficient Low-Rank Formulation}
\label{B:lowrank}

Evaluating the Gaussian PDFs directly requires inverting the $C\times C$ covariance matrix at a cost of $O(C^3)$ per evaluation, giving $O(C^3PN)$ over all objects and particles. The storage cost is $O(C^2PN)$. Both are unacceptable when $C$ or $P$ is large. The approach proceeds in three steps:

First, we approximate the covariance of each Gaussian (from \eqref{B:lphi_exact} to \eqref{B:leta_exact}) by a rank-one outer product (Section~\ref{B:rank1}).
Second, we exploit this structure via the Woodbury lemma and the Sherman--Morrison formula to compute matrix--vector products involving the inverse without ever forming the $C\times C$ inverse explicitly (Section~\ref{B:woodbury}).
Third, we cast the particle weight updates as log-likelihood ratios (Section~\ref{B:loglik}), which, combined with the rank-one structure, reduce each likelihood evaluation to two scalars per particle (Section~\ref{B:messages}).
Table~\ref{B:tab_lowrank} summarizes the key quantities introduced in Sections~\ref{B:lowrank}--\ref{B:messages}.

\begin{table}[htbp]
	\centering
	\small
	\caption{Key quantities for the low-rank formulation (Sections~\ref{B:lowrank}--\ref{B:messages}).}
	\label{B:tab_lowrank}
	\begin{tabular}{@{}lll@{}}
		\toprule
		Symbol & Description & Dimension \\
		\midrule
		$\bar{\V{a}}_n$ & Weighted mean array response & $C \times 1$ \\
		$\V{u}_n$ & $\sqrt{\bar\gamma_n}\,\bar{\V{a}}_n$; object contribution & $C \times 1$  \\
		$\M{U}$ & $[\V{u}_1,\dots,\V{u}_N]$ & $C \times N$  \\
		$\M{B}^{-1}$ & $(\bar\eta\,\M{I}_C + \M{U}\M{U}^{\mathrm{H}})^{-1}$; full covariance & $C \times C$  \\
		$\M{G}$ & Gram matrix $\M{U}^{\mathrm{H}}\M{U}$ & $N \times N$  \\
		$\M{L}$ & $(\bar\eta\,\M{I}_N + \M{G})^{-1}$; inverse kernel & $N \times N$  \\
		$\M{A}_n$ & $\M{B} - \V{u}_n\V{u}_n^{\mathrm{H}}$; leave-one-out cov. & $C \times C$  \\
		$\V{h}_{n,p}$ & Particle-specific perturbation & $C \times 1$   \\
		$s_{n,p},\, q_{n,p}$ & Log-likelihood ratio scalars & scalar   \\
		\bottomrule
	\end{tabular}
\end{table}

\subsubsection{Rank-One Approximation}
\label{B:rank1}
Recall from Section~\ref{B:mc} that $\M{R}_n = \sum_p \omega_{\V{x},n,p}\,\V{a}(\V{x}_{n,p})\V{a}(\V{x}_{n,p})^{\mathrm{H}}$ is a weighted sum of $P$ rank-one outer products, so it has rank up to $\min(P,C)$.
Storing or inverting this $C\times C$ matrix is infeasible when $C$ is large. However, in high-dimensional sensing, the measurement likelihood is generally sharply peaked, causing the particle cloud to concentrate as shown in Fig.~\ref{fig:radar_particles_full}. This motivates the following approximation:
\begin{align}
	\M{R}_n \approx&~ \left(\sum_p
	\omega_{\V{x}, n,p}\V{a}(\V{x}_{n,p})\right)\left(\sum_p
	\omega_{\V{x}, n,p}\V{a}(\V{x}_{n,p})\right)^{\mathrm{H}}
	= \bar {\V{a}}_{n}\bar {\V{a}}^{\mathrm{H}}_{n}.
	\label{B:rank1_approx}
\end{align}
The approximation error equals the matrix $\M{R}_n - \bar{\mathbf a}_n\bar{\V{a}}_n^{\mathrm{H}}$, which is positive semidefinite. In practice, because our framework treats the background noise variance $\eta$ as an unknown random variable rather than a fixed parameter, the Bayesian inference process partly absorbs this residual error. Another way is to manually add a diagonal correction matrix to the background noise term $\bar \eta \M{I}_C$, which will not influence the overall derivations following.
The approximation is accurate if the particles of object $n$ are concentrated, so that the phases of the array responses $\V{a}(\V{x}_{n,p})$ are nearly aligned. For a widespread particle cloud, e.g., for a newborn object whose particles are drawn from the birth PDF, these phases average out, and $\|\bar{\V{a}}_n\|^2$ can be much smaller than $\sum_p \omega_{\V{x},n,p}\|\V{a}(\V{x}_{n,p})\|^2$.

\subsubsection{Notation for the Low-Rank Model}
Under the rank-one approximation, we introduce the following quantities:
\begin{itemize}
	\item ${\mathbf a}_{n,p}\in \mathbb{C}^{C}$ is the array response of object $n$ for particle $p$. It is a simplified notation of $\mathbf a(\mathbf x_{n,p})$ in Section~\ref{B:mc_covariance}. The weighted mean is $\bar {\mathbf a}_{n} = \sum_{p}\omega_{\V{x},n,p}\, {\mathbf a}_{n,p}$.
	\item ${\gamma}_{n,p}\in \mathbb{R}$ is the power particle $p$ of object $n$. The effective mean signal power is $\bar {\gamma}_{n} = f(r_n\!=\!1)\sum_{p}\omega_{\gamma,n,p}\, {\gamma}_{n,p}$.
	\item  ${\eta_p} \in \mathbb{R}$ is the noise variance particle $p$. The weighted mean is $\bar \eta = \sum_{p}\omega_{\eta,p}\, {\eta_p}$.
\end{itemize}
We use $P$ to denote the total number of particles, $C$ the dimension of the channel, and $N$ the number of objects. We usually have the following relationships:
\begin{align*}
	C \gg N^2 \qquad P \gg N^2
\end{align*}
The values of dimension $C$ and $P$ depend on the specific problem, but both are typically very large; for the complexity analysis we treat them as being of the same order.

\vspace{0.2mm}
The contribution of object $n$ to the measurement is represented by the $C$-dimensional vector
\begin{align}
	{\mathbf u}_{n} &= \sqrt{\bar \gamma_n} \cdot \bar {\mathbf a}_{n} \in \mathbb{C}^{C}
\end{align}
We collect them into a matrix
\begin{equation}
	{\mathbf U} =
	\bigl[ {\mathbf u}_{1}, \dots, {\mathbf u}_{N} \bigr] \in \mathbb{C}^{C \times N}.
\end{equation}
The covariance of the measurement ${\mathbf z}$ is
\begin{equation}
	{\mathbf B} = \bar\eta\, {\mathbf I}_C + {\mathbf U} {\mathbf U}^{\mathrm{H}} \in \mathbb{C}^{C\times C}.
	\label{B:B_def}
\end{equation}
For the object $n$, we define the covariance when excluding this object:
\begin{equation}
	{\mathbf A}_{n} = {\mathbf B} - {\mathbf u}_{n} {\mathbf u}_{n}^{\mathrm{H}}.
	\label{B:A_n_def}
\end{equation}

\paragraph{The leave-one-out structure}
The definition of ${\M A}_n$ comes directly from the sum-product algorithm. Recall from Section~\ref{B:cov_comp} that the covariance of the likelihood to object~$n$'s state or signal power always takes the form
\[
\M{C}_{n,p} = \underbrace{\sum_{m \neq n} \bar\gamma_m\,\M{R}_m + \bar\eta\,\M{I}_C}_{\text{contributions of all other objects + noise}} + \underbrace{\V{h}_{n,p}\V{h}_{n,p}^{\mathrm{H}}}_{\text{particle-specific term of object } n}.
\]
That is, when computing the likelihood directed at one variable, the sum-product rule marginalizes out all remaining variables, so object~$n$'s own contribution is naturally separated from the rest.

Under the rank-one approximation~\eqref{B:rank1_approx}, the first part is exactly ${\mathbf A}_n = {\mathbf B} - {\mathbf u}_n {\mathbf u}_n^{\mathrm{H}}$, the covariance of the ``background'' as seen by object~$n$.

The second part, $\V{h}_{n,p}\V{h}_{n,p}^{\mathrm{H}}$, varies across particles and captures the specific hypothesis about object~$n$ in that particle. Because ${\mathbf A}_n$ is shared by all particles of the same object, it only needs to be inverted once per object --- an identity that will be used in Section~\ref{B:loglik} to reduce each likelihood evaluation to two scalars.

\subsubsection{Woodbury Lemma}
\label{B:woodbury}
The computation of Gaussian PDFs will repeatedly require products of the form ${\mathbf B}^{-1}{\mathbf X}$, where ${\mathbf X}$ varies across objects and particles. By using the Woodbury lemma, the complexity of this matrix multiplication would reduce to $O(CNK)$ where $K$ is the number of columns of~${\mathbf X}$.

For an arbitrary matrix ${\mathbf X} \in \mathbb{C}^{C\times K}$, we have the following.

The inverse of ${\mathbf B}$ is given by the Woodbury lemma
\begin{equation}
	{\mathbf B}^{-1} =  \frac{1}{\bar\eta}{\mathbf I}_{C} - \frac{1}{\bar\eta}{\mathbf U}({\bar\eta}{\mathbf I}_{N} + {\mathbf U}^{\mathrm{H}}{\mathbf U})^{-1}{\mathbf U}^{\mathrm{H}}.
	\label{B:woodbury_eq}
\end{equation}
Then
\begin{align}
	{\mathbf B}^{-1} {\mathbf X} = \frac{1}{\bar\eta} {\mathbf X}
	-\frac{1}{\bar\eta}{\mathbf U}({\bar\eta}{\mathbf I}_{N} + {\mathbf U}^{\mathrm{H}}{\mathbf U})^{-1}{\mathbf U}^{\mathrm{H}}{\mathbf X}
\end{align}
This is implemented as:
\begin{align}
	{\mathbf L} &= ({\bar\eta}{\mathbf I}_{N} + {\mathbf U}^{\mathrm{H}}{\mathbf U})^{-1}, \\
	{\mathbf T} &= {\mathbf L} {\mathbf U}^{\mathrm{H}}{\mathbf X}, \\
	{\M B}^{-1} {\mathbf X} &= \frac{1}{\bar\eta} ({\mathbf X} - {\mathbf U}{\mathbf T}).
\end{align}
The two shared quantities ${\mathbf U}^{\mathrm{H}}{\mathbf U}\in\mathbb{C}^{N\times N}$ and ${\mathbf L}=({\bar\eta}{\mathbf I}_{N}+{\mathbf U}^{\mathrm{H}}{\mathbf U})^{-1}\in\mathbb{C}^{N\times N}$ are computed once and reused throughout all subsequent computations.
Their one-time cost is $O(CN^2)$ for the Gram matrix and $O(N^3)$ for the inversion.
After this pre-computation, the cost of each product ${\mathbf B}^{-1}{\mathbf X}\in\mathbb{C}^{C\times K}$ is only $O(CNK)$ if $K > N$.

\paragraph{Sherman--Morrison formula}
The marginal likelihood for each object~$n$ involves the ``leave-one-out'' covariance ${\mathbf A}_n$, obtained by removing object~$n$'s own contribution from~${\mathbf B}$.
Rather than recomputing the inverse from scratch for each~$n$, we apply the Sherman--Morrison formula to compute ${\mathbf A}_n^{-1}$. We have
\begin{equation}
	{\mathbf A}_{n}^{-1}
	= {\mathbf B}^{-1}
	+ \frac{{\mathbf B}^{-1}{\mathbf u}_{n}{\mathbf u}_{n}^{\mathrm{H}}{\mathbf B}^{-1}}
	{1 - {\mathbf u}_{n}^{\mathrm{H}}{\mathbf B}^{-1}{\mathbf u}_{n}},
	\label{B:sherman_morrison}
\end{equation}
provided $1 - {\mathbf u}_{n}^{\mathrm{H}}{\mathbf B}^{-1}{\mathbf u}_{n} \neq 0$. This formula can be proved via the Woodbury lemma. The condition always holds, because ${\mathbf A}_n=\bar\eta\,{\mathbf I}_C+\sum_{m\neq n}{\mathbf u}_m{\mathbf u}_m^{\mathrm{H}}$ is positive definite and, by the matrix determinant lemma, $1-{\mathbf u}_{n}^{\mathrm{H}}{\mathbf B}^{-1}{\mathbf u}_{n}=\det({\mathbf A}_n)/\det({\mathbf B})>0$.

\subsection{Log-Likelihood Ratio}
\label{B:loglik}

In the Monte Carlo implementation, the weight of each particle is updated by multiplying the prior weight with a likelihood value and renormalizing.
Therefore, any factor common to all particles of a given object can cancel out.
Hence, it suffices to compute likelihood ratios with respect to a particle-independent reference.

Each particle-specific covariance $\mathbf C$ differs from a common factor $\mathbf A$ by a single rank-one term: $\mathbf C = \mathbf A + \mathbf h\mathbf h^{\mathrm{H}}$. Direct evaluation of a $C$-dimensional Gaussian PDF requires evaluating $\log\det(\mathbf C)$ and $\mathbf z^{\mathrm{H}}\mathbf C^{-1}\mathbf z$. Both require $O(C^3)$ operations.
However, the \emph{ratio of the likelihood} evaluated with $\mathbf C$ to that evaluated with $\mathbf A$ can be expressed entirely in terms of \textbf{two scalars}. Since $\mathbf A$ is the same for all particles, this ratio is all that is needed for weight updates.

The object state and signal-power computation share this common structure, so we derive the log-likelihood ratio once here and apply it repeatedly in Section~\ref{B:messages}.

Consider two covariance matrices that differ by a rank-one term:
\begin{equation}
	\mathbf C = \mathbf A + \mathbf h\mathbf h^{\mathrm{H}},
\end{equation}
where $\mathbf A\in\mathbb{C}^{C\times C}$ is Hermitian positive definite and $\mathbf h\in\mathbb{C}^C$ can be any vector.
Define the shorthand
\begin{equation}
	s = \mathbf h^{\mathrm{H}} \mathbf A^{-1}\mathbf h,\qquad
	q = \mathbf z^{\mathrm{H}} \mathbf A^{-1}\mathbf h.
	\label{B:sq_def}
\end{equation}

\paragraph{Log-determinant}
By the matrix determinant lemma,
\begin{equation}
	\det(\mathbf C) = (1+s)\,\det(\mathbf A),
\end{equation}
hence $\log\det(\mathbf C) - \log\det(\mathbf A) = \log(1+s)$.

\paragraph{Quadratic form}
By the Sherman--Morrison formula \eqref{B:sherman_morrison},
\begin{equation}
	\mathbf C^{-1}
	= \mathbf A^{-1}
	- \frac{\mathbf A^{-1}\mathbf h\mathbf h^{\mathrm{H}}\mathbf A^{-1}}{1+s},
\end{equation}
so that
\begin{equation}
	\mathbf z^{\mathrm{H}}\mathbf A^{-1}\mathbf z
	- \mathbf z^{\mathrm{H}}\mathbf C^{-1}\mathbf z
	= \frac{|q|^2}{1+s}.
\end{equation}

\paragraph{Log-likelihood ratio}
Since the zero-mean complex Gaussian log-PDF is $\log\mathcal{CN}(\mathbf z;\mathbf 0,\mathbf C) = -\log\det(\mathbf C) - \mathbf z^{\mathrm{H}}\mathbf C^{-1}\mathbf z + \mathrm{const}$, the log-ratio becomes
\begin{equation}
	\log\frac{\mathcal{CN}(\mathbf z;\mathbf 0,\mathbf C)}{\mathcal{CN}(\mathbf z;\mathbf 0,\mathbf A)}
	= -\log(1+s) + \frac{|q|^2}{1+s}.
	\label{B:loglik_ratio}
\end{equation}
This identity is applied in the sequel with $\mathbf A = \mathbf A_n$ (the excluded covariance) and $\mathbf h = \mathbf h_{n,p}$ (the particle-specific contribution, which will be discussed in detail in the next subsection).

\subsection{Efficient Likelihood Computation}
\label{B:messages}
With the log-likelihood ratio identity~\eqref{B:loglik_ratio} and the efficient inversion tools from Section~\ref{B:lowrank} in hand, we now derive the three types of marginal likelihoods. The key step is to identify the common covariance~$\mathbf A$ and the rank-one perturbation~$\mathbf h$, and then evaluate the two scalars $s$ and $q$ from~\eqref{B:sq_def}.

\subsubsection{Object Update}
For each object particle, the particle-specific covariance is $\mathbf C_{\V{x},n,p} = \mathbf A_n + \mathbf h_{n,p}\mathbf h_{n,p}^{\mathrm{H}}$, where
\begin{equation}
	{\mathbf h}_{n,p} = \sqrt{\bar \gamma_n}\, {\mathbf a}_{n,p}.
\end{equation}
Applying \eqref{B:loglik_ratio} with $s_{n,p} = {\mathbf h}_{n,p}^{\mathrm{H}} {\mathbf A}_{n}^{-1} {\mathbf h}_{n,p}$ and $q_{n,p} = {\mathbf z}^{\mathrm{H}} {\mathbf A}_{n}^{-1} {\mathbf h}_{n,p}$, the log-likelihood message for object $n$ with particle $p$ is
\begin{equation}
	\log \ell_{\V{x},n,p}
	= -\log\bigl(1 + s_{n,p}\bigr)
	+ \frac{\lvert q_{n,p}\rvert^2}{1 + s_{n,p}} + \mathrm{const}.
	\label{B:log_ell_x}
\end{equation}

\paragraph{Formula for $s_{n,p}$}
Applying \eqref{B:sherman_morrison},
\begin{align}
	s_{n,p} = {\mathbf h}_{n,p}^{\mathrm{H}} {\mathbf A}_{n}^{-1} {\mathbf h}_{n,p}
	&= {\mathbf h}_{n,p}^{\mathrm{H}} {\mathbf B}^{-1} {\mathbf h}_{n,p}
	+ {\mathbf h}_{n,p}^{\mathrm{H}} {\mathbf B}^{-1}{\mathbf u}_{n}
	\bigl(1 - {\mathbf u}_{n}^{\mathrm{H}} {\mathbf B}^{-1}  {\mathbf u}_{n}\bigr)^{-1}
	{\mathbf u}_{n}^{\mathrm{H}} {\mathbf B}^{-1}{\mathbf h}_{n,p} \notag\\
	&= {\mathbf h}_{n,p}^{\mathrm{H}}{\mathbf B}^{-1}{\mathbf h}_{n,p}
	+ \frac{ \bigl| {\mathbf u}_{n}^{\mathrm{H}} {\mathbf B}^{-1}{\mathbf h}_{n,p} \bigr|^2  }{ 1 - {\mathbf u}_{n}^{\mathrm{H}} {\mathbf B}^{-1} {\mathbf u}_{n}}.
\end{align}
\paragraph{Formula for $q_{n,p}$}
Likewise, applying \eqref{B:sherman_morrison},
\begin{align}
	q_{n,p} = {\mathbf z}^{\mathrm{H}} {\mathbf A}_{n}^{-1} {\mathbf h}_{n,p} = {\mathbf z}^{\mathrm{H}}{\mathbf B}^{-1}{\mathbf h}_{n,p}
	+ \frac{
		{\mathbf z}^{\mathrm{H}}{\mathbf B}^{-1}{\mathbf u}_{n} \cdot {\mathbf u}^{\mathrm{H}}_n {\mathbf B}^{-1}{\mathbf h}_{n,p}
	}{
		1 - {\mathbf u}_{n}^{\mathrm{H}} {\mathbf B}^{-1} {\mathbf u}_{n}
	}.
\end{align}
\paragraph{Complexity Analysis}
Using the pre-computed ${\mathbf U}^{\mathrm{H}}{\mathbf U}$ and ${\mathbf L}$ from Section~\ref{B:woodbury}, the remaining steps are:
\begin{enumerate}
	\item Compute ${\mathbf B}^{-1}{\mathbf u}_{n}$ at cost $O(C N)$ per object; for all $N$ objects: $O(C N^2)$.
	\item Compute ${\mathbf B}^{-1}{\mathbf h}_{n,p}$ at cost $O(C N)$ per particle; for all objects and particles: $O(CP N^2)$.
\end{enumerate}
The dominant cost is $O(CPN^2)$.

\subsubsection{Signal-Power Update}
The particle-specific covariance is $\mathbf C_{{\boldsymbol{\phi}},n,p} = \mathbf A_n + r_n\,\mathbf h_{n,p}\mathbf h_{n,p}^{\mathrm{H}}$, where the existence variable $r_n$ appears explicitly because ${\boldsymbol{\phi}}_n=(r_n,\gamma_n)$ is the message target.

\paragraph{Case $r_n = 0$} The rank-one term vanishes, so $\mathbf C_{{\boldsymbol{\phi}},n,p} = \mathbf A_n$ and the log-likelihood ratio \eqref{B:loglik_ratio} is zero (i.e., $s_{n,p}=q_{n,p}=0$).

\paragraph{Case $r_n = 1$}
The particle-specific vector $\mathbf h_{n,p}$ is a scaled version of $\mathbf u_n$:
\begin{equation}
	\alpha_{n,p} = \frac{\sqrt{\gamma_{n,p}}}{\sqrt{\bar \gamma_n}},\qquad
	{\mathbf h}_{n,p} = \alpha_{n,p}\, {\mathbf u}_{n}.
\end{equation}
Since ${\mathbf h}_{n,p}$ is proportional to ${\mathbf u}_{n}$, the quantities in \eqref{B:sq_def} simplify to
\begin{align}
	s_{n,p} &= \alpha_{n,p}^2\, {\mathbf u}_{n}^{\mathrm{H}} {\mathbf A}_{n}^{-1} {\mathbf u}_{n}, \\
	q_{n,p} &= \alpha_{n,p}\, {\mathbf z}^{\mathrm{H}} {\mathbf A}_{n}^{-1} {\mathbf u}_{n}.
\end{align}
Applying \eqref{B:loglik_ratio}, the log-likelihood message is
\begin{equation}
	\log \ell_{{\boldsymbol{\phi}},n,p}
	=  -\log\bigl(1 + s_{n,p}\bigr)
	+ \frac{\lvert q_{n,p}\rvert^2}{1 + s_{n,p}} + \mathrm{const}.
\end{equation}

\paragraph{Formula for $s_{n,p}$}
Applying \eqref{B:sherman_morrison} and dividing by $\alpha_{n,p}^2$,
\begin{align}
	\frac{s_{n,p}}{\alpha_{n,p}^2} = {\mathbf u}^H_{n}{\mathbf B}^{-1}{\mathbf u}_{n}
	+ \frac{ \bigl| {\mathbf u}_{n}^{\mathrm{H}} {\mathbf B}^{-1}{\mathbf u}_{n} \bigr|^2  }{ 1 - {\mathbf u}_{n}^{\mathrm{H}} {\mathbf B}^{-1} {\mathbf u}_{n}}.
\end{align}
\paragraph{Formula for $q_{n,p}$}
Likewise, dividing by $\alpha_{n,p}$,
\begin{align}
	\frac{q_{n,p}}{\alpha_{n,p}}  = {\mathbf z}^{\mathrm{H}}{\mathbf B}^{-1}{\mathbf u}_{n}
	+ \frac{
		{\mathbf z}^{\mathrm{H}}{\mathbf B}^{-1}{\mathbf u}_{n} \cdot {\mathbf u}^{\mathrm{H}}_n {\mathbf B}^{-1}{\mathbf u}_{n}
	}{
		1 - {\mathbf u}_{n}^{\mathrm{H}} {\mathbf B}^{-1} {\mathbf u}_{n}
	}.
\end{align}

With $\beta_n={\mathbf u}_{n}^{\mathrm{H}}{\mathbf B}^{-1}{\mathbf u}_{n}$, these expressions simplify to $s_{n,p}/\alpha_{n,p}^2=\beta_n/(1-\beta_n)$ and $q_{n,p}/\alpha_{n,p}={\mathbf z}^{\mathrm{H}}{\mathbf B}^{-1}{\mathbf u}_{n}/(1-\beta_n)$. To avoid a division by $\sqrt{\bar\gamma_n}$, which vanishes if $f(r_n\!=\!1)=0$, the vector ${\mathbf h}_{n,p}=\sqrt{\gamma_{n,p}}\,\bar{\mathbf a}_n$ can also be used directly.

\paragraph{Complexity Analysis}
The pre-computed quantities from Section~\ref{B:woodbury} are reused. The remaining steps are:
\begin{enumerate}
	\item Compute ${\mathbf B}^{-1}{\mathbf u}_{n}$ at cost $O(CN)$ per object; for all $N$ objects: $O(CN^2)$.
	\item Perform $O(PN)$ scalar multiplications for $\alpha_{n,p}$.
\end{enumerate}
The dominant cost is $O(CN^2)$.

\subsubsection{Noise Update}
Unlike the object and signal-power messages, the noise update does not fit the rank-one ratio framework of~\eqref{B:loglik_ratio}.
Instead, we use the eigendecomposition of the Gram matrix $\mathbf G = \mathbf U^{\mathrm{H}}\mathbf U$ to express the log-likelihood entirely in terms of its $N$ eigenvalues. In this way, all $C$-dimensional matrix operations are reduced to scalar operations.

The covariance we need for each noise particle is
\begin{align}
	\M{C}_{p} = \eta_{p}{\mathbf {I}_C} + \M{U} \M{U}^{\mathrm{H}}.
\end{align}
Consider the $N\times N$ Gram matrix
\begin{equation}
	\M{G}=
	{\M{U}}^{\mathrm H} {\M{U}} \in \mathbb{C}^{N\times N}.
	\label{B:G_def}
\end{equation}
Since $\M{G}$ is Hermitian positive semidefinite, we have the eigen-decomposition
\begin{equation}
	\M{G}
	= \M{V}\,\boldsymbol{\Lambda}\,\M{V}^{\mathrm H},
	\qquad
	\M{V}\in\mathbb{C}^{N\times N}\ \text{unitary},\quad
	\boldsymbol{\Lambda} = \operatorname{diag}(\lambda_1,\dots,\lambda_N),\ \lambda_i \ge 0.
\end{equation}
Let $r=\operatorname{rank}(\M{U})=\operatorname{rank}(\M{G})\le N$.
Without loss of generality, assume $\lambda_1,\dots,\lambda_r>0$ and $\lambda_{r+1}=\cdots=\lambda_N=0$.
Partition
\[
\M{V} = [\M{V}_r\ \M{V}_0],\qquad
\boldsymbol{\Lambda}=
\begin{bmatrix}
	\boldsymbol{\Lambda}_r & \M{0}\\
	\M{0} & \M{0}
\end{bmatrix},
\]
where $\M{V}_r\in\mathbb{C}^{N\times r}$, $\boldsymbol{\Lambda}_r=\operatorname{diag}(\lambda_1,\dots,\lambda_r)\in\mathbb{R}^{r\times r}$.

Define
\begin{equation}
	\M{W} =
	\M{U}\,\M{V}_r\,\boldsymbol{\Lambda}_r^{-1/2}
	\in\mathbb{C}^{C\times r},
\end{equation}
where $\boldsymbol{\Lambda}_r^{-1/2} = \operatorname{diag}(\lambda_1^{-1/2},\dots,\lambda_r^{-1/2})$.
It is easy to verify that $\M{W}^{\mathrm H}\M{W} = \M{I}_r$.
Therefore, we write
\begin{align}
	\M{S} = \M{U}\M{U}^{\mathrm H} = \M{W}\,\boldsymbol{\Lambda}_r\,\M{W}^{\mathrm H}.
\end{align}
Hence $\M{S}$ has at most $r$ nonzero eigenvalues $\lambda_1,\dots,\lambda_r$ with corresponding eigenvectors given by the columns of $\M{W}$. Using the eigenvalues (recalling that $\M{C}_{p} = \eta_{p}{\mathbf {I}_C} + \M{S}$), we can easily obtain the determinant of $\M{C}_p$
\begin{equation}
	\det\bigl(\M{C}_p\bigr)
	= \left(\prod_{i=1}^{r} (\lambda_i + \eta_p)\right)\eta_p^{C-r},
\end{equation}
hence
\begin{equation}
	\log\det\bigl(\M{C}_p\bigr)
	= \sum_{i=1}^{r}\log(\lambda_i + \eta_p) + (C-r)\log\eta_p.
\end{equation}
We use the Woodbury lemma to compute the inverse of $\M{C}_p$,
\begin{align}
	\M{C}^{-1}_{p}
	&=   \frac{1}{\eta_p}\M{I}_{C} - \frac{1}{\eta_p}\M{W}\operatorname{diag}\!\left(\frac{\lambda_i}{\lambda_i + \eta_p}\right)\M{W}^{\mathrm H}.
\end{align}
Define
\begin{align*}
	\boldsymbol{\kappa} =&~ \M{W}^{\mathrm H} \M{z} \in \mathbb{C}^r \\
	=&~ \boldsymbol{\Lambda}_r^{-1/2} \M{V}_r^{\mathrm H}\M{U}^{\mathrm H}\V{z}.
\end{align*}

Then
\begin{align}
	\mathbf{z}^{\mathrm H} \mathbf{C}_p^{-1} \mathbf{z}
	&= \frac{1}{\eta_p}\|\mathbf{z}\|^2 - \frac{1}{\eta_p}\sum_{i=1}^{r} \frac{|\kappa_{i}|^2\lambda_i}{\lambda_i + \eta_p},
\end{align}
where $\kappa_{i}$ is the $i$-th entry of $\boldsymbol{\kappa}$.

\paragraph{Complexity Analysis}
The main steps are:
\begin{enumerate}
	\item Eigen-decomposition of $\mathbf{G}$: $O(N^3)$ (done once).
	\item Compute $\boldsymbol{\kappa}$ via $\mathbf{v} = \mathbf{U}^{\mathrm H}\mathbf{z}$, and then $\mathbf{s}=\mathbf{V}_r^H\mathbf{v}$, and scaling by $\boldsymbol{\Lambda}_r^{-1/2}$:
	\begin{align*}
		O(CN) + O(Nr) + O(r) \approx O(CN).
	\end{align*}
	\item For each noise particle, evaluate $\log\det(\mathbf{C}_p)$ and $\mathbf{z}^{\mathrm H}\mathbf{C}_p^{-1}\mathbf{z}$:
	\begin{align*}
		O(r)\ \text{per particle},\quad O(Pr)\ \text{total}.
	\end{align*}
\end{enumerate}
Thus the total cost is $O(CN+PN)$.

\subsubsection{Total Complexity}
Combining all three messages, the total computational complexity per measurement vector is $O(CPN^2)$.
The dominant storage cost comes from the collection $\{{\mathbf h}_{n,p} \in \mathbb{C}^C\}_{n,p}$, giving $O(CPN)$.

\subsection{Kronecker-Structured Array Responses}
\label{B:kronecker}
In many applications the array manifold $\mathbf{a} \in \mathbb{C}^C$ admits a Kronecker product decomposition,
\[
\mathbf{a}=\mathbf{a}^{(1)} \otimes \mathbf{a}^{(2)} \otimes \cdots,
\]
where each factor $\mathbf{a}^{(i)} \in \mathbb{C}^{C_i}$ ($C = \prod_i C_i$) captures information in a distinct domain. For example, in radar ADC data cubes, the fast-time (range), slow-time (Doppler), and antenna (angle) responses are often approximately separable, although certain waveforms can introduce coupling across dimensions.

The low-rank formulation of the preceding sections already achieves $O(CPN^2)$ computational complexity, but it still requires storing and manipulating full $C$-dimensional vectors such as $\mathbf{h}_{n,p}$ and $\mathbf{u}_n$.
When $C$ is the product of several moderate dimensions---as in radar ADC data cubes where $C = C_{\text{range}}\times C_{\text{Doppler}}\times C_{\text{antenna}}$---the Kronecker structure of the manifold lets us replace every $C$-dimensional operation with operations on the smaller factors, cutting storage from $O(CPN)$ to $O(C_{\max}NP)$ where $C_{\max} = \max_i C_i$, without increasing the computational order.

This section is organized as follows.
(1) After introducing the Kronecker notation and two key identities, we describe the shared pre-computation that replaces all $C$-dimensional matrix operations with factor-level operations (Section~\ref{B:kron_precomp}).
(2) We then derive how each of the three message types---object update (Section~\ref{B:kron_object}), signal-power update (Section~\ref{B:kron_intensity}), and noise update (Section~\ref{B:kron_noise})---is computed entirely in terms of the Kronecker factors, without forming any $C$-dimensional vector explicitly.
Table~\ref{B:tab_kron} summarizes the key quantities introduced in this section.

\begin{table}[t]
	\centering
	\small
	\caption{Key quantities for the Kronecker-structured formulation (Section~\ref{B:kronecker}).}
	\label{B:tab_kron}
	\begin{tabular}{@{}lll@{}}
		\toprule
		Symbol & Description & Dimension  \\
		\midrule
		$\mathbf{a}_{n,p}^{(i)}$ & Kronecker factor of $\mathbf{a}_{n,p}$ & $C_i \times 1$  \\
		$\mathbf{u}_n^{(i)}$ & Kronecker factor of $\mathbf{u}_n$ & $C_i \times 1$  \\
		$\mathbf{U}^{(i)}$ & $[\mathbf{u}_1^{(i)},\dots,\mathbf{u}_N^{(i)}]$ & $C_i \times N$  \\
		$\mathbf{G}$ & Hadamard of factor Gram matrices & $N \times N$  \\
		$\mathbf{v}$ & Measurement projection $\mathbf{U}^{\mathrm{H}}\mathbf{z}$ & $N \times 1$\\
		$\tilde{\mathbf v}$ & Measurement projection $\mathbf{L}\mathbf{v}$ & $N \times 1$  \\
		$\mathbf{y}_{n,p}$ & Cross-correlation $\mathbf{U}^{\mathrm{H}}\mathbf{h}_{n,p}$ & $N \times 1$  \\
		\bottomrule
	\end{tabular}
\end{table}

For simplicity, we consider the two-factor case. For three or more factors, e.g., the fast-time, slow-time, and antenna factors of radar data, the derivation applies analogously: the Gram matrix becomes the Hadamard product of all factor Gram matrices, the inner products with the measurement tensor are computed by successive mode products, and the storage scales with $\sum_i C_i$ instead of $C_{\max}$. The measurement vector ${\mathbf z}$ is the column-wise vectorization of a matrix ${\mathbf Z} \in \mathbb{C}^{C_1 \times C_2}$, i.e., ${\mathbf z} = \operatorname{vec}({\mathbf Z}) \in \mathbb{C}^{C}$ with $C = C_1 C_2$. Each manifold decomposes as
\begin{equation}
	{\mathbf a}_{n,p} = {\mathbf a}_{n,p}^{(1)} \otimes{\mathbf a}_{n,p}^{(2)},\qquad
	{\mathbf a}_{n,p}^{(1)} \in \mathbb{C}^{C_2}, \quad
	{\mathbf a}_{n,p}^{(2)} \in \mathbb{C}^{C_1},
\end{equation}
where ${\mathbf a}_{n,p}^{(i)}$ is the $i$-th factor evaluated at $\V{x}_{n,p}$.
Note that, in the particle implementation, the weighted mean $\bar{\mathbf a}_n$ is in general not a Kronecker product, it is approximated as $\bar{\mathbf a}_n \approx \bar{\mathbf a}^{(1)}_n\otimes\bar{\mathbf a}^{(2)}_n$ with $\bar{\mathbf a}^{(i)}_n=\sum_p \omega_{\V{x},n,p}\,{\mathbf a}^{(i)}_{n,p}$. This approximation is exact if all particles of object $n$ share the same factor in all but one domain and it is accurate for concentrated particle clouds.
Correspondingly, ${\mathbf u}_n \approx {\mathbf u}_n^{(1)} \otimes {\mathbf u}_n^{(2)}$, where
\begin{align*}
	{\mathbf u}_{n}^{(1)} =&~ {\bar \gamma_n}^{\frac{1}{4}} \cdot \bar {\mathbf a}^{(1)}_{n},\\
	{\mathbf u}_{n}^{(2)} =&~ {\bar \gamma_n}^{\frac{1}{4}} \cdot \bar {\mathbf a}^{(2)}_{n}.
\end{align*}

\textbf{Kronecker identities:}
All subsequent derivations rely on two standard properties of the Kronecker product:
\begin{enumerate}
	\item For any $\mathbf a = \mathbf a^{(1)}\otimes\mathbf a^{(2)}$,
	\begin{equation}
		\|\mathbf a\|^2 = \|\mathbf a^{(1)}\|^2\,\|\mathbf a^{(2)}\|^2.
		\label{B:kron_norm}
	\end{equation}
	\item For $\mathbf a^{(1)}\in\mathbb{C}^{C_2}$, $\mathbf a^{(2)}\in\mathbb{C}^{C_1}$, and $\mathbf Z\in\mathbb{C}^{C_1\times C_2}$,
	\begin{equation}
		\operatorname{vec}(\mathbf Z)^{\mathrm{H}} (\mathbf a^{(1)}\otimes\mathbf a^{(2)})
		= (\mathbf a^{(1)})^{\mathrm{T}}\,\mathbf Z^{\mathrm{H}}\,\mathbf a^{(2)}.
		\label{B:vec_kron}
	\end{equation}
\end{enumerate}

\subsubsection{Pre-computation for Shared Terms}
\label{B:kron_precomp}
We first express the shared building blocks from Sections~\ref{B:lowrank}--\ref{B:messages} in terms of the Kronecker factors. The quantities below are computed once and reused by all three message types.

\textbf{Gram matrix $\mathbf{G}$:}
Each entry of ${\mathbf G} = {\mathbf U}^{\mathrm{H}}{\mathbf U} \in \mathbb{C}^{N \times N}$ factorizes by the mixed-product property of the Kronecker product:
\begin{align}
	[{\mathbf G}]_{mn}
	&= ({\mathbf u}_m^{(1)} \otimes {\mathbf u}_m^{(2)})^{\mathrm{H}}({\mathbf u}_n^{(1)} \otimes {\mathbf u}_n^{(2)})
	= ({\mathbf u}_m^{(1)\mathrm{H}} {\mathbf u}_n^{(1)})\,({\mathbf u}_m^{(2)\mathrm{H}}{\mathbf u}_n^{(2)}),
\end{align}
which can be written compactly as the Hadamard (element-wise) product of two small Gram matrices:
\begin{equation}
	{\mathbf G}  = \bigl( {\mathbf U}^{(1)\mathrm{H}} {\mathbf U}^{(1)} \bigr) \odot \bigl( {\mathbf U}^{(2)\mathrm{H}} {\mathbf U}^{(2)} \bigr).
	\label{B:gram_hadamard}
\end{equation}
Dimension: $N \times N$. Complexity: $O((C_1+C_2)N^2)$.

\textbf{Inverse kernel $\mathbf{L}$:}
We compute the inverse of the regularized Gram matrix once:
\begin{equation}
	{\mathbf L} = (\bar \eta {\mathbf I}_N + {\mathbf G})^{-1}.
\end{equation}
Dimension: $N \times N$. Complexity: $O(N^3)$.

\textbf{Measurement projection $\mathbf{v}$ and $\tilde{\mathbf v}$:}
Applying~\eqref{B:vec_kron}, each entry of ${\mathbf v} = {\mathbf U}^{\mathrm{H}}{\mathbf z} \in \mathbb{C}^{N}$ can be written as
\begin{equation}
	v_n = ({\mathbf u}_n^{(1)\mathrm{H}} \otimes {\mathbf u}_n^{(2)\mathrm{H}})\,\operatorname{vec}(\mathbf Z)
	= ({\mathbf u}_n^{(2)})^{\mathrm{H}}\,\mathbf Z\,({\mathbf u}_n^{(1)})^{*},
\end{equation}
Dimension: $N \times 1$. Complexity: $O(CN)$.
The filtered projection is then
\begin{equation}
	{\tilde{\mathbf v}} = {\mathbf L} {\mathbf v} \in \mathbb{C}^{N}.
\end{equation}
Dimension: $N \times 1$. Complexity: $O(N^2)$.

\textbf{Per-object scalar ${\mathbf u}_{n}^{\mathrm{H}} {\mathbf B}^{-1} {\mathbf u}_{n}$:}
This scalar appears in the Sherman--Morrison expansion for both the object and signal-power updates.
Since ${\mathbf u}_n$ is the $n$-th column of ${\mathbf U}$, the Woodbury lemma yields
\begin{equation}
	{\mathbf u}_{n}^{\mathrm{H}} {\mathbf B}^{-1} {\mathbf u}_{n} = \frac{1}{\bar \eta} \Bigl( [{\mathbf G}]_{nn} - [{\mathbf G} {\mathbf L} {\mathbf G}]_{nn} \Bigr).
\end{equation}
The matrix ${\mathbf G} {\mathbf L} {\mathbf G}$ is pre-computed once in $O(N^3)$, after which extracting diagonal entries is $O(1)$ per object.

\textbf{Per-object scalar ${\mathbf z}^{\mathrm{H}} {\mathbf B}^{-1} {\mathbf u}_{n}$:}
Similarly, using ${\mathbf v} = {\mathbf U}^{\mathrm{H}}{\mathbf z}$ and ${\tilde{\mathbf v}} = {\mathbf L}{\mathbf v}$,
\begin{equation}
	{\mathbf z}^{\mathrm{H}} {\mathbf B}^{-1} {\mathbf u}_{n} = \frac{1}{\bar \eta} \Bigl( v_n - [{\mathbf G} {\tilde{\mathbf v}}]_n \Bigr)^*.
\end{equation}
The product ${\mathbf G}{\tilde{\mathbf v}}$ is computed once in $O(N^2)$, making the cost $O(N^2)$ for all $N$ objects combined.

\subsubsection{Object Update with Kronecker Factors}
\label{B:kron_object}
Recall from Section~\ref{B:messages} that the object message for particle~$p$ of object~$n$ requires the two scalars $s_{n,p}$ and $q_{n,p}$~\eqref{B:sq_def}, which---via the Sherman--Morrison expansion~\eqref{B:sherman_morrison}---depend on ${\mathbf h}_{n,p}^{\mathrm{H}} {\mathbf B}^{-1} {\mathbf h}_{n,p}$, ${\mathbf z}^{\mathrm{H}} {\mathbf B}^{-1} {\mathbf h}_{n,p}$, and ${\mathbf u}_{n}^{\mathrm{H}} {\mathbf B}^{-1} {\mathbf h}_{n,p}$.
The per-object scalars ${\mathbf u}_{n}^{\mathrm{H}} {\mathbf B}^{-1} {\mathbf u}_{n}$ and ${\mathbf z}^{\mathrm{H}} {\mathbf B}^{-1} {\mathbf u}_{n}$ are already available from Section~\ref{B:kron_precomp}.
We now derive the remaining particle-specific terms without forming any $C$-dimensional vector.

All three terms share the cross-correlation vector
\begin{equation}
	{\mathbf y}_{n,p} = {\mathbf U}^{\mathrm{H}} {\mathbf h}_{n,p} = \bigl( {\mathbf U}^{(1)\mathrm{H}} {\mathbf h}_{n,p}^{(1)} \bigr) \odot \bigl( {\mathbf U}^{(2)\mathrm{H}} {\mathbf h}_{n,p}^{(2)} \bigr),
\end{equation}
by the same Hadamard factorization as~\eqref{B:gram_hadamard}, where ${\mathbf h}_{n,p}^{(1)} = {\bar \gamma_n}^{{1}/{4}} {\mathbf a}^{(1)}_{n,p}$.
Complexity: $O((C_1+C_2)N)$ per particle; $O(C_{\max}N^2P)$ over all objects and particles.

\textbf{Term ${\mathbf h}_{n,p}^{\mathrm{H}} {\mathbf B}^{-1} {\mathbf h}_{n,p}$:}
Expanding ${\mathbf B}^{-1}$ via the Woodbury lemma~\eqref{B:woodbury_eq} and using the norm factorization~\eqref{B:kron_norm} for the first term gives
\begin{equation}
	{\mathbf h}_{n,p}^{\mathrm{H}} {\mathbf B}^{-1} {\mathbf h}_{n,p} = \frac{1}{\bar \eta} \Bigl( \|{\mathbf h}_{n,p}^{(1)}\|^2\, \|{\mathbf h}_{n,p}^{(2)}\|^2 - {\mathbf y}_{n,p}^{\mathrm{H}} {\mathbf L}\, {\mathbf y}_{n,p} \Bigr).
\end{equation}
The first term equals $\|{\mathbf h}_{n,p}\|^2$ but is computed from the $C_1$- and $C_2$-dimensional factors, avoiding any $C$-dimensional operation.
Complexity: $O(N^2)$ per particle; $O(N^3P)$ over all objects and particles.

\textbf{Term ${\mathbf z}^{\mathrm{H}} {\mathbf B}^{-1} {\mathbf h}_{n,p}$:}
Expanding ${\mathbf B}^{-1}$ and applying the vec-Kronecker identity~\eqref{B:vec_kron} to the first inner product,
\begin{align}
	{\mathbf z}^{\mathrm{H}} {\mathbf B}^{-1} {\mathbf h}_{n,p}
	&= \frac{1}{\bar \eta} \Bigl( \operatorname{vec}({\mathbf Z})^{\mathrm{H}}({\mathbf h}_{n,p}^{(1)} \otimes{\mathbf h}_{n,p}^{(2)} ) - {\mathbf v}^{\mathrm{H}} {\mathbf L}\, {\mathbf y}_{n,p} \Bigr)\notag\\
	&=\frac{1}{\bar \eta} \Bigl( ({\mathbf h}_{n,p}^{(1)})^{\mathrm{T}}\, {\mathbf Z}^{\mathrm{H}}\,{\mathbf h}_{n,p}^{(2)} - {\mathbf v}^{\mathrm{H}} {\mathbf L}\, {\mathbf y}_{n,p} \Bigr).
\end{align}
The matrix product ${\mathbf Z}^{\mathrm{H}}\,{\mathbf h}_{n,p}^{(2)}\in\mathbb{C}^{C_2}$ costs $O(C_1 C_2)=O(C)$, while the second term reuses ${\mathbf L}$ and ${\mathbf y}_{n,p}$.
Complexity: $O(C)+O(N^2)$ per particle; $O(CNP)$ over all objects and particles.

\textbf{Term ${\mathbf u}_{n}^{\mathrm{H}} {\mathbf B}^{-1} {\mathbf h}_{n,p}$:}
Noting that ${\mathbf U}^{\mathrm{H}}{\mathbf h}_{n,p} = {\mathbf y}_{n,p}$,
\begin{equation}
	{\mathbf u}_{n}^{\mathrm{H}} {\mathbf B}^{-1} {\mathbf h}_{n,p} = \frac{1}{\bar \eta} \Bigl( [{\mathbf y}_{n,p}]_n - [{\mathbf G} {\mathbf L}\, {\mathbf y}_{n,p}]_n \Bigr).
\end{equation}
Complexity: $O(N^2)$ per particle; $O(N^3P)$ over all objects and particles.

\textbf{Term $s_{n,p}$ and $q_{n,p}$:}
Substituting the terms above into the Sherman--Morrison expansion from Section~\ref{B:messages}, the two scalars that determine the object log-likelihood message~\eqref{B:log_ell_x} are
\begin{align}
	s_{n,p}
	&= {\mathbf h}_{n,p}^{\mathrm{H}}{\mathbf B}^{-1}{\mathbf h}_{n,p}
	+ \frac{ \bigl| {\mathbf u}_{n}^{\mathrm{H}} {\mathbf B}^{-1}{\mathbf h}_{n,p} \bigr|^2  }{ 1 - {\mathbf u}_{n}^{\mathrm{H}} {\mathbf B}^{-1} {\mathbf u}_{n}},\\[1.5mm]
	q_{n,p} &= {\mathbf z}^{\mathrm{H}}{\mathbf B}^{-1}{\mathbf h}_{n,p}
	+ \frac{
		{\mathbf z}^{\mathrm{H}}{\mathbf B}^{-1}{\mathbf u}_{n} \cdot {\mathbf u}^{\mathrm{H}}_n {\mathbf B}^{-1}{\mathbf h}_{n,p}
	}{
		1 - {\mathbf u}_{n}^{\mathrm{H}} {\mathbf B}^{-1} {\mathbf u}_{n}
	},
\end{align}
where every building block on the right-hand side has been expressed in terms of the Kronecker factors.
The dominant cost for the object update is reduced to $O(CNP)$.

\subsubsection{Signal-Power Update with Kronecker Factors}
\label{B:kron_intensity}
Recall from Section~\ref{B:messages} that the signal-power message uses ${\mathbf h}_{n,p} = \alpha_{n,p}\, {\mathbf u}_{n}$, where $\alpha_{n,p} = \sqrt{\gamma_{n,p}/\bar\gamma_n}$.
Because ${\mathbf h}_{n,p}$ is proportional to ${\mathbf u}_{n}$, the scalars $s_{n,p}$ and $q_{n,p}$ reduce to scaled versions of the per-object quantities already computed in Section~\ref{B:kron_precomp}:
\begin{align}
	\frac{s_{n,p}}{\alpha_{n,p}^2} &= {\mathbf u}^{\mathrm{H}}_{n}{\mathbf B}^{-1}{\mathbf u}_{n}
	+ \frac{ \bigl| {\mathbf u}_{n}^{\mathrm{H}} {\mathbf B}^{-1}{\mathbf u}_{n} \bigr|^2  }{ 1 - {\mathbf u}_{n}^{\mathrm{H}} {\mathbf B}^{-1} {\mathbf u}_{n}},\\
	\frac{q_{n,p}}{\alpha_{n,p}} &= {\mathbf z}^{\mathrm{H}}{\mathbf B}^{-1}{\mathbf u}_{n}
	+ \frac{
		{\mathbf z}^{\mathrm{H}}{\mathbf B}^{-1}{\mathbf u}_{n} \cdot {\mathbf u}^{\mathrm{H}}_n {\mathbf B}^{-1}{\mathbf u}_{n}
	}{
		1 - {\mathbf u}_{n}^{\mathrm{H}} {\mathbf B}^{-1} {\mathbf u}_{n}
	}.
\end{align}
Both right-hand sides depend only on ${\mathbf u}_{n}^{\mathrm{H}} {\mathbf B}^{-1} {\mathbf u}_{n}$ and ${\mathbf z}^{\mathrm{H}} {\mathbf B}^{-1} {\mathbf u}_{n}$, which are the per-object scalars from Section~\ref{B:kron_precomp}.
No additional $C$- or $C_{\max}$-dimensional operations are needed; the remaining work is $O(PN)$ scalar multiplications.

\subsubsection{Noise Update with Kronecker Factors}
\label{B:kron_noise}
The noise update messages involve the eigendecomposition of the Gram matrix $\mathbf G$ and do not depend on the Kronecker structure. The complexity therefore remains $O(CN+PN)$, identical to the non-Kronecker formulation.

\subsubsection{Total Complexity}
Collecting all terms, the total complexity is
\begin{itemize}
	\item Shared pre-computation (Section~\ref{B:kron_precomp}):
	\begin{align*}
		\underbrace{O(C_{\max}N^2)}_{\mathbf G} + \underbrace{O(N^3)}_{\mathbf L} + \underbrace{O(CN)}_{\mathbf v} + \underbrace{O(N^2)}_{\tilde{\mathbf v}} + \underbrace{O(N^3)}_{\text{per-object scalars}} \approx O(CN).
	\end{align*}
	\item Object update (Section~\ref{B:kron_object}):
	\begin{align*}
		\underbrace{O(C_{\max}N^2P)}_{\mathbf y} + \underbrace{O(N^3P)}_{{\mathbf h}^{\mathrm{H}}{\mathbf B}^{-1}{\mathbf h}} + \underbrace{O(CNP)}_{{\mathbf z}^{\mathrm{H}}{\mathbf B}^{-1}{\mathbf h}} + \underbrace{O(N^3P)}_{{\mathbf u}^{\mathrm{H}}{\mathbf B}^{-1}{\mathbf h}} \approx O(CNP).
	\end{align*}
	\item Signal-power update (Section~\ref{B:kron_intensity}): $O(PN)$ scalar operations.
	\item Noise update (Section~\ref{B:kron_noise}): $O(CN+PN)$.
\end{itemize}
Therefore, the total computational complexity is $O(CPN)$. The key advantage is in storage: only the Kronecker factors ${\mathbf h}_{n,p}^{(1)}\in\mathbb{C}^{C_2}$ and ${\mathbf h}_{n,p}^{(2)}\in\mathbb{C}^{C_1}$ need to be stored, reducing memory from $O(CNP)$ to $O(C_{\max}NP)$ where $C_{\max} = \max(C_1,C_2)$.

\subsection{Example for Complexity Analysis}
\label{B:example}
\begin{figure*}[htbp]
	\centering
	\includegraphics[scale=0.32]{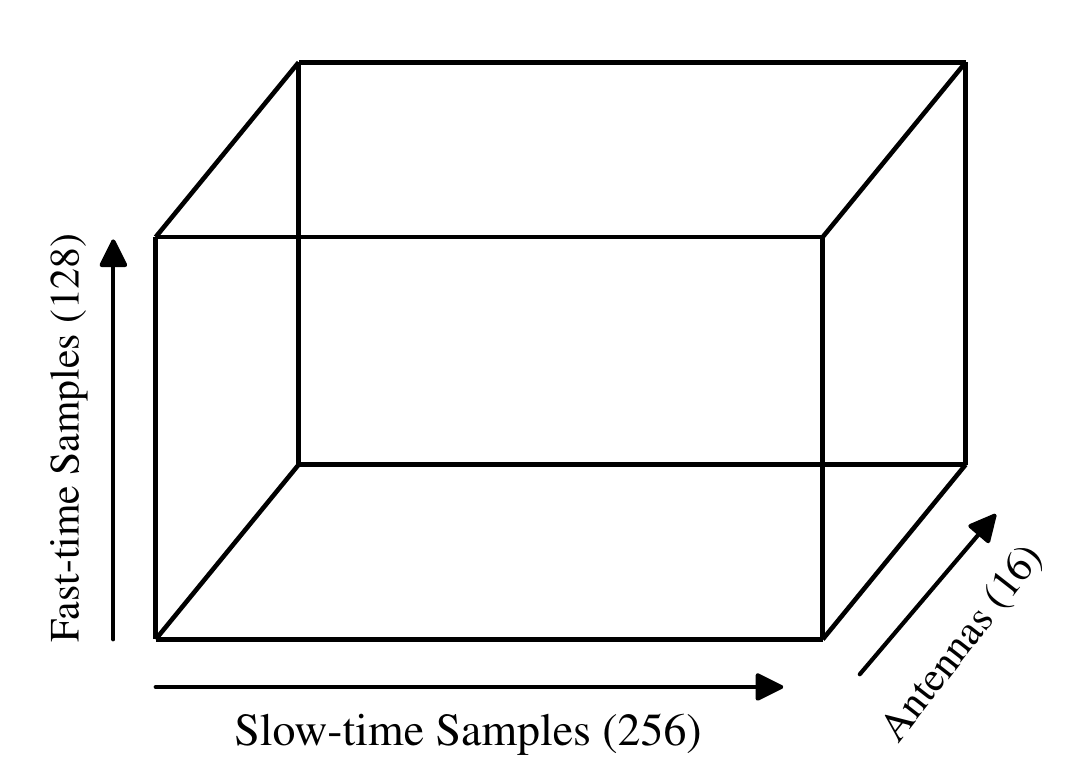}
	\caption{Raw ADC data of an FMCW radar arranged as a 3-D tensor with fast-time, slow-time, and antenna dimensions}
	\label{B:radar_fig}
\end{figure*}
A typical mmWave radar sensing setup consists of 128 fast-time samples, 256 slow-time samples, and 16 antennas (i.e., $C = 128 \times 256 \times 16$). Consider a 2-D coordinate localization task, where the object state is represented by approximately $P = 10^4$ particles, which is generally sufficient for accurate estimation.
Assume there are about $N = 10$ objects in the scene.
All computations are performed using complex64/float32 precision.
At each time step, a detailed comparison of the three approaches is provided below.

\subsubsection{Storage}
\begin{itemize}
	\item \textbf{A direct computation of Gaussian PDF}: a single full covariance matrix in $\mathbb{C}^{C\times C}$ requires approximately 2~TiB; storing the covariance matrices of all objects and particles simultaneously would require over $2\times10^{5}\,\text{TiB}$.
	\item \textbf{The low-rank formulation}: the dominant storage is $O(CNP)$, which is around $5\times 10^{10} \times 8\,\text{bytes} \approx 400\,\text{GiB}$.
	\item \textbf{With Kronecker decomposition}: the dominant storage is reduced to $O((C_1+C_2+C_3)NP)$, i.e., $(128+256+16)\times N P\times 8$\,bytes $\approx 0.3\,\text{GiB}$.
\end{itemize}
For a single likelihood evaluation, i.e., one object and one particle, the full covariance matrix requires approximately 2~TiB, whereas the Kronecker factors of one array response comprise $128+256+16=400$ complex values, i.e., approximately 3.1~KiB; these values are reported in Table~\ref{M-tab:complexity_example} of the main paper.

\subsubsection{Computation Complexity}
\begin{itemize}
	\item \textbf{A direct computation of Gaussian PDF}: The complexity is $O(C^3NP)$, around $10^{22}$.
	\item \textbf{The low-rank formulation}: the complexity is $O(CN^2P)$, around $5\times10^{11}$.
	\item \textbf{With Kronecker decomposition}: the complexity is $O(CNP)$, around $5\times10^{10}$.
\end{itemize}

\FloatBarrier

\section{Additional Results for the Simulated Multisensor Radar Scenario}
\label{app:sim}
This section complements Section~\ref{M-sec:exp_sim} of the main paper.

\begin{figure}[htbp]
	\centering
	\includegraphics[width=0.4\linewidth]{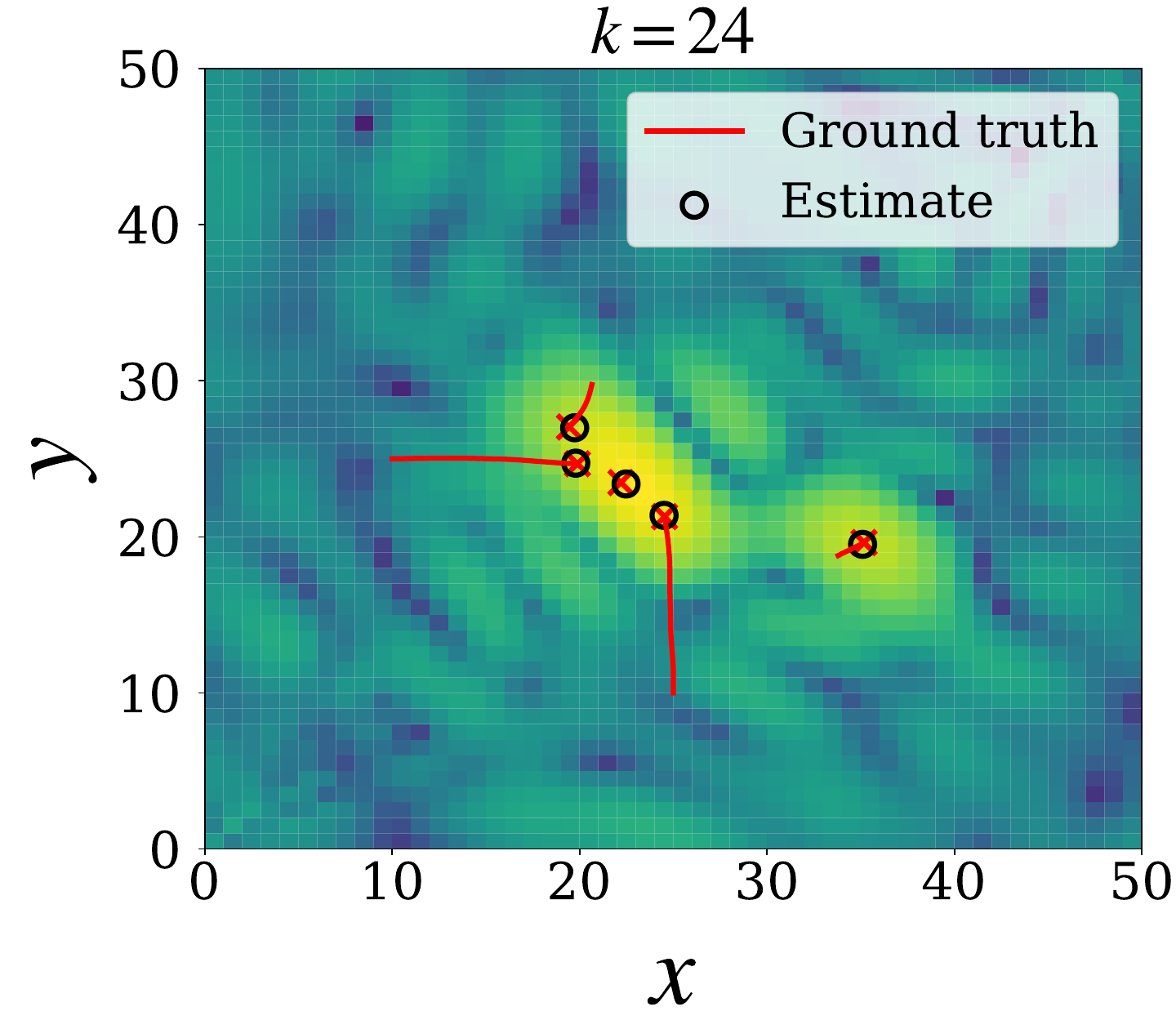}\hspace{6mm}
	\includegraphics[width=0.4\linewidth]{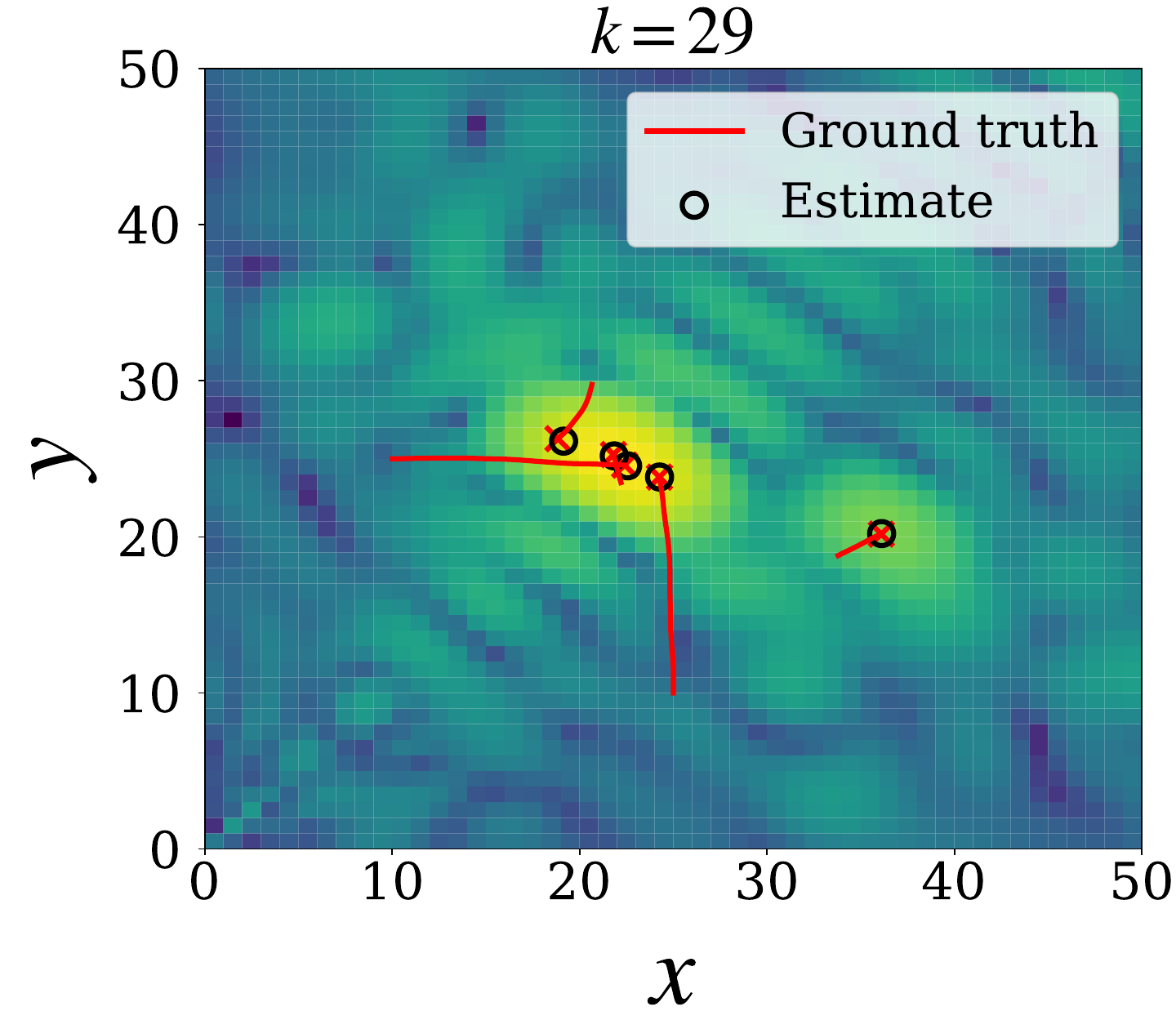}\\[3mm]
	\includegraphics[width=0.4\linewidth]{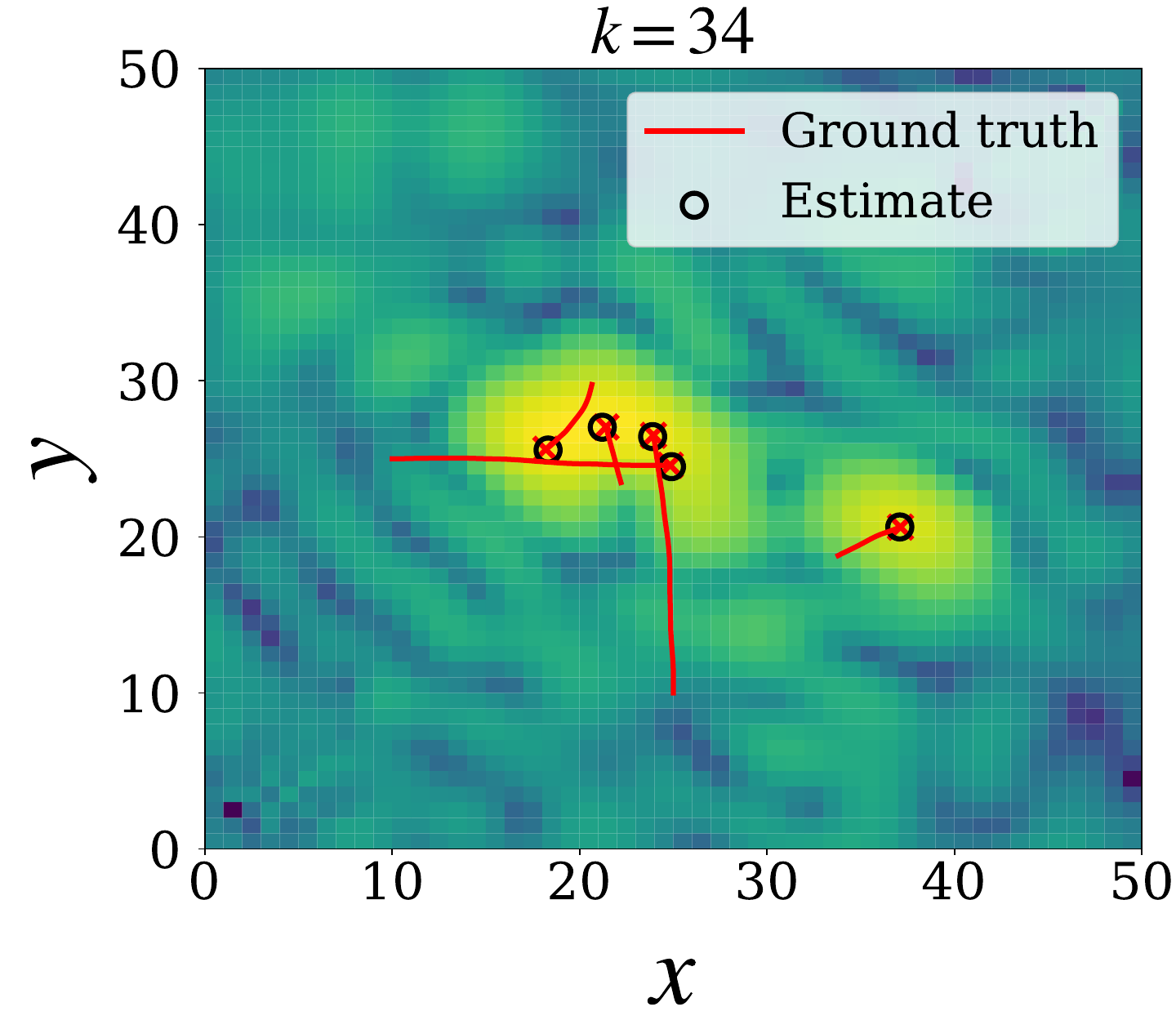}\hspace{6mm}
	\includegraphics[width=0.4\linewidth]{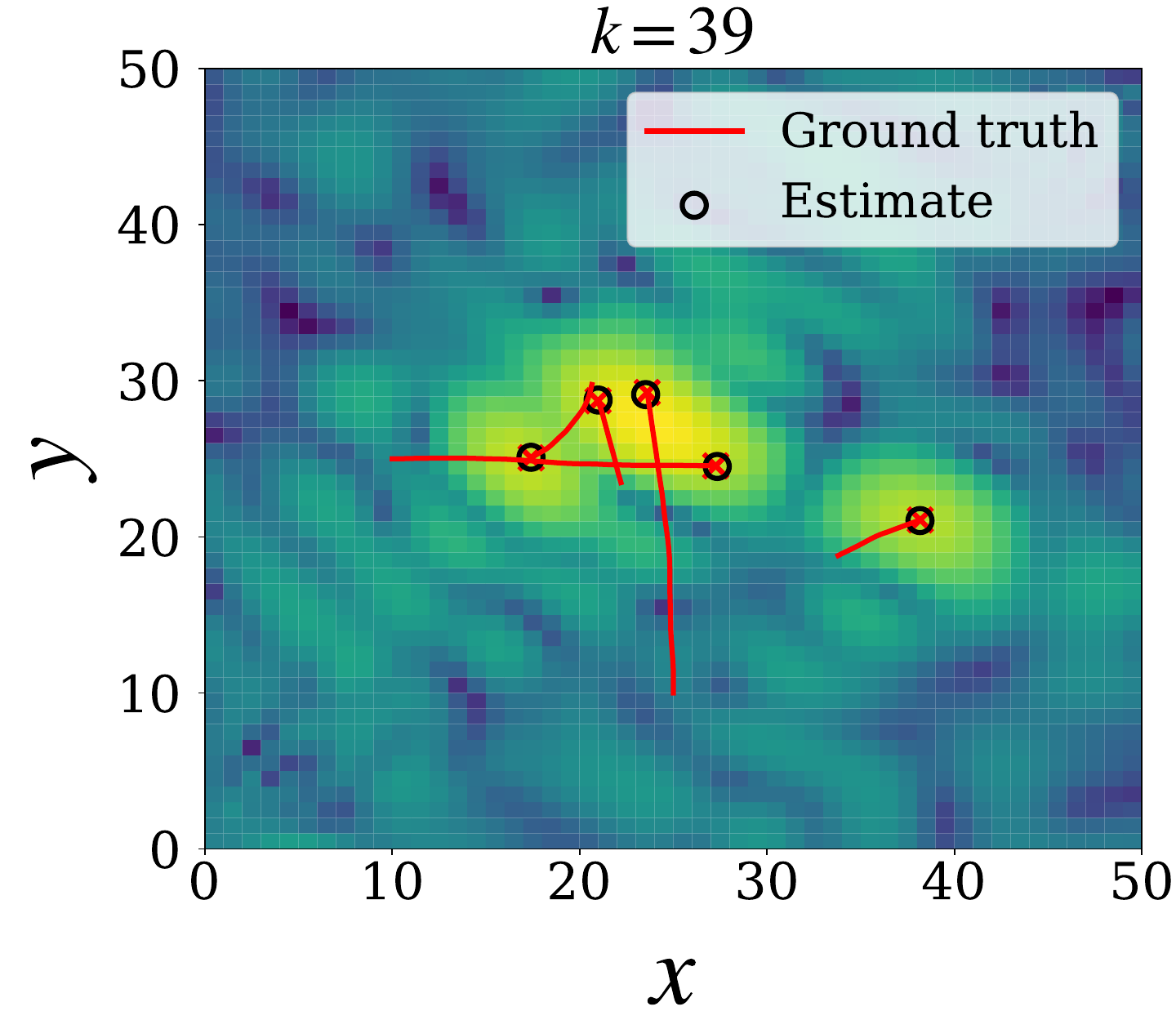}
	\caption{Tracking results obtained by the proposed method in the simulated multisensor radar scenario at time steps $k=24$, $29$, $34$, and $39$, shown from top left to bottom right. True trajectories are drawn as red curves, true current object positions as red crosses, and estimated object positions as black circles. Matched filtering yields the Bartlett spectrum depicted in the background. Positions are given in meters.}
	\label{fig:sim_tracks}
\end{figure}

Representative tracking results of the proposed method are depicted in Fig.~\ref{fig:sim_tracks}. Owing to the low sensor resolution, the responses of closely spaced objects overlap in the Bartlett spectrum obtained by matched filtering, as can be seen, e.g., at $k=24$. Nevertheless, the proposed method resolves these objects and maintains their tracks.

\FloatBarrier
\section{Additional Results for the Active Radio Sensing Experiment}
\label{app:radio}
This section complements Section~\ref{M-sec:exp_radio} of the main paper.

\subsection{Posterior PDF}
\label{app:radio_posterior}
\begin{figure}[htbp]
	\centering
	\subfloat[Observed power]{\includegraphics[width=0.4\linewidth]{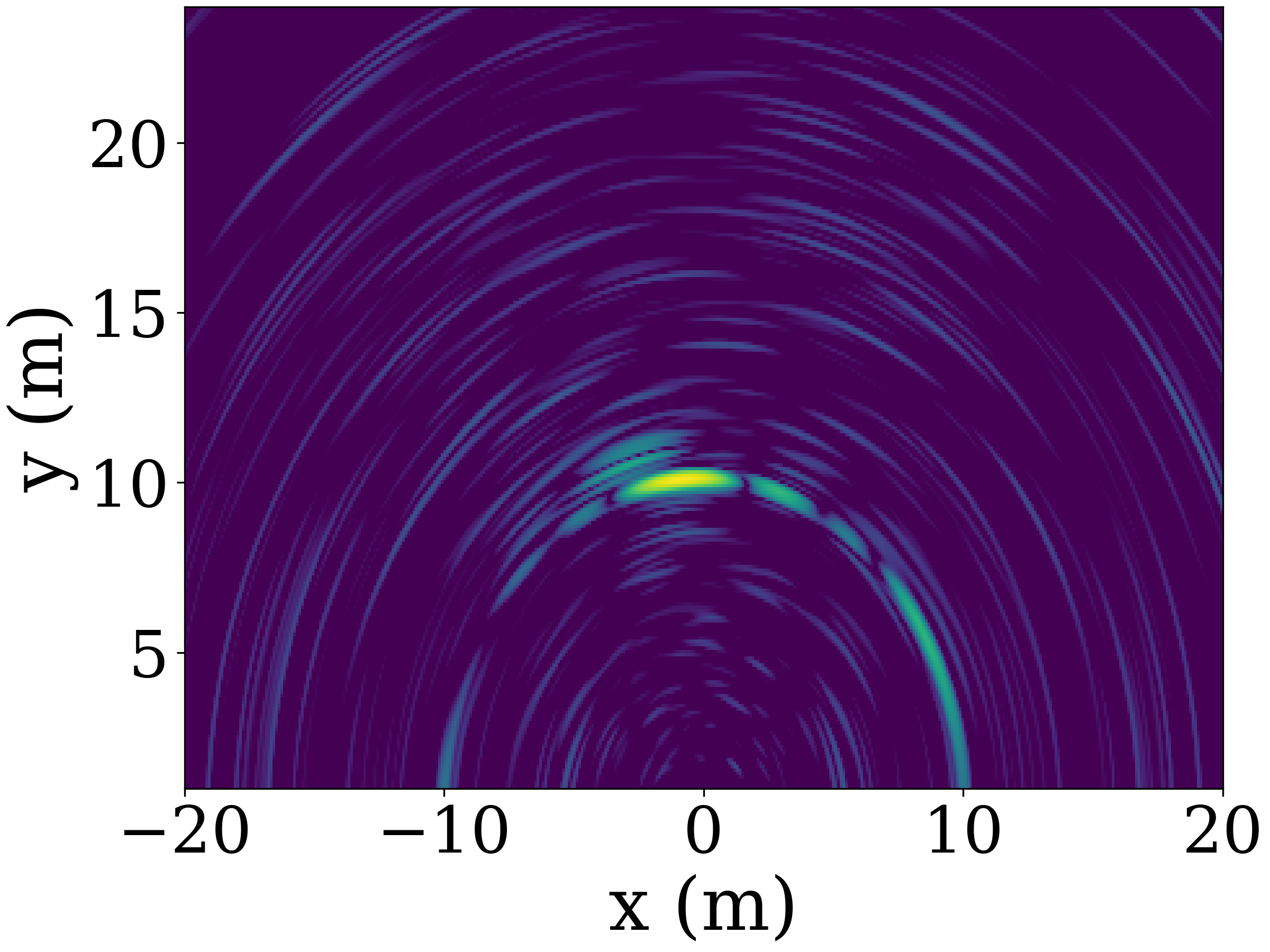}\label{fig:mf_290_true}}\hspace{6mm}
	\subfloat[Posterior PDF]{\includegraphics[width=0.4\linewidth]{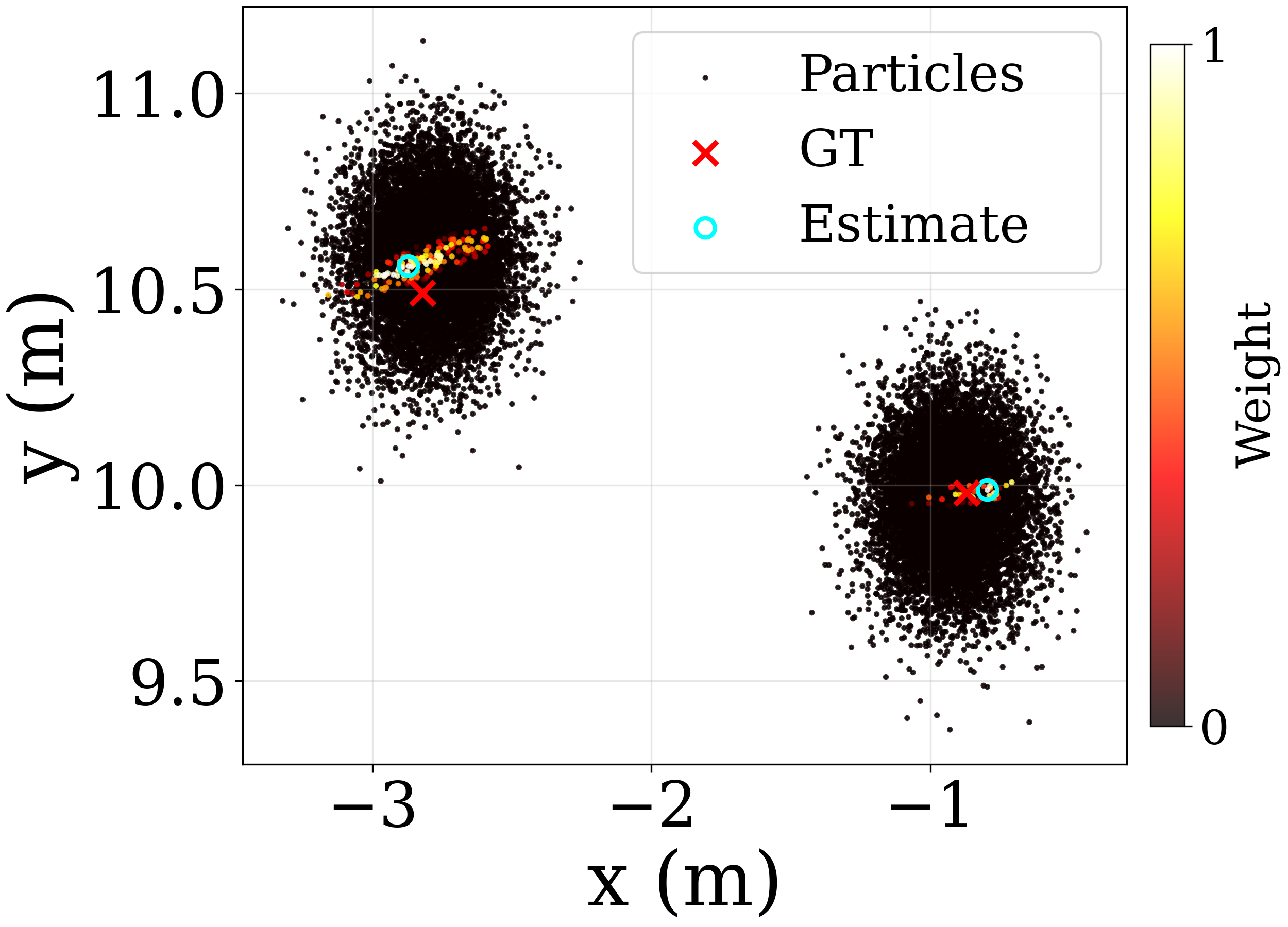}\label{fig:radar_particles}}
	\caption{(a) Observed spatial power computed from the received complex measurement in a parking-lot scene. (b) Posterior PDF of object positions at the considered time step.}
	\label{fig:radar_particles_full}
\end{figure}
Fig.~\ref{fig:radar_particles_full}(a) displays the observed spatial power computed from the received measurement via beamforming, which contains both object-related responses and background variations. Fig.~\ref{fig:radar_particles_full}(b) shows a particle representation of the posterior PDF of the object positions at the same time step. The likelihood function of the proposed signal model is highly informative and leads to sharply concentrated posterior PDFs. This concentration motivates the rank-one approximation in Section~\ref{B:rank1}.

\subsection{Evolution of the Learned Signal-Power PDF}
\label{app:radio_flow}
\begin{figure}[htbp]
	\centering
	\includegraphics[width=\linewidth]{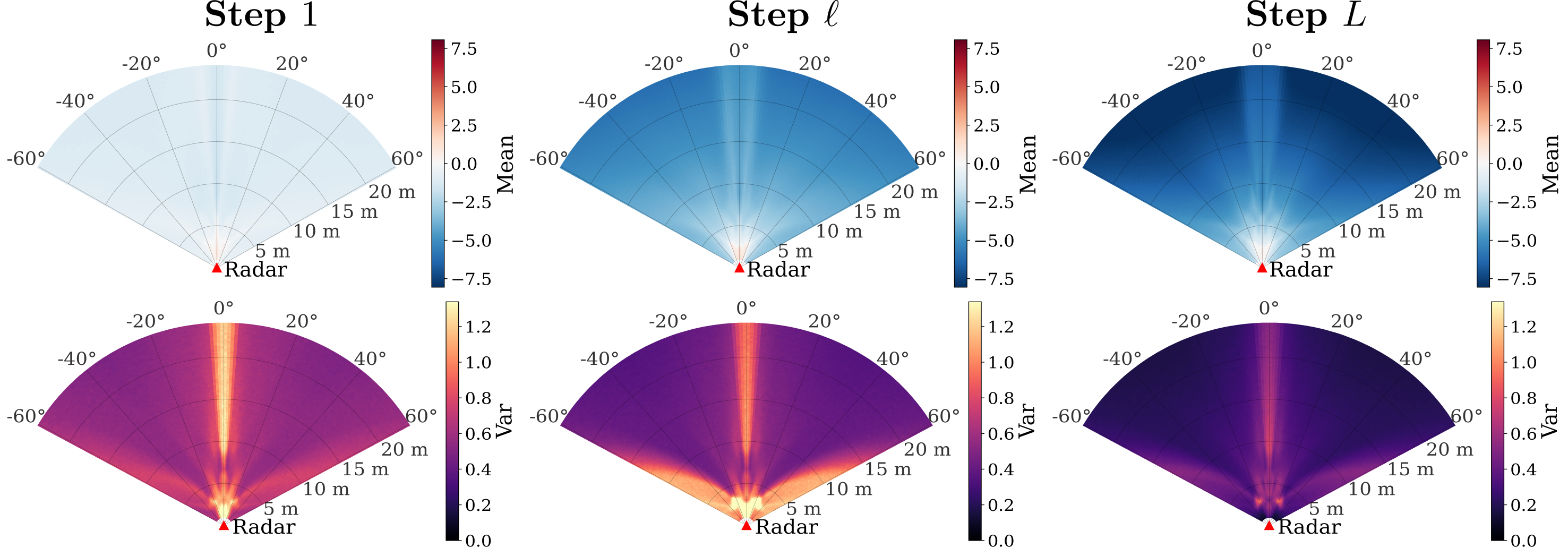}
	\caption{Transformation of the radio latent field at intermediate flow steps, with the mean shown in the top row and the variance shown in the bottom row.}
	\label{fig:flow_med}
\end{figure}
Fig.~\ref{fig:flow_med} shows how the learned signal-power PDF evolves through the normalizing flow transformations, using the positional context in~\eqref{M-eq:radar_visualization_context} of the main paper. At the early stage, the mean field is nearly uniform and the variance is high over most of the scene. As the samples pass through successive transformations, the mean develops the mainlobe and range-dependent attenuation visible in Fig.~\ref{M-fig:coverage_intensity} of the main paper, while the variance decreases in regions supported by the observations. This evolution provides a distributional view of how the spatial latent field is acquired rather than merely displaying its final mean field.

\subsection{Sensitivity to Birth Initialization}
\label{app:radio_birth}
Because the 3-D FFT is used only to initialize new potential objects, we test whether the final performance depends strongly on this coarse proposal. We replace the 3-D FFT peak used as the mean of each Gaussian birth PDF with detections from MVDR or RAMP-CNN and report one run for each strategy in Table~\ref{tab:init_ablation}.
\begin{table}[htbp]
	\centering
	\caption{GOSPA in meters for different radio birth-initialization strategies, each obtained from a single run.}
	\label{tab:init_ablation}
	\begin{tabular}{l c}
		\toprule
		Initialization strategy & GOSPA \\
		\midrule
		3-D FFT peak & 0.546 \\
		MVDR & 0.550 \\
		RAMP-CNN & 0.548 \\
		\bottomrule
	\end{tabular}
\end{table}
The three strategies yield nearly identical GOSPA values; since each value is obtained from a single run, the value for the 3-D FFT peak differs slightly from the average over Monte Carlo runs in Table~\ref{M-tab:performance_uwcr} of the main paper. The birth PDF therefore needs only to identify a plausible search region. Subsequent posterior updates based on the raw measurements dominate the initial point estimate.

\FloatBarrier
\section{Additional Results for the Passive Acoustic Sensing Experiment}
\label{app:sonar}
\begin{figure}[htbp]
	\centering
	\subfloat[Tones at 49, 79, and 112~Hz]{\includegraphics[width=0.8\linewidth]{experiment/Intensity_Map_Sonar.pdf}\label{fig:sonar_intmap_low}}\\
	\subfloat[Tones at 148, 201, and 283~Hz]{\includegraphics[width=0.8\linewidth]{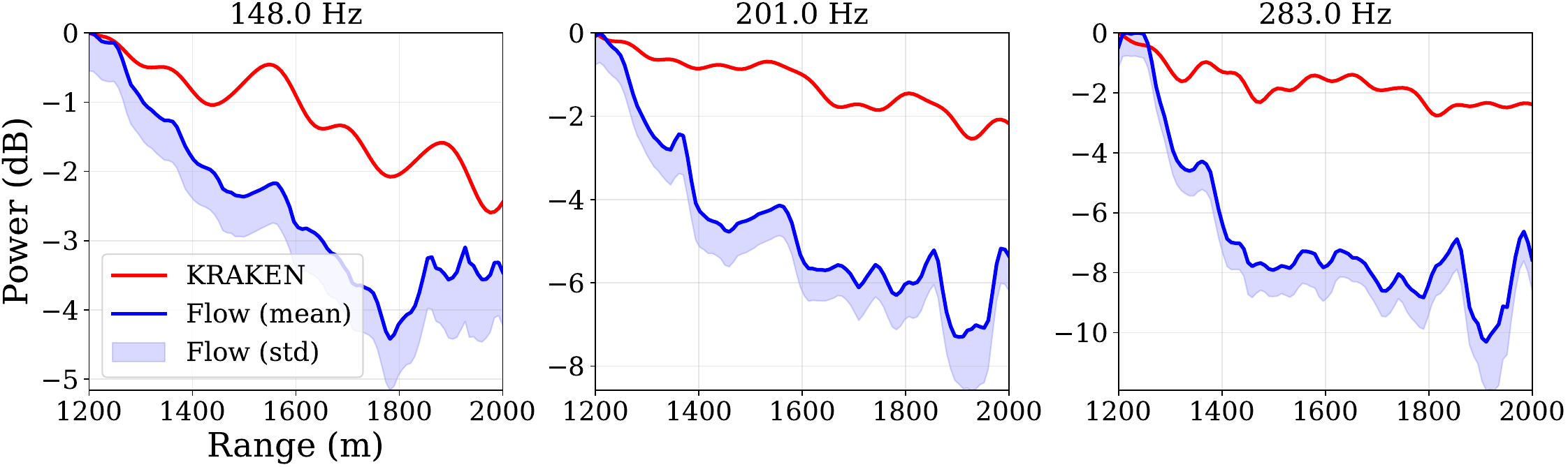}\label{fig:sonar_intmap_high}}\\
	\subfloat[Tone at 388~Hz]{\includegraphics[width=0.265\linewidth]{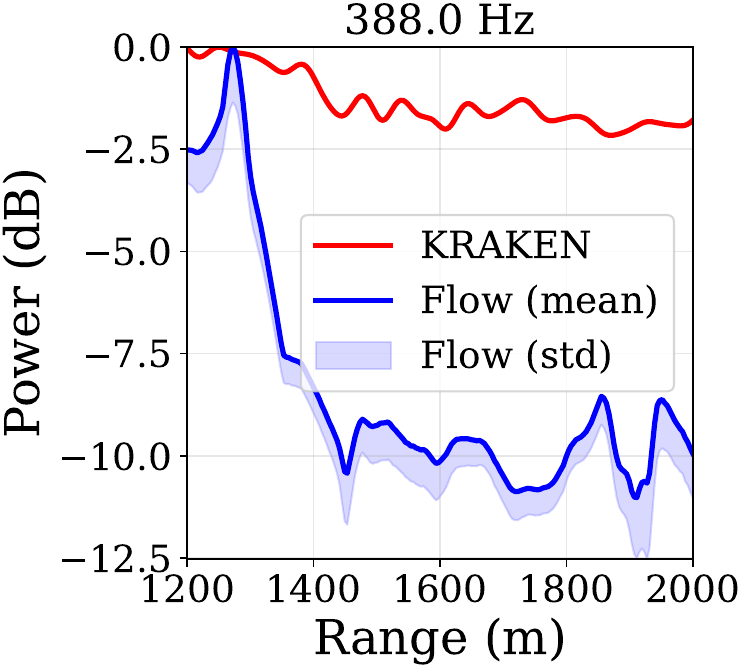}\label{fig:sonar_intmap_388}}
	\caption{Normalized power as a function of range for Event S59 according to the KRAKEN normal-mode program and to the learned latent field at a source depth of 70~m. For the latent field, the mean field $m_{\bm{\theta}}(\V{c})$ and, for readability, only the lower half of the $\pm1$ standard-deviation band (shaded) are shown.}
	\label{fig:sonar_intmap}
\end{figure}

This section complements Section~\ref{M-sec:exp_sonar} of the main paper. Fig.~\ref{fig:sonar_intmap} compares the mean field $m_{\bm{\theta}}(\V{c})$ in~\eqref{M-eq:mean_field} of the main paper, i.e., the mean of the learned latent field, with the power predicted by the KRAKEN normal-mode program~\cite{Por:91} as a function of range for Event S59 and the fixed estimated source depth of 70~m, for the tones at 49, 79, and 112~Hz in Fig.~\ref{fig:sonar_intmap_low}, at 148, 201, and 283~Hz in Fig.~\ref{fig:sonar_intmap_high}, and at 388~Hz in Fig.~\ref{fig:sonar_intmap_388}; the curves in Fig.~\ref{fig:sonar_intmap_low} are also shown in Fig.~\ref{M-fig:sonar_curves} of the main paper. Each curve is normalized to a maximum of 0~dB, so the comparison concerns the range dependence of the power rather than its absolute level. Possible causes of the growing gap between the curves at high frequencies are discussed in Section~\ref{M-sec:exp_sonar} of the main paper.

\FloatBarrier
\section{Ablation Study of the Signal-Power Prior}
\label{app:ablation}
This section provides the details of the ablation study summarized in Section~\ref{M-sec:latent_field_ablation} of the main paper.
We perform a controlled acoustic simulation in which KRAKEN generates the measurements and the propagation conditions are known. This experiment compares different spatial and temporal parameterizations while keeping the measurement model, inference procedure, and object trajectories fixed.
\begin{table*}[htbp]
	\centering
	\caption{Ablation study of the signal-power prior in the controlled environment. GOSPA values with $c=100$\,m, $\alpha=2$, and $p=1$ are given in meters and averaged over $100$ Monte Carlo runs.}
	\label{tab:ablation_study}
	\begin{tabular}{lcccccc}
		\toprule
		\textbf{Method} & \textbf{Temporal} & \textbf{Spatial} &
		\textbf{GOSPA} $\downarrow$ & \textbf{Loc.} $\downarrow$ &
		\textbf{Miss} $\downarrow$ & \textbf{False} $\downarrow$ \\
		\midrule
		Flow-G  & No  & Yes & 22.13 & 12.64 & 4.47 & 5.02 \\
		Flow-B  & Yes & Yes & 23.57 & 13.14 & 4.78 & 5.66 \\
		Flow-T  & Yes & Yes & \textbf{21.44} & 11.78 & 4.53 & 5.12 \\
		Gamma   & No  & Yes & 27.16 & 15.42 & 6.32 & 5.42 \\
		Gamma-R & Yes & No  & 31.92 & 16.27 & 3.78 & 11.87 \\
		\bottomrule
	\end{tabular}
\end{table*}

\paragraph{Controlled environment}
For this ablation study, we simulate a shallow-water waveguide with a depth of approximately 216.5~m and a 21-element VLA. The sound-speed profile and the waveguide and seabed parameters used in KRAKEN, as well as the array geometry, are those of the SWellEx-96 environmental model, which is available on the SWellEx-96 website~\cite{swellex:Env96}. Between two and six moving sources are distributed over ranges from 500 to 1500~m and depths from 20 to 100~m. Each source transmits the same seven tones from 49 to 388~Hz used in the real-data experiment. The time-varying source number and trajectories produce complex acoustic measurements from which both the object states and latent field must be inferred. The motion and birth models are the same as in the acoustic experiment in Section~\ref{M-sec:exp_sonar} of the main paper. Each simulation run comprises $60$ time steps, and the results are averaged over $100$ Monte Carlo runs.

\paragraph{Compared latent fields and priors}
We compare different normalizing flow designs and conventional parametric alternatives:
\begin{enumerate}
	\item \textbf{Flow-G} uses a stationary Gaussian base and spatial context $\V{c}^{\mathrm{S}}_{k,n}$ containing the estimated source position and tone frequency:
	\begin{equation}
		\gamma_{k,n}
		=
		G\!\left(T_{\bm{\theta}}
		(\varepsilon;\V{c}^{\mathrm{S}}_{k,n})\right),
		\qquad
		\varepsilon\sim\mathcal{N}(0,1).
		\label{eq:ablation_flow_g}
	\end{equation}
	\item \textbf{Flow-B} introduces temporal information by shifting the base PDF according to the mean $\mu_{q,k-1}$ and variance $\sigma_{q,k-1}^{2}$ of the preceding posterior of $G^{-1}(\gamma_{k-1,n})$:
	\begin{equation}
		\varepsilon
		\sim
		\mathcal{N}(\mu_{q,k-1},\sigma_{q,k-1}^{2}).
		\label{eq:ablation_flow_b}
	\end{equation}
	\item \textbf{Flow-T} retains the stationary base $\mathcal{N}(0,1)$ but augments the spatial context $\V{c}^{\mathrm{S}}_{k,n}$ with the mean $\mu_{q,k-1}$ and variance $\sigma_{q,k-1}^{2}$ of the preceding posterior of $G^{-1}(\gamma_{k-1,n})$, i.e., with the same statistics that Flow-B uses to shift its base PDF. Temporal dependence is therefore represented by the conditional transformation rather than by a time-varying reference PDF.
	\item \textbf{Gamma} replaces the flow with a context-conditioned Gamma PDF whose parameters are regressed from $\V{c}^{\mathrm{S}}_{k,n}$. The derivations are presented in Section~\ref{A:parametric}.
	\item \textbf{Gamma-R} models the power of each source and tone by a Gamma random walk with transition PDF
	\begin{equation}
		f(\gamma_{k,n}\mid\gamma_{k-1,n})
		=
		\mathrm{Gamma}\!\left(\gamma_{k,n};\,\nu,\,\nu/\gamma_{k-1,n}\right)
		\label{eq:ablation_gamma_r}
	\end{equation}
	where the shape $\nu=100$ is fixed and the rate is obtained by dividing $\nu$ by the preceding power. The predicted power therefore has mean $\gamma_{k-1,n}$ and variance $\gamma_{k-1,n}^{2}/\nu$. This variant has temporal dependence but no explicit spatial context and is representative of a conventional fluctuating-target model \cite{LiaMey:J25}.
\end{enumerate}

\paragraph{Ablation results}
Table~\ref{tab:ablation_study} shows that Flow-T achieves the lowest GOSPA and localization error. Compared with Flow-G, adding temporal posterior statistics to the context reduces GOSPA from 22.13 to 21.44. However, temporal information is not uniformly beneficial: Flow-B increases GOSPA to 23.57. Shifting the base PDF creates a time-varying latent reference on top of which the flow must learn a conditional transport map, adding non-stationarity to the ELBO optimization during runtime. Flow-T instead preserves a fixed Gaussian reference and lets the conditional transformation adapt to temporal information, resulting in a more stable and flexible prior.

The parametric alternatives further demonstrate the value of the flow. The Gamma and Gamma-R priors obtain GOSPA values of 27.16 and 31.92, respectively, compared with 21.44 for Flow-T. A single Gamma PDF cannot represent multimodal or heavy-tailed power statistics caused by shallow-water multipath. Gamma-R performs worst overall, with a GOSPA of 31.92 and a particularly large false-object component of 11.87. Its low miss component reflects persistent power hypotheses rather than better overall localization; without spatial context, the random walk cannot follow rapid position-dependent power fluctuations. Together, these results show that both spatial conditioning and a flexible density model are important, and that temporal information is most effective when supplied through the flow context rather than through a moving base PDF.

\FloatBarrier

\bibliographystyle{IEEEtran}
\bibliography{IEEEabrv,StringDefinitions,SALBooks,SALPapers,temp}